\documentclass[%
 reprint,
 amsmath,amssymb,
 aps,
]{revtex4-2}

\usepackage{graphicx}
\usepackage{dcolumn}
\usepackage{bm}
\usepackage{hyperref}

\usepackage[utf8]{inputenc}

\usepackage{color}
\usepackage{comment}

\newcommand{\ini}{\text{i}}
\newcommand{\fin}{\text{f}}
\newcommand{\mean}[1]{\left\langle #1 \right\rangle}
\newcommand{\st}{\text{s}}
\newcommand{\HCS}{\text{HCS}}
\newcommand{\ext}{\text{ext}}
\newcommand{\vect}{\bm{v}}
\newcommand{\tot}{\text{tot}}

\begin{document}


\title{Optimal control of uniformly heated granular fluids in linear response}

\author{Natalia Ruiz-Pino}
\author{Antonio Prados}%
 \email{prados@us.es}
\affiliation{%
 Física Teórica, Universidad de Sevilla, Apartado de
  Correos 1065, E-41080 Sevilla, Spain
}%


\date{\today}

\begin{abstract}
We present a detailed analytical investigation of the optimal control of uniformly heated granular gases in the linear regime. The intensity of the stochastic driving is therefore assumed to be bounded between two values that are close, which limits the possible values of the granular temperature to a correspondingly small interval. Specifically, we are interested in minimising the connection time between the non-equilibrium steady states (NESSs) for two different values of the granular temperature, by controlling the time dependence of the driving intensity.  The closeness of the initial and target NESSs make it possible to linearise the evolution equations and rigorously---from a mathematical point of view---prove that the optimal controls are of bang-bang type, with only one switching in the first Sonine approximation. We also look into the dependence of the optimal connection time on the bounds of the driving intensity. Moreover, the limits of validity of the linear regime are investigated.
\end{abstract}

\maketitle


\section{Introduction}
\label{sec:intro}

The study of granular media, beyond its own theoretical interest, is particularly important for industrial applications such as improving their transport or storage. Granular materials are discrete clusters of macroscopic particles that exhibit two fundamental features. First, collisions between particles are inelastic, so that energy is not conserved: it monotonically decreases with time if there is no external mechanism that injects energy into the system. Second, thermal energy is many orders of magnitude lower than the characteristic potential energy, making thermal fluctuations largely irrelevant for the behaviour of granular systems~\cite{jaeger_granular_1996}.

In the simplest model for granular fluids, particles are $d$-dimensional smooth hard spheres of mass $m$ that undergo inelastic binary collisions. In each collision, the tangential component of the relative velocity is unchanged, whereas the normal component is reversed and shrunk by a factor $\alpha$, $0\leq\alpha\leq 1$, which is termed the restitution coefficient. Energy is only kinetic and the energy dissipated in each collision is thus proportional to $1-\alpha^2$---the elastic limit corresponds to $\alpha=1$. In the undriven system, after a few collisions per particle the so-called \textit{homogeneous cooling state} (HCS) is reached~\cite{haff_grain_1983,goldshtein_mechanics_1995,brey_homogeneous_1996,huthmann_dynamics_2000,brey_steady-state_2004,brey_scaling_2007}, in which the system remains homogeneous and the granular temperature $T$---basically the average kinetic temperature---monotonically decreases following an algebraic decay, the Haff law~\cite{haff_grain_1983}.

In order to allow the system to reach a stationary state, an energy injection mechanism is needed. A simple but also relevant situation is the uniformly heated granular fluid~\cite{van_noije_velocity_1998,montanero_computer_2000} that we consider throughout this work. Therein, independent white noise forces act on the particles of the granular fluid, the intensity of which is characterised by a parameter $\chi\geq 0$ related to the variance of the stochastic force.  The granular fluid reaches a  non-equilibrium steady state (NESS) in the long-time limit, in which the system remain homogeneous.  Therein, the energy injected by the stochastic thermostat balances---in average---the energy loss in collisions and the value of the granular temperature depends on the intensity of the driving, whereas higher-order cumulants of the velocity are independent thereof. The uniformly heated granular gas has been extensively studied, both its properties at the NESS~\cite{van_noije_velocity_1998,van_noije_randomly_1999,montanero_computer_2000,garcia_de_soria_energy_2009} and its dynamical evolution~\cite{maynar_fluctuating_2009,garcia_de_soria_universal_2012,sanchez-rey_linear_2021}.

Granular systems are intrinsically out-of-equilibrium systems. Their dissipative dynamics entail that their velocity distribution function (VDF) is non-Gaussian, even in the long-time limit in which a hydrodynamic, independent of the initial condition, state is reached. This is true for both the HCS in the undriven case and the NESS in the uniformly heated situation. The non-Gaussianities of the VDF are essential to understand the behaviour of granular fluids and are  incorporated to the picture by implementing a Sonine expansion~\cite{goldshtein_mechanics_1995} of the Enskog-Fokker-Planck equation. This leads to an infinite hierarchy of equations for the cumulants, which is typically closed by introducing the so-called first Sonine approximation: only the fourth cumulant or excess kurtosis $a_2$ is retained---higher order cumulants are neglected. Therein, the granular temperature and the excess kurtosis obey a system of two coupled ordinary differential equations, the accuracy of which for describing the dynamical evolution of the granular fluid has been validated in many works, e.g.~\cite{van_noije_velocity_1998,montanero_computer_2000,garcia_de_soria_energy_2009,garcia_de_soria_universal_2012,prados_kovacs-like_2014,trizac_memory_2014,lasanta_when_2017,sanchez-rey_linear_2021}. In this context, especially relevant are those analysing memory effects such as the Kovacs hump or the Mpemba crossing, in which non-Gaussianities are key to facilitate their emergence~\cite{prados_kovacs-like_2014,trizac_memory_2014,lasanta_when_2017,sanchez-rey_linear_2021}.     

Only very recently has the possibility of controlling the dynamical evolution of granular systems been analysed~\cite{prados_optimizing_2021}. This might be surprising at first sight, since the control of physical systems has been considered for some time in different physical contexts, such as quantum mechanics~\cite{chen_fast_2010,chen_shortcut_2010,deffner_quantum_2017,guery-odelin_shortcuts_2019} and statistical mechanics~\cite{schmiedl_optimal_2007,aurell_optimal_2011,machta_dissipation_2015,martinez_brownian_2016,muratore-ginanneschi_application_2017,van_vu_thermodynamic_2020}. A paradigmatic case of control of a mesoscopic system is that of an optically trapped colloidal particle~\cite{schmiedl_optimal_2007,aurell_optimal_2011,martinez_engineered_2016,plata_optimal_2019,muratore-ginanneschi_application_2017,li_shortcuts_2017,chupeau_engineered_2018,albay_thermodynamic_2019,albay_realization_2020,plata_finite-time_2020}. When the confining potential is harmonic, the time dependence of the stiffness of the trap $\kappa(t)$ can be externally controlled and one aims at optimising the connection between two given equilibrium states, corresponding to different values of the stiffness of the trap---i.e. the colloidal particle is being confined or deconfined. Here, optimising means that some relevant physical observable (irreersible work, entropy production, connection time,\ldots) is minimised.  The time-dependent stiffness $\kappa(t)$ plays the role of the control function---sometimes together with the temperature of the bath, which can be changed in an effective way by adding a random force~\cite{martinez_effective_2013,ciliberto_experiments_2017}. The control problem is greatly simplified by the following three features. First, the initial and target states are equilibrium states, so that their corresponding probability distribution functions (PDFs) are perfectly known. Second, the PDF is Gaussian for all times, so that it is completely characterised by its average and variance. Third, the evolution equations for the average and the variance are exactly solvable in closed form.

%

The delay in posing the problem of controlling granular systems probably stems from the challenging character of the control problem in this case, both at the conceptual and mathematical level. None of the three simplifying features above, holding for the harmonically trapped Brownian particle, is present in granular fluids. First, the initial and target states are NESS, and their PDFs are only approximately known. Second,  the PDF is non-Gaussian for all times. Third, the evolution equations are non-linear and thus not exactly solvable. 
It is interesting to compare the situation in the granular case described above with the one appearing in other paradigmatic system, the Brownian gyrator~\cite{filliger_brownian_2007,argun_experimental_2017,chiang_electrical_2017}. Although the initial and final states are also NESSs in that case, the PDF is Gaussian for all times and the evolution equations for the relevant moments can be exactly solved. Thus, the control problem of this system is simpler, although only non-optimal connections have been worked out, to the best of our knowledge~\cite{baldassarri_engineered_2020}.

One may thus pose the problem of connecting two NESSs of the granular fluid corresponding to different values of the driving intensity $\chi$, $\chi_{\ini}$ and $\chi_{\fin}$, i.e. to different values of the granular temperature $T_{\ini}$ and $T_{\fin}$. The control function here is the intensity of the driving $\chi(t)$. We are interested in the \textit{time optimisation} problem, i.e. to  find the  protocol $\chi(t)$---starting from (and ending at) the desired initial (and target) NESS---that minimises the connection time between the initial and final states. This kind of time optimisation problem is important from a fundamental point of view, and has also relevance for applications. For the connection between equilibrium states, related problems emerge in the optimisation of irreversible heat engines~\cite{plata_building_2020}, the analysis of  the Mpemba effect~\cite{lu_nonequilibrium_2017,lasanta_when_2017,baity-jesi_mpemba_2019,santos_mpemba_2020}, and the optimisation of the relaxation route to equilibrium~\cite{gal_precooling_2020,kumar_exponentially_2020,lapolla_faster_2020}.

The limiting situation in which all the power of the stochastic thermostat is available, i.e. $0\leq \chi<\infty$, was 
investigated in Ref.~\cite{prados_optimizing_2021} within the first Sonine approximation. Despite the challenges mentioned above, the unboundedness of the control makes it possible to give analytical predictions for the connecting time---the evolution equations are heavily simplified in the limiting cases $\chi=0$ and $\chi=\infty$.  

In this work, we analytically investigate the more realistic case in which the driving intensity is bounded between two values, $\chi_{\min}\leq\chi\leq\chi_{\max}$.
In order to make analytical progress, we consider the linear response regime, in which $\chi_{\min}$ and $\chi_{\max}$ are close: this allows us to linearise the evolution equations and make exact---in the linear response limit---predictions for the optimal connecting time as a function of the bounds $(\chi_{\min},\chi_{\max})$. The linearisation of the equations also allows us to employ rigorous mathematical results of optimal control theory (OCT) and to check that the underlying hypothesis are fulfilled, a program that was unattainable in the non-linear case~\cite{prados_optimizing_2021}. Moreover, we also explore the limits of validity of the linear response regime, by taking the double limit $(\chi_{\min}\ll 1,\chi_{\max}\gg 1)$ and comparing the obtained behaviour with those for the non-linear case with unbounded driving~\cite{prados_optimizing_2021}. 

The structure of this paper is as follows. In Sec.~\ref{sec:model}, we put forward the model, write the evolution equations for the temperature and the excess kurtosis, and linearise them around the final NESS. Section~\ref{sec:optimal-control} is devoted to the derivation of the optimal controls, in the sense of minimising the connection time. The trajectories of the temperature and the excess kurtosis---both as functions of time and in the phase plane---for the optimal controls are analysed in Sec.~\ref{sec:trajectories}. The dependence of the minimum connection time on the bounds of the driving is the subject of study of Sec.~\ref{sec:connection-time}. We investigate the limits of validity of the linear response approximation, as the bounds in the driving are loosened, in Sec.~\ref{sec:validity}. Finally, a discussion of the obtained results is presented in Sec.~\ref{sec:discussion}. The Appendices deal with some technicalities that are omitted in the main text.


\section{The model}
\label{sec:model}

Our system is a granular fluid with number density $n$, comprising $N$ $d$-dimensional ($d=2,3$) hard-spheres of mass $m$ and diameter $\sigma$ (hard discs in $d=2$). Specifically, we consider smooth inelastic hard spheres, collisions between them are binary and the post-collisional velocities $(\bm{v}_1',\bm{v}_2')$ are given in terms of the pre-collisional ones $(\bm{v}_1,\bm{v}_2)$ by
\begin{equation}
   {v}_1' = \bm{v}_1 - \frac{1+\alpha}{2}(\bm{v}_{12}\cdot \widehat{\bm{\sigma}})\widehat{\bm{\sigma}}, \quad  \bm{v}_2' = \bm{v}_2 +\frac{1+\alpha}{2} (\bm{v}_{12}\cdot \widehat{\bm{\sigma}})\widehat{\bm{\sigma}} ,
\end{equation}
where $\widehat{\bm{\sigma}}$ is the unit vector along the direction joining the center of the particles and $\alpha$ is the restitution coefficient, $0\leq\alpha\leq 1$. In addition, the system is heated by a stochastic thermostat, i.e. a white-noise force $\bm{F}_i$ independently acts on every particle verifying $\mean{\bm{F}_i(t)}=0$, $\mean{\bm{F}_i(t)\bm{F}_j(t)}=m^2\xi^2\delta_{ij} \delta (t-t')$, $\forall i,j=1,\ldots,N$ and $\forall (t,t')$.

In the first Sonine approximation that we employ thorughout, the system is described  by two variables, the granular temperature $T$ and the excess kurtosis $a_2$. Their definitions in terms of moments of the velocity are
\begin{equation}\label{eq:T-a2-defs}
\mean{v^2}=\frac{d  T}{m}, \quad a_2=\frac{d}{d+2}\frac{\mean{v^4}}{\mean{v^2}^2}-1.
\end{equation}
As stated  in the introduction, the system reaches a NESS in the long-time limit, due to the balance---in average--of the energy input and dissipation. The stationary values of $T$ and $a_2$ are given by
\begin{equation}
     T_{\st}^{3/2} = \frac {m {\xi}^2} {\zeta_0(1+\frac {3}{16}a_2^{\st})}\equiv \chi, \; \zeta_0 = \frac {2n {\sigma}^{d-1}(1-\alpha^2) {\pi}^{\frac {d-1}{2}}}{\sqrt m d \Gamma(d/2)},
 \end{equation}
 \begin{equation}
     a_2^{\st} = \frac {16(1-\alpha)(1-2\alpha^2)} {73+56d-24 d \alpha-105\alpha+30(1-\alpha)\alpha^2}, 
 \end{equation}
Note that $a_2^{\st}$ is independent of the thermostat intensity, as measured by $\chi$, it only depends on $(d, \alpha)$.  From the kinetic equation, the following coupled system of ordinary differential equations (ODEs) are obtained---see e.g.~\cite{garcia_de_soria_universal_2012,prados_kovacs-like_2014},
\begin{subequations}
\begin{equation}
 \label{dt}
     \dot{T} = \zeta_0 \left[\chi (1+ \frac {3} {16} a_2^{\st})-T^{3/2} (1+ \frac {3} {16} {a_2})\right], \quad
 \end{equation}
 \begin{equation}
 \label{da}
     \dot{a_2} = \frac {2 \zeta_0}{T} \left[ (T^{3/2} -\chi)a_2 +BT^{3/2}(a_2^{\st}-a_2)\right],
 \end{equation}
 \end{subequations}
where the parameter $B$ is given by~\cite{garcia_de_soria_universal_2012,prados_kovacs-like_2014}
\begin{align}
     B&=\frac {a_2^{\HCS}}{a_2^{\HCS}-a_2^{\st}}, \\ a_2^{\HCS} &=\frac{16(1-\alpha)(1-2\alpha^{2})} {25
  +2\alpha^{2}(\alpha-1)+ 24d+\alpha(8d-57)};
\end{align}
$a_2^{\HCS}$ is the value of the excess kurtosis in the HCS~\cite{van_noije_velocity_1998,montanero_computer_2000}.



\section{Optimal control in linear response}
\label{sec:optimal-control}

Above, we have considered that the driving intensity of the thermostat $\chi$ is constant. In general, it may be time-dependent, a certain given function of time $\chi(t)$ that determines the externally enforced driving program. Looking at the evolution equations \eqref{dt} and \eqref{da} for $(T,a_2)$ in the light of OCT, this means that $\chi(t)$ is the control function. In this paper, we consider the following control problem: the connection of two NESS, i.e. bringing the system from an initial state $(T_{\ini},a_{2 \ini}=a_2^{\st})$, to a target, final, one $(T_{\fin},a_{2 \fin}=a_2^{\st})$, by engineering a suitable driving program $\chi(t)$. Moreover, we would like to do this connection in the shortest possible time. The case in which all the power of the thermostat is available has been considered in Ref.~\cite{prados_optimizing_2021}. Here, we analyse the more realistic case in which the driving intensity is bounded between two limiting values, $\chi_{\min}\leq\chi(t)\leq\chi_{\max}$, with $\chi_{\min}\geq 0$ and $\chi_{\max}<\infty$.   

In order to solve the control problem analytically, we restrict ourselves to the linear response regime, i.e $\chi_{\min}$ and $\chi_{\max}$ are close---and so are $T_{\ini}$ and $T_{\fin}$.  To look into the dynamics of the system, it is preferable to introduce scaled variables as follows,
\begin{equation}\label{eq:scaled-vars}
      t^* = \zeta_0 T_{\fin}^{1/2}t,\quad T^*= \frac{T}{T_{\fin}},\quad {\chi}^* =\frac{\chi}{T_{\fin}^{3/2}},\quad A_2 =\frac{a_2}{a_2^{\st}}.
\end{equation}
In this way, we have defined dimensionless time $t^*$, granular temperature $T^*$, and driving $\chi^*$; moreover, scaling the excess kurtosis with its steady value simplifies our analysis~\footnote{Both $a_2^{\HCS}$ and $a_2^{\st}$ change sign for $ \alpha=1/ \sqrt{2}$, so that $a_2$ typically changes sign with the inelasticity. On the other hand, the scaled variable $A_2$ always remains positive.}. For these scaled variables, we have the evolution equations
\begin{subequations}\label{eq:evol-eqs-non-linear}
\begin{equation}
\label{dta}
\dot{T}= \chi (1+ \frac {3} {16} a_2^{\st})-T^{3/2} (1+ \frac {3} {16} a_2^{\st} A_2),
\end{equation}
\begin{equation}
\label{daa}
\dot{A_2} = \frac {2}{T} \left[ (T^{3/2} - \chi)A_2 +BT^{3/2}(1-A_2) \right].
\end{equation}
\end{subequations}
We have omitted the superscript $*$ in the dimensionless variables in order to simplify the notation since, from now on, these are the variables used. The term $ \chi (1+ \frac {3} {16} a_2^{\st})$ on the right hand side (rhs) of \eqref{dta} represents the energy injection due to the action of the thermostat, while the term $- T^{3/2} (1+ \frac {3} {16} a_2^{\st} A_2)$ collects the energy losses due to the inelastic collisions. Of course, if $\chi$ is kept constant and equal to its target value, i.e. $\chi(t)=1$ $\forall t\geq 0$, the system reaches the NESS $(T_{\st}=1,A_2^{\st}=1)$ in the long-time limit, consistently with our discussion in the previous section.

In linear response, we thus write
\begin{equation}
T=1 +  \delta T,\quad A_2 = 1 +  \delta A_2,\quad \chi= 1 +  \delta \chi,
\end{equation}
with $\delta T \ll 1$, $\delta A_2 \ll 1$, and $\delta\chi\ll 1$. Note that, to be consistent we must assume that  $\delta \chi_{\min},\delta \chi_{\max} \ll 1$. This allows us to linearise the evolution equations of $T$ and $A_2$
\begin{equation}
\label{eq:sl}
\frac{d}{dt}\begin{pmatrix}
\delta T\\ \delta A_2
\end{pmatrix}= \begin{pmatrix}
\beta \\ -2 \end{pmatrix}\delta \chi + \begin{pmatrix}
-\frac{3}{2}  \beta & 1-\beta\\ 3   & -2B \end{pmatrix} \begin{pmatrix}
\delta T \\ \delta A_2 \end{pmatrix},
\end{equation}
where we have defined 
\begin{equation}
    \beta \equiv 1 + \frac {3} {16} a_2^{\st}.
\end{equation} 
Now it is $\delta\chi$ that plays the role of the control function, $\delta\chi_{\min}\leq\delta\chi\leq\delta\chi_{\max}$.

Once the evolution equations are linearised, the control problem is stated as follows: we would like to bring the system from the initial NESS corresponding to
\begin{equation}\label{eq:ini-conds-deltas}
    \delta T(t=0)=\delta T_{\ini}, \quad \delta A_2(t=0)=0,
\end{equation}
to the target NESS
\begin{equation}\label{eq:fin-conds-deltas}
    \delta T(t_{\fin})=0, \quad \delta A_2(t_{\fin})=0,
\end{equation}
in the minimum possible time $t_{\fin}$. Moreover, the system remains stationary for $t<0$ and $t>t_{\fin}$: this means that for $t<0$ we have prepared the system in the NESS with the initial value of the temperature, by driving it with the corresponding intensity, and that for $t\geq t_{\fin}$ the driving intensity for the target temperature is applied, i.e.
\begin{equation}
    \delta\chi(t)=\delta\chi_{\ini}=\frac{3}{2}\delta T_{\ini}, \; t<0, \quad \delta\chi(t)=0, \; t\geq t_{\fin}.
\end{equation}

Equation~\eqref{eq:sl} is linear in both the variables $(\delta T,\delta A_2)$ and the control function $\delta\chi$, and therefore the rigorous theorems for linear control systems are applicable---see, for example, chapter III of Ref.~\cite{pontryagin_mathematical_1987}. For our specific situation in which $\delta\chi_{\min}\leq\delta\chi\leq\delta\chi_{\max}$, these theorems ensure that the optimal protocol that minimises the connection time $t_{\fin}$ is of bang-bang type with at most one change. That is, $\delta\chi(t)$ is piece-wise continuous, taking either the value $\delta\chi_{\max}$ or $\delta\chi_{\min}$ and presenting, at most, one jump between these two values  in the time window $(0,t_{\fin})$~\footnote{More specifically, this result stems from theorem 10 in section 17 of Ref.~\cite{pontryagin_mathematical_1987}, We check that the hypotheses of this theorem are fulfilled in Appendix~\ref{sec:verifying-hypothesis}.}. This kind of bang-bang optimal protocols arise in different physical situations~\cite{liberzon_calculus_2012,chen_fast_2010,ding_smooth_2020,martikyan_comparison_2020,prados_optimizing_2021}. In general, bang-bang protocols emerge as the optimal ones when Pontryagin's Hamiltonian is linear in the controls---i.e. when the evolution equations are linear in the controls although they may be non-linear in the relevant physical variables~\cite{martinez_brownian_2016,chupeau_thermal_2018,kourbane-houssene_exact_2018,manacorda_lattice_2017}.

To be able to determine the optimal protocol we must distinguish two cases according to the initial temperature, a global cooling process ($T_{\ini}>T_{\fin}=1$, $\delta T_{\ini}>0$) and a global heating process ($T_{\ini}<T_{\fin}=1$, $\delta T_{\ini}<0$):
\begin{itemize}
    \item {For $\delta T_{\ini} >0$, \textit{CH protocol}: In the time window $[0,t_J)$, the driving $\delta\chi$ is set to its minimum value $\delta\chi_{\min}$ (cooling), whereas in the time window $[t_J,t_{\fin})$, it is set to its maximum $\delta\chi_{\max}$ (heating). }
\begin{equation}
\label{ch_pro}
  \delta\chi(t) = \left\{ \begin{array}{ll}
             \delta\chi_{\ini},   & t < 0 ,
             \\ \delta\chi_{\min}, &  0 \leq t < t_{J}, 
             \\\delta\chi_{\max},  &   t_{J}\leq t<t_{\fin},
             \\ 0, &  t\geq t_{\fin}.
             \end{array}
   \right.  
\end{equation}
\end{itemize}
\begin{itemize}
 \item For $\delta T_{\ini}<0$, \textit{HC protocol}: In the time window $[0,t_J)$, the driving is set to its maximum value $\chi_{\max}$, whereas in the time window $[t_J,t_{\fin})$, it is set to its minimum $\chi_{\min}$.
\begin{equation}
\label{hc_pro}
  \chi(t) = \left\{ \begin{array}{ll}
             \delta\chi_{\ini}, & t < 0 ,\\
             \delta\chi_{\max}, & 0  \leq t < t_{J}, \\
             \delta\chi_{\min},  & t_{J}\leq t<t_{\fin},
             \\ 0, & t \geq t_{\fin}.
             \end{array}
   \right.  
\end{equation}
\end{itemize}
The switching time $t_{J}$, $0\leq t_J\leq t_{\fin}$, will be determined later as a function of the parameters of the problem, i.e. as a function of $(\delta\chi_{\ini},\delta\chi_{\min},\delta\chi_{\max})$. As already stated above, the values of $\delta\chi$ for $t<0$ and for  $t \geq t_{\fin}$ ensure that the system starts from the NESS with $T=T_{\ini}$ and, after the application of the bang-bang protocol, remains in the target NESS with $T=T_{\fin}=1$, for both the CH and  HC protocols.

At first, there is no clear reason to assign  the CH protocol to the case $\delta T_{\ini}>0$ and the HC protocol to the case $\delta T_{\ini}<0$. (Aside from the analogy with the full-thermostat-power case analysed in Ref.~\cite{prados_optimizing_2021}.) In order to show that this is indeed the case, one needs to study the behaviour of the trajectories swept by the point in the phase plane $(\delta A_2,\delta T)$. We defer this analysis until Sec.~\ref{sec:trajectories}.

\subsection{$T_{\ini}>T_{\fin}=1$. Cooling-heating bang-bang}\label{sec:C-H}
In this section we integrate the solution of the system in two time windows: the first one, $[0,t_{J})$, when $\delta\chi_{\min}$ is applied, and a second one $[t_{J},t_{\fin})$, when $ \delta\chi_{\max}$ is applied. We also determine the time $t_{J}$  as well as the value of the variables $ \delta T_{J}$ and $ \delta A_{2J}$ at that time. The point $(\delta A_{2J},\delta T_{J})$ constitutes the set of initial conditions for the control system in the second window.

Equation \eqref{eq:sl} is inhomogeneous, due to the term proportional to $\delta\chi$ on its rhs. (It is only homogeneous when $\delta\chi=0$, i.e. when the control is set to the constant value $\chi_{\fin}=T_f^{3/2}$ corresponding to the final temperature.) Over each time window, $\delta\chi(t)=\delta\chi_{\ext}$, where the subscript ``$\ext$''  includes both bangs, $\delta\chi_{\ext}=\delta\chi_{\min}$ (first  window) and  $\delta\chi_{\ext}=\delta\chi_{\max}$ (second window). The inhomogeneity can be thus understood as the system being relaxing towards the NESS corresponding to $\delta\chi_{\ext}$. Let us denote by $T_{\ext}=1+\delta T_{\ext}$ the temperature corresponding to the NESS reached when the system is driven with constant intensity $\chi_{\ext}=1+\delta\chi_{\ext}$. Since the steady value of the excess kurtosis does not depend on the driving intensity, we have only to subtract
\begin{equation}
   \delta T_{\ext} = (\chi_{\ext})^{2/3}-1 =\frac {2} {3} \delta
    \chi_{\ext}+O(\delta\chi_{\ext})^2
\end{equation}
from $\delta T$ to make the system homogeneous. Thus, we define 
\begin{equation}
 \delta\tilde{T}\equiv \delta T - \delta T_{\ext}.
 \label{var_hom}
\end{equation}

The homogeneous system for $\delta\tilde{T}$ and $\delta A_2$ reads
\begin{equation}
   \frac{d}{dt} \begin{pmatrix}  {\delta\tilde{T}}\\
 {\delta A_{2}} \end{pmatrix}= \begin{pmatrix}-\frac {3} {2} \beta & 1-\beta \\
3 & -2B \end{pmatrix} \begin{pmatrix}  {\delta\tilde{T}}\\
 {\delta A_{2}} \end{pmatrix}.
\end{equation}
The eigenvalues $(-\lambda_1,-\lambda_2)$ and eigenvectors $(\vect_1,\vect_2)$ of this system are given by 
\begin{subequations}\label{eq:eigenvalues}
 \begin{align}
     \lambda_1 & =\frac {1}{2}\left(\frac{3}{2} \beta+ 2B+k\right)>0, & \vect_1 = \frac{1}{6}\begin{pmatrix} 2\lambda_2-3\beta \\
    6 \end{pmatrix},\\
    \lambda_2 &=\frac {1}{2}\left(\frac{3}{2} \beta+ 2B-k\right)>0, & \vect_2 = \frac{1}{6} \begin{pmatrix} 2\lambda_1-3\beta \\
    6 \end{pmatrix},
 \end{align}
\end{subequations}
where we have introduced the parameter
\begin{equation}\label{eq:k-def}
k\equiv \lambda_1-\lambda_2= \sqrt{\left(\frac{3}{2}\beta-2B\right)^2+12(1-\beta)}>0. 
\end{equation} 
With the definitions above, both $\lambda_1$ and $\lambda_2$ are positive, and $\lambda_1>\lambda_2$. 

Now we can write the solution in both time windows, separately because $\delta\tilde{T}$ is different over each one. In the first step of the bang-bang, $t \in [0,t_{J})$, where $\delta\chi(t)=\delta\chi_{\min}$,
\begin{equation}
 \label{s1}
\begin{pmatrix}  {\delta\tilde{T}}\\
{\delta A_{2}} \end{pmatrix} =\begin{pmatrix}  {\delta T-\frac{2}{3}\delta\chi_{\min}}\\
{\delta A_{2}} \end{pmatrix}= C_1 \vect_1 e^{-\lambda_1 t}+C_2 \vect_2 e^{-\lambda_2 t}.
\end{equation}
In the second step of the bang-bang, $t \in [t_{J},t_{\fin})$, where $\delta\chi(t)=\delta\chi_{\max}$,
\begin{equation}
\label{s2}
\begin{pmatrix}  {\delta\tilde{T}}\\
{\delta A_{2}} \end{pmatrix} = \begin{pmatrix}  {\delta T-\frac{2}{3} \delta\chi_{\max}}\\
{\delta A_{2}} \end{pmatrix} =  C_3 \vect_1 e^{-\lambda_1 t}+C_4 \vect_2 e^{-\lambda_2 t}.
\end{equation}
The constants $(C_1,C_2,C_3,C_4)$ are obtained by imposing the initial conditions in each time window. For $t=0$ we have the initial condition \eqref{eq:ini-conds-deltas}, which determines $C_1$ and $C_2$, 
\begin{equation*}
         C_2 =-C_1=3\frac {\delta T_{\ini}-\delta T_{\min}}{k}=2\frac{ \delta\chi_{\ini}-\,\delta \chi_{\min}}{k}. 
\end{equation*}
The point at the final time $t_{J}$ of the first time window is
\begin{subequations}\label{eq:cooling-CH}   
\begin{align}
    \delta T_{J} &= \delta\tilde{T}_J + \delta T_{\min}\nonumber \\
    & = C_1 \left[\vect_1(1) e^{-\lambda_1 t_{J}} -\vect_2(1)  e^{-\lambda_2 t_{J}}\right] +\frac {2} {3} \delta \chi_{\min},\\
         \delta A_{2J}& =   C_1 \left(e^{-\lambda_1 t_{J}} - e^{-\lambda_2 t_{J}}
         \right).
\end{align}
\end{subequations}
The initial conditions for Eq.~\eqref {s2} are supplied by $( \delta T_{J}, \delta A_{2J})$. Therefore, we can obtain $(C_3,C_4)$ as a function of the switching time $t_J$, 
\begin{equation}\label{eq:CH-sol-part1}
\begin{pmatrix}  {\delta T_{J} - \delta T_{\max}}\\ {\delta A_{2J}} \end{pmatrix} = C_3 \vect_1 e^{-\lambda_1 t_{J}}
         +C_4 \vect_2 e^{-\lambda_2 t_{J}}.
\end{equation}
Note that $(C_3,C_4)$ also depend on the bounds of the driving $(\delta\chi_{\min},\delta\chi_{\max})$ through $\delta T_{\min}$ and $\delta T_{\max}$. By imposing that we have to reach the target state, i.e. Eq.~\eqref{eq:fin-conds-deltas}, we write
\begin{equation}\label{eq:CH-sol-part2}
\begin{pmatrix}  {-\delta T_{\max}}\\
{0} \end{pmatrix}=C_3 \vect_1 e^{-\lambda_1 t_{\fin}}+C_4 \vect_2 e^{-\lambda_2 t_{\fin}}.
\end{equation}
Equations \eqref{eq:CH-sol-part1} and \eqref{eq:CH-sol-part2} are four equations for the four unknowns $(C_3,C_4,t_{J},t_{\fin})$, which thus provide us with the solution to the control problem. Solving for $C_3$ and $C_4$,
\begin{equation}
    C_3= \frac{2\delta\chi_{\max}}{ k}\!\!\left(\frac{\frac {\delta\chi_{\tot}}{\delta \chi_{\ini}-\delta \chi_{\min}}e^{\lambda_1 t_{J}}-1}{
        \frac{\delta\chi_{\tot} }{\delta \chi_{\ini}-\delta \chi_{\min}}e^{\lambda_2 t_{J}}-1}  \right)^{\!\!\!\frac{\lambda_1}{k}}\!\!\!\!,
        \; C_4 =-C_3 e^{-k t_{\fin}}.
\end{equation}
We have introduced the total amplitude of the allowed interval for the driving,
\begin{equation}
    \delta\chi_{\tot}\equiv \delta\chi_{\max}-\delta\chi_{\min}=\chi_{\max}-\chi_{\min}>0.
\end{equation}
In this way, the final time $t_{\fin}$ is given as a function of the switching time $t_J$,
\begin{align}
        t_{\fin}= &\frac{1}{k} \ln {\left(\frac  {\frac {\delta\chi_{\tot}}{\delta\chi_{\ini}-\delta\chi_{\min}}e^{\lambda_1 t_{J}}-1}{\frac {\delta\chi_{\tot}}{\delta\chi_{\ini}-\delta\chi_{\min}}e^{\lambda_2 t_{J}}-1} \right)} \nonumber \\
        =& t_J+\frac{1}{k}\ln {\left(\frac  {1-\frac{\delta\chi_{\ini}-\delta\chi_{\min}}{\delta\chi_{\tot}}e^{-\lambda_1 t_{J}}}{1-\frac{\delta\chi_{\ini}-\delta\chi_{\min}}{\delta\chi_{\tot}}e^{-\lambda_2 t_{J}}} \right)}\, ,
        \label{1}
\end{align}
which is in turn given by the solution of the implicit equation
     \begin{align}
    \delta\chi_{\tot} &\left(1-\frac{\delta\chi_{\ini}-\delta\chi_{\min}}{\delta\chi_{\tot}}
    e^{-\lambda_2 t_{J}}
      \right)^{\!\!\frac{\lambda_1}{k}} \nonumber \\&= \delta \chi_{\max} \left(1-\frac{\delta\chi_{\ini}-\delta\chi_{\min}}{\delta\chi_{\tot}}
      e^{-\lambda_1 t_{J}}
      \right)^{\!\!\frac{\lambda_2}{k}}.
       \label{1.2}
    \end{align}
The set of equations \eqref{1} and \eqref{1.2} provides an analytical solution for the minimum connection time $t_{\fin}$ in the CH protocol, which is valid in the linear approximation we are considering in this paper.


\subsection{$T_{\ini}<T_{\fin}=1$. Heating-cooling bang-bang}\label{sec:H-C}

Let us now start from an initial state with $\delta T_{\ini}<0$. The analysis of this case is similar to that just carried out for $\delta T_{\ini}>0$, but the order of the bangs is reversed. In the first time window, $[0,t_J)$, the maximum driving $\delta\chi_{\max}$ is applied, whereas in the second time window, $[t_J,t_{\fin})$, the minimum driving $\delta\chi_{\min}$ is applied. Therefore, the homogenisation procedure for the temperature $\delta T$ is also reversed. In the first window $t \in [0,t_{J})$ we have 
\begin{equation}
 \label {s3}
\begin{pmatrix}  {\delta\tilde{T}}\\
{\delta A_{2}} \end{pmatrix} =\begin{pmatrix}  {\delta T-\frac{2} {3} \delta\chi_{\max}}\\
{\delta A_{2}} \end{pmatrix}=  C_5 \vect_1 e^{-\lambda_1 t}+C_6 \vect_2 e^{-\lambda_2 t},
\end{equation}
whereas in the second window $t \in [t_{J},t_{\fin})$ it is
\begin{equation}
\label {s4}
\begin{pmatrix}  {\delta\tilde{T}}\\
{\delta A_{2}} \end{pmatrix} = \begin{pmatrix}  {\delta T-\frac{2} {3}\delta\chi_{\min}}\\
{\delta A_{2}} \end{pmatrix}= C_7 v_1 e^{-\lambda_1 t}+C_8 v_2 e^{-\lambda_2 t}.
\end{equation}
The initial conditions are given by Eq.~\eqref{eq:ini-conds-deltas}. Inserting them into Eq.~\eqref {s3}, we get 
\begin{equation}
             C_5 = -C_6= 3\frac{\delta T_{\max}-\delta T_{\ini}} {k}=2\frac {\delta \chi_{\max}- \delta\chi_{\ini}}{k},
\end{equation}   
The evaluation of Eq.~\eqref{s3} at time $t_{J}$ gives the initial condition for the second time window. Taking into account the difference in the $\delta\tilde{T}$ variables in Eqs.~\eqref{s3} and \eqref{s4}, due to our switching the value of the driving intensity at $t=t_J$, one gets $(C_7,C_8)$ in terms of $t_J$, and also of the bounds $(\delta\chi_{\min},\delta\chi_{\max}$), in complete analogy with the CH protocol. Also, at the final time $t_{\fin}$ one must impose that the system reaches the target NESS, i.e. Eq.~\eqref{eq:fin-conds-deltas}, which provides the two extra equations needed to determine the switching time $t_J$ and the connection time $t_{\fin}$ as functions of the system parameters.  The result is
 \begin{equation}
       C_7= \frac{2 \delta\chi_{\min}}{k}\!\!\!\left(\frac{\frac {\delta\chi_{\tot}}{\delta\chi_{\max}-\delta\chi_{\ini}}e^{\lambda_1 t_{J}}-1} {\frac {\delta\chi_{\tot}  }{\delta\chi_{\max}-\delta\chi_{\ini}}e^{\lambda_2 t_{J}}-1} \right)^{\!\!\!\frac{\lambda_1}{k}}\!\!\!\!,
         \;   C_8=-C_7 e^{-k t_{\fin}},
    \end{equation}
for $(C_7,C_8)$ in terms of $(t_J,t_{\fin})$,
\begin{align}
       &t_{\fin}= \frac {1} {k} \ln {\left(\frac {\frac {\delta\chi_{\tot}}{\delta\chi_{\max}-\delta \chi_{\ini}}e^{\lambda_1 t_{J}}-1}  {\frac {\delta\chi_{\tot} }{\delta\chi_{\max}-\delta \chi_{\ini}} e^{\lambda_2 t_{J}}-1} \right)}\\ &=t_J+\frac{1}{k}\ln {\left(\frac  {1-\frac{\delta\chi_{\max}-\delta\chi_{\ini}}{\delta\chi_{\tot}}e^{-\lambda_1 t_{J}}}{1-\frac{\delta\chi_{\max}-\delta\chi_{\ini}}{\delta\chi_{\tot}}e^{-\lambda_2 t_{J}}} \right)}\,,
       \label{2} 
    \end{align}
for the minimum connection time in terms of the switching time, and the following implicit equation,
\begin{align}
      \delta \chi_{\tot}  &\left(1-\frac{ \delta\chi_{\max}-\delta\chi_{\ini}} {\delta\chi_{\tot}}e^{-\lambda_2 t_{J}}\right)^{\frac {\lambda_1}{k}} \nonumber \\&=
       -\delta\chi_{\min}\left(1-\frac{\delta \chi_{\max}-\delta\chi_{\ini}} {\delta\chi_{\tot}}e^{-\lambda_1 t_{J}}\right)^{\frac {\lambda_2}{k}},
       \label{2.2}
 \end{align}
for $t_J$. In complete analogy with the CH case, the set of equations \eqref{2} and \eqref{2.2} gives the minimum connection time for the HC protocol in the linear approximation. Note that exchanging $\delta\chi_{\min}\leftrightarrow\delta\chi_{\max}$ (which entails $\delta\chi_{\tot}\to-\delta\chi_{\tot}$) leads from Eqs.~\eqref{2} and \eqref{2.2} to Eqs.~\eqref{1} and \eqref{1.2}---and vice versa.

Above we have derived analytical expressions for different physical variables of interest, for both the CH and HC bang-bang protocols, in the linear response approximation. More specifically, we have (i) the complete description of the trajectory followed by the system in the phase plane, i.e. the time evolution of the point $(\delta A_2(t),\delta T(t))$, and (ii) the switching time $t_J$ and the minimum connection time $t_{\fin}$. The linear response approximation have allowed us to obtain analytical predictions as functions of all the relevant physical parameters: not only of the initial temperature $T_{\ini}$, as measured by $\delta\chi_{\ini}=3\delta T_{\ini}/2$, but also of the bounds of the driving $(\chi_{\min},\chi_{\max})$, as measured by $(\delta\chi_{\min},\delta\chi_{\max})$. Therefore, it is interesting to inspect the behaviour of the obtained expressions as a function of $(\delta\chi_{\ini},\delta\chi_{\min},\delta\chi_{\max})$, in order to understand the response of the system to the optimal control designed.


\section{Trajectories for the temperature and the excess kurtosis}
\label{sec:trajectories}

\begin{figure} 
  \begin{minipage}{1\linewidth}
    \includegraphics[width=1\linewidth]{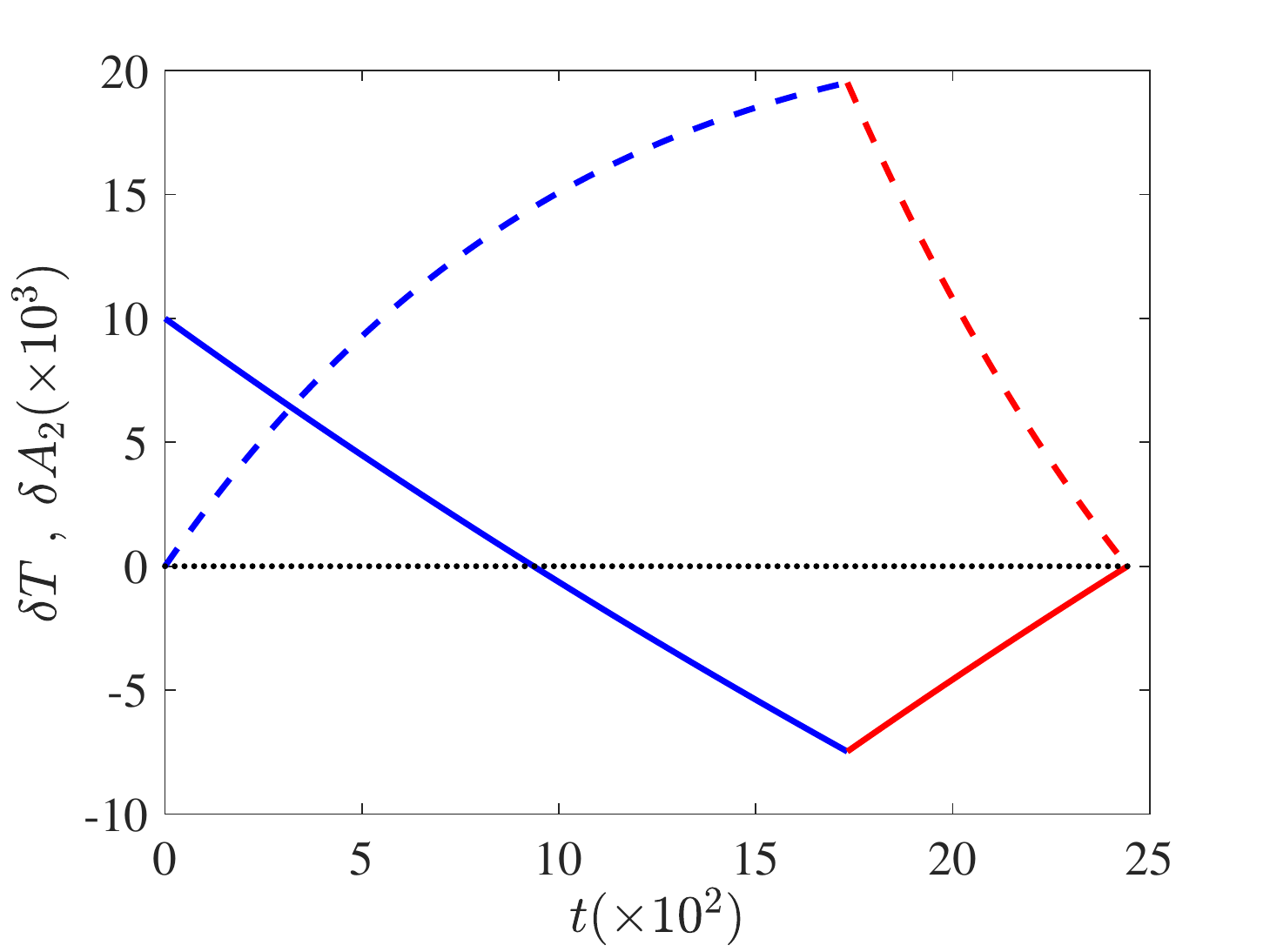} 
  \end{minipage} 
  \begin{minipage}{1\linewidth}
    \includegraphics[width=1\linewidth]{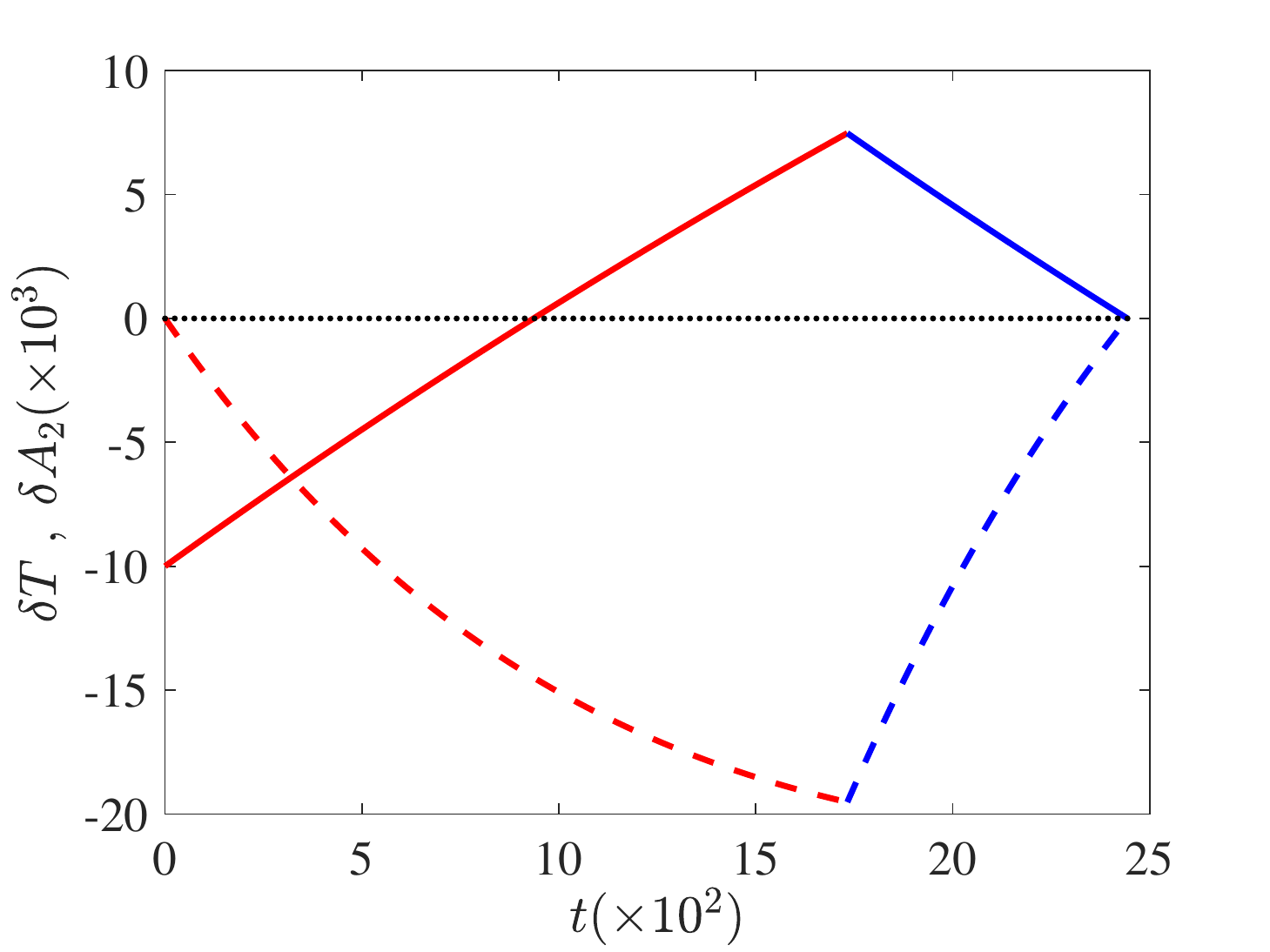} 
  \end{minipage} 
 \caption{Time evolution of the temperature and the excess kurtosis. Specifically, we plot $\delta T$ (solid line) and $\delta A_2$ (dashed line), both for the CH protocol (upper panel) and for the HC protocol (lower panel). Dotted line represents the horizontal axis. The bounds for the driving intensity are $\delta\chi_{\max}=0.1$ and $\delta\chi_{\min}=-0.1$, and the initial temperature is $\delta T_{\ini}=0.01$ for CH and $\delta T_{\ini}=-0.01$ for HC. The evolution under the action of $\delta\chi_{\min}$ is shown in blue and  the evolution under $\delta\chi_{\max}$ in red. Other parameters are $\alpha=0.9$ and $d=2$.
 }
\label{evoltch}
\end{figure}

In this section, we look into the trajectories of the temperature and the excess kurtosis, to understand the need for a two-step bang-bang protocol on a physical basis. The time evolution of $\delta T$ and $\delta A_2$ is presented in Fig.~\ref{evoltch}, for both the CH  and the HC  cases ($\delta T$ solid lines, $\delta A_2$ dashed lines). First, let us analyse the CH protocol (upper panel), i.e. $\delta T_{\ini}>0$ (also $\delta\chi_{\ini}>0$). Therein, $\delta T$ relaxes to $\delta T_{\min}$ under the action of $\delta\chi_{\min}$ in the time window $[0,t_J)$ (without reaching it, since the relaxation at constant driving lasts for an infinite time). Simultaneously, $\delta A_2$ starts to increase from its steady value, equal to zero, because Eq.~\eqref{eq:sl} implies that
\begin{equation}
   \left. \frac{d}{dt}\delta A_2 \right|_{t=0^+}=-2\delta\chi_{\min}+3\delta T_{\ini}=2\left(\delta \chi_{\ini}-\delta \chi_{\min}\right),
\end{equation}
which is non-negative because $\delta\chi_{\min}\leq 0 \leq \delta\chi_{\ini}$. (Otherwise the connection of the two NESS would be impossible, as rigorously proved in the next section---and in agreement with physical intuition.) A decrease (an increase) in the granular temperature makes the VDF separate from (closer to) the Gaussian shape, i.e the scaled excess kurtosis $A_2$ correspondingly increases (decreases). Once the target temperature $T_{\fin}=1$ is reached inside this first time window, i.e. the temperature curve crosses the horizontal axis $\delta T=0$, the action of $\delta\chi_{\min}$ cannot be interrupted by setting the thermostat intensity to unity,  because $\delta A_2 > 0$ and the system is not in the target NESS. This is why we must let $\delta T$ continue to drop to a value $\delta T_J$ such that $\delta T_J<0$, associated with a kurtosis value $\delta A_{2J}>0$. This point $(\delta A_{2J},\delta T_J)$ is determined by the condition that, at the end of the subsequent relaxation with $\delta\chi_{\max}$ in the time window $[t_J,t_{\fin})$, $\delta A_2$ and $\delta T$ must simultaneously reach  their target value (zero). Second, we analyse the HC case (lower panel), the discussion is completely analogous and thus we summarise it in the following. In the first time window $[0,t_J)$ with $\delta\chi_{\max}$, the horizontal axis $\delta T=0$ is crossed at some time smaller than $t_J$ but it is necessary to continue applying $\delta\chi_{\max}$ to overshoot it, since  $\delta A_2 < 1$ for that time and the system has not reached the target NESS. Once more, the point $(\delta A_{2J},\delta T_J)$ is determined by the condition that, at the end of the second time window with $\delta\chi_{\min}$, both variables simultaneously vanish.

\begin{figure}
     \begin{minipage}{1\linewidth}
    \includegraphics[width=1\linewidth]{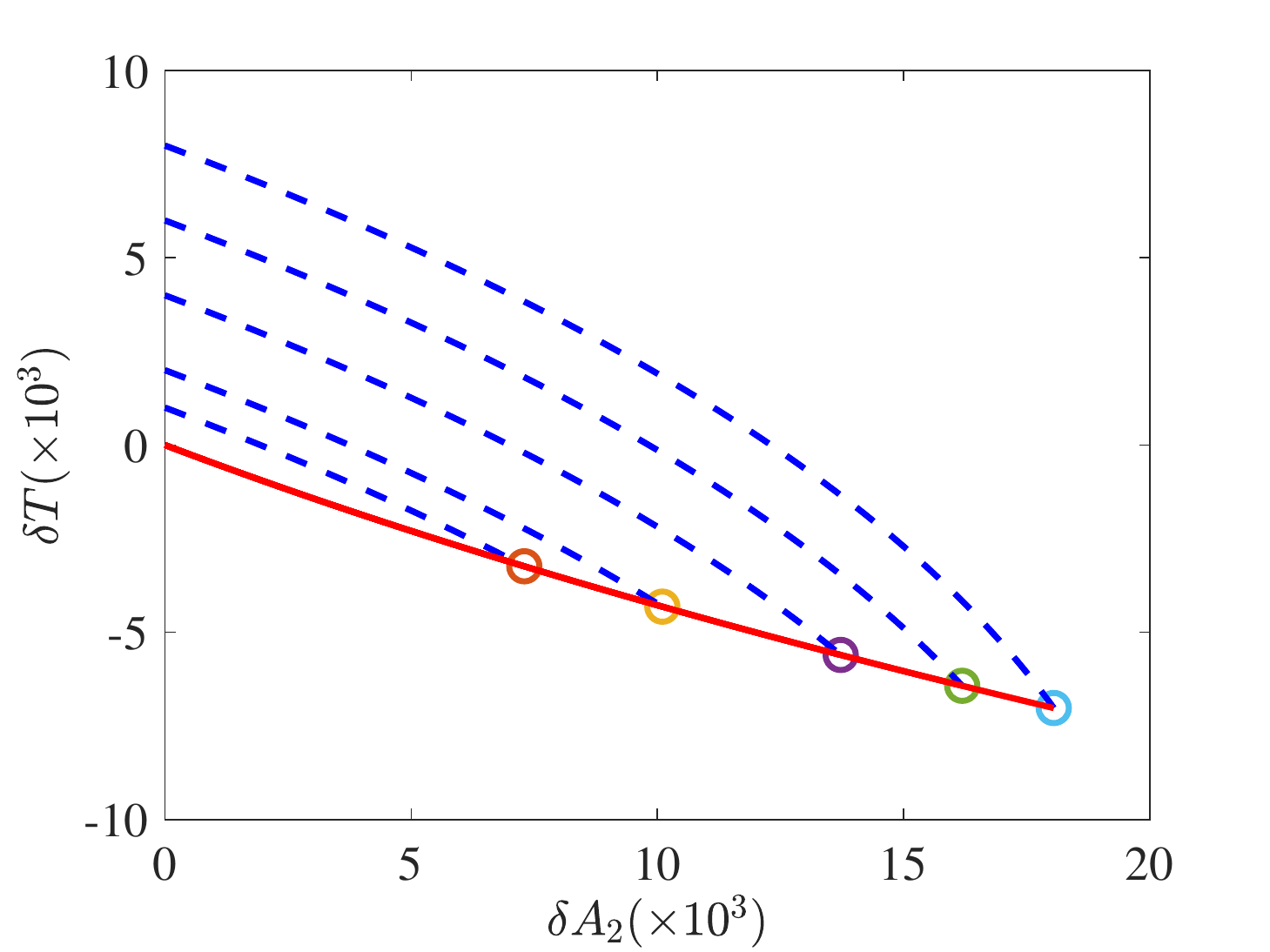}
\end{minipage}
     \begin{minipage}{1\linewidth}
    \includegraphics[width=1\linewidth]{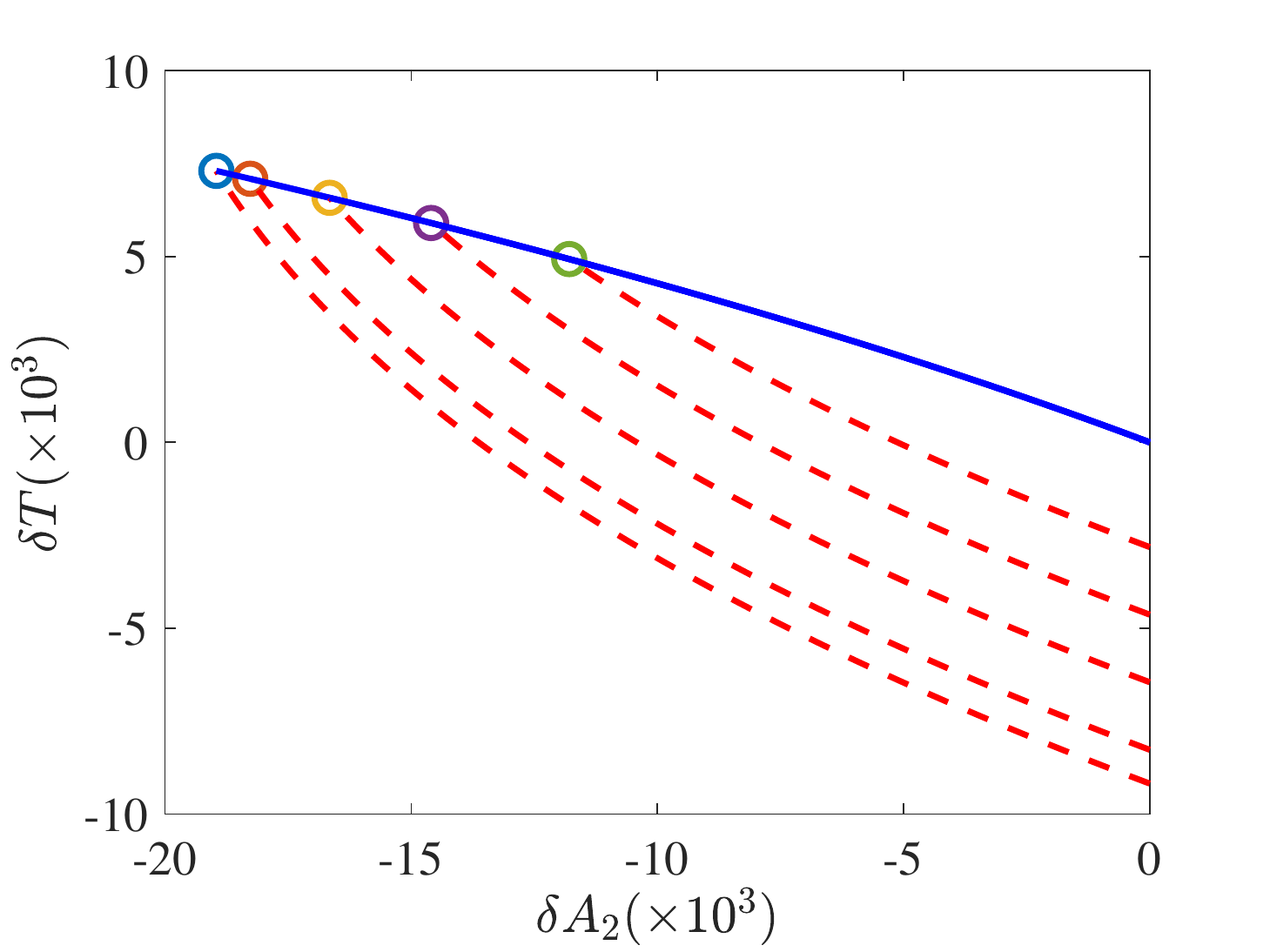}
\end{minipage}
    \caption{Phase plane trajectories. The CH case is illustrated in the upper panel and the HC case in the lower panel. Several trajectories are shown for different initial temperatures $\delta T_{\ini}\in [0,0.01]$ ($\delta T_{\ini}\in [-0.01,0]$) for the CH (HC) protocol. The remainder of the system parameters are the same as in Fig.~\ref{evoltch}. In each panel, the solid line (red in the upper, blue in the lower) represents the second part of the phase trajectory, arriving at the target NESS---the origin $(\delta A_2=0,\delta T=0)$. As in the previous figure, red (blue) lines correspond to $\delta\chi_{\max}$ ($\delta\chi_{\min}$). Again in each panel, the dashed lines represent the first part of the phase trajectory, starting from the initial points $(0,\delta T_{\ini})$. These curves end up at the points $(\delta A_{2J},\delta T_{J})$, marked with circles, at which the dashed and solid lines intersect.
    }
    \label{fases}
\end{figure}

The need of a two-step protocol, and the order of the bangs, can also be understood---maybe more clearly---by looking at the trajectories in phase space. The trajectories of the phase space $(\delta A_2,\delta T)$ for the CH and the HC cases are shown in Fig.~\ref{fases}. The target state is the origin $(0,0)$, so the optimal trajectories must end up thereat. Since the optimal protocols are of bang-bang type, with at most one switch, there are two possibilities: the system approaches the origin following either the heating curve with $\delta\chi_{\max}$ (red solid line in the upper panel) or the cooling curve with $\delta\chi_{\min}$ (blue solid line in the lower panel). These two curves are uniquely defined, because the origin is not a fixed point of the evolution equations for $\delta\chi_{\max}$, nor for $\delta\chi_{\min}$. 
The initial NESS does not lie on either of these two curves---they do not contain any NESS apart from the target state $(0,0)$, thus the necessity of having a two-step bang-bang is clear. Recall that, for the linear case, there is a theorem ensuring that there is at most one switching.

In Appendix~\ref{sec:order-bangs}, we rigorously show that the CH (HC) protocol is the one making it possible to connect the initial NESS with $\delta T_{\ini}>0$ ($\delta T_{\ini}<0)$. 


\section{Minimum connection time as a function of the bounds in the driving intensity}\label{sec:connection-time}

This section is devoted to studying the behaviour of the minimum connection time $t_{\fin}$ (and also of the switching time $t_J$) as a function of the bounds in the driving intensity. The analysis is carried out for both the CH ($\delta T_{\ini}>0$) and HC ($\delta T_{\ini}<0$) protocols.  We will use the variables without $'\delta'$ in order to keep the discussion clearer. 

A first question that naturally arises is the range of values of $\chi_{\min}$ and $\chi_{\max}$ allowing to connect the initial and target states. In the non-linear regime and in the limit case $(\chi_{\min}=0,\chi_{\max}= \infty)$, 
it is always possible to connect two NESS corresponding to temperatures $T_{\ini}$ and $T_{\fin}$~\cite{prados_optimizing_2021}. However, it is not obvious at all that this is possible when not all the power of the thermostat is available, i.e. in our case with bounds in the driving: $\chi_{\min}>0$ and $ \chi_{\max}<\infty$.  For example, given $T_{\ini}>1$, it is unclear whether there appears some change in the behaviour of the connecting time when the upper bound $\chi_{\max}$ crosses the value $\chi_{\ini}>1$. Accordingly with our approach throughout, we intend to study this problem within the linear response approximation. 

\subsection{CH protocol}

\begin{figure} 
  \begin{minipage}[b]{1\linewidth}
    \includegraphics[width=1\linewidth]{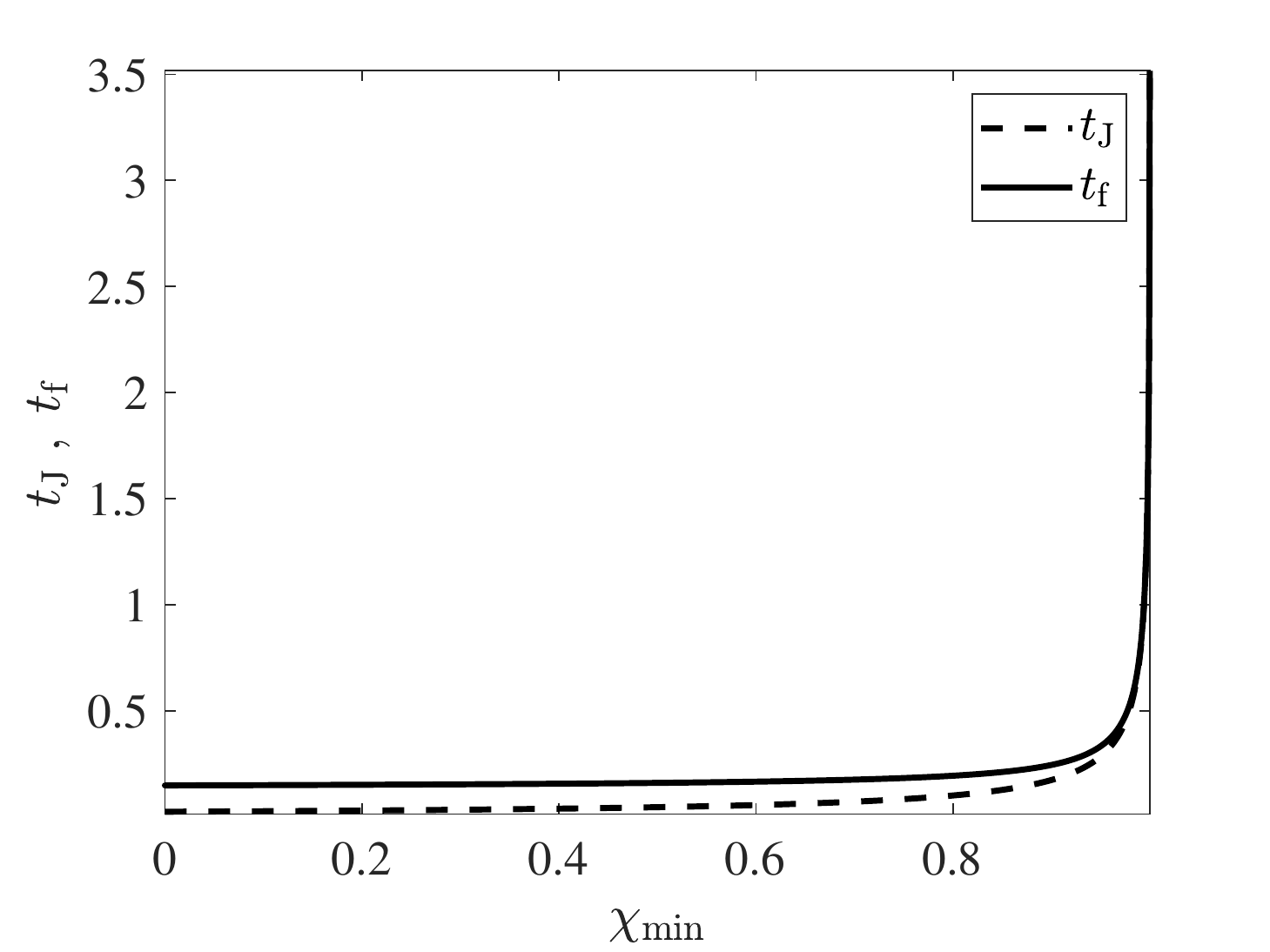} 
  \end{minipage} 
   \hfill
  \begin{minipage}[b]{1\linewidth}
    \includegraphics[width=1\linewidth]{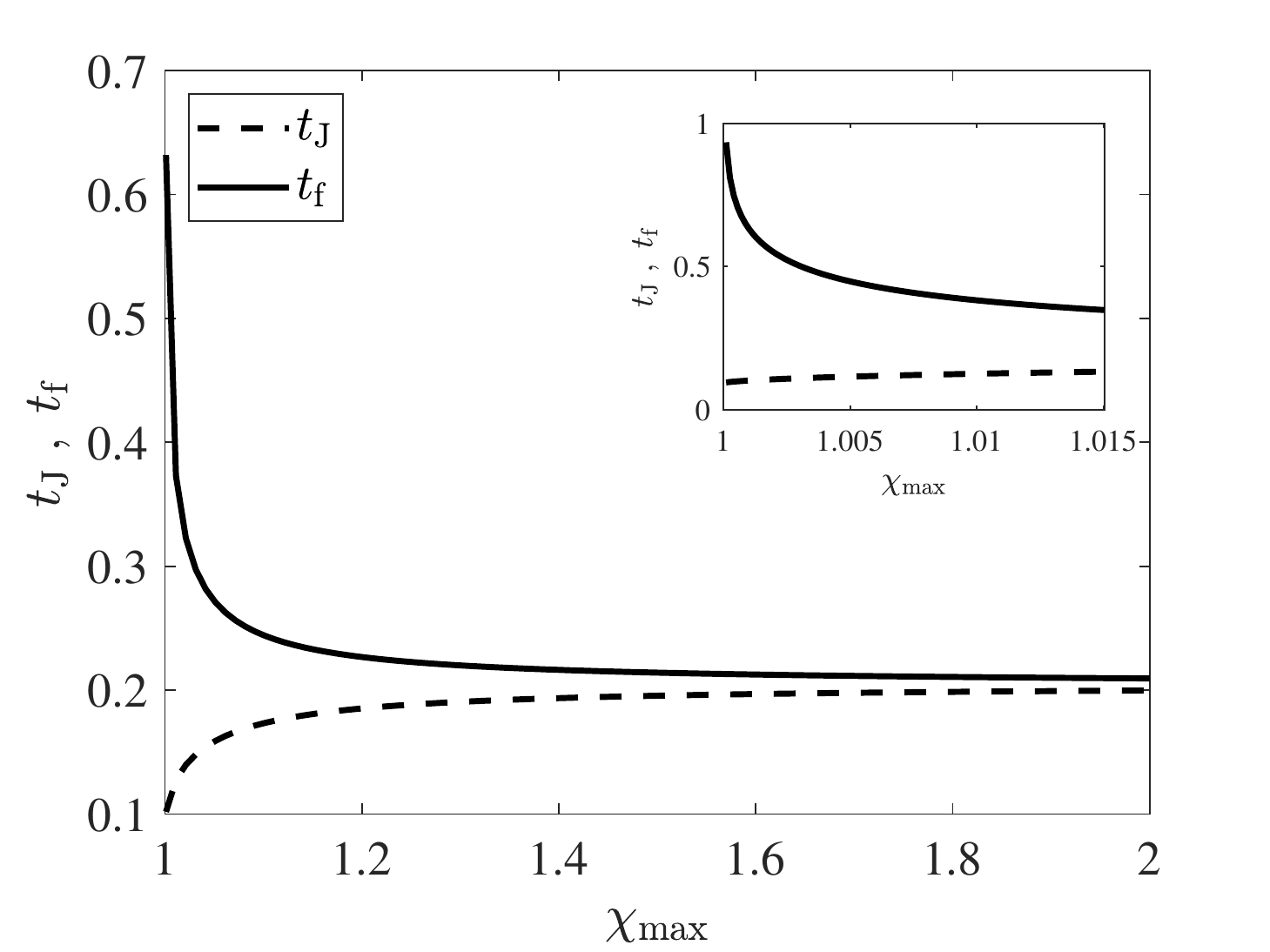} 
  \end{minipage}
   \caption{Switching time $t_J$ and minimum connection time $t_{\fin}$ as functions of the thermostat limit values for the CH protocol. Specifically, we have chosen the initial temperature $T_{\ini}=1.01>1$. In the upper panel, $t_J$ (dashed line) and $t_{\fin}$ (solid line) are plotted as functions of the lower bound $\chi_{\min}$, for a fixed value of the upper bound, namely $\chi_{\max}=1.1$. In the lower panel, they are plotted as functions of the upper bound $\chi_{\max}$, for a fixed value of the upper bound, namely $\chi_{\min}=0.9$. Additional parameters are $\alpha=0.9$ and $d=2$. There are no qualitative changes for other values of $(\alpha,d)$, aside from an increase of the connecting time as $\alpha$ decreases. The inset shows a zoom of the panel for $1\leq\chi_{\max}\leq\chi_{\ini}$, $\chi_{\ini}=1.015$ for $T_{\ini}=1.01$.}
	\label{fig:tj-tf-ch}
\end{figure}

First, we consider the CH protocol, $T_{\ini}>1$ or $\delta T_{\ini}>0$. Figure~\ref{fig:tj-tf-ch} illustrates the dependence of $t_J$ and $t_{\fin}$ on the bounds in the driving. Fixing the value of $\chi_{\max}$, we can look into their behaviour as functions of $\chi_{\min}$ (upper panel). As the cooling capacity of the thermostat decreases, i.e. as $\chi_{\min}$ increases, the minimum connection time $t_{\fin}$ increases. This is logical, since the class of admissible control functions is being shrunk and the optimal connection thus lasts longer. Also, the switching time $t_J$ increases: the cooling step of the bang-bang must be longer to compensate for the decrease of cooling power. Both times diverge in the limit as $\chi_{\min}\to 1^-$, where the cooling power of the thermostat is vanishingly small and thus the cooling step of the bang-bang process takes an infinite time. Now we fix the value of $\chi_{\min}$ and study the behaviour as functions of $\chi_{\max}$ (lower panel). Analogously, as the heating capacity of the thermostat decreases, i.e. as $\chi_{\max}$ decreases, $t_{\fin}$ increases, because the class of admissible controls becomes smaller. On the other hand, the behaviour of $t_J$ is reversed, $t_J$ increases with $\chi_{\max}$. This is also logical, since the first step of the bang-bang is the cooling one and, as the heating capacity of the thermostat is increased, the cooling step must take a longer time. In this case, it is only $t_{\fin}$ that diverges in the limit as $\chi_{\max}\to 1^+$. The lack of heating capacity makes the duration of the second (heating) step diverge, since the time needed to relax towards $T_{\fin}=1$ is infinity for a constant value of the driving $\chi=\chi_{\fin}=1$. There is no change of behaviour in the connection time when $\chi_{\max}$ crosses the value $\chi_{\ini}$, the driving intensity corresponding to the initial value of the temperature. This is neatly observed in the inset, where a zoom of the graph for drivings $\chi_{\max}\in[1,\chi_{\ini}]$ is plotted.

An important point is the divergence of the connection time as $\chi_{\min}\to 1^-$ (for fixed $\chi_{\max}$), and as $\chi_{\max}\to 1^+$ (for fixed $\chi_{\min}$). Therefore, if $\chi_{\fin}=1$ lies outside the interval $[\chi_{\min},\chi_{\max}]$, the target NESS is unreachable. In other words, the bounds in the driving must verify  $(\chi_{\max}\geq 1,\chi_{\min}\leq 1)$, i.e. $(\delta\chi_{\max}\geq 0,\delta\chi_{\min}\leq 0)$, to make it possible to connect the initial and target states. In other words, $T_{\fin}$ must belong in the interval $[T_{\min},T_{\max}]$. In fact, it is possible to rigorously show that the connecting time only diverges when either $\chi_{\min}\to 1^-$ or $\chi_{\max}\to 1^+$, see Appendix \ref{sec:limit-values} for details.

\subsection{HC protocol}

\begin{figure}
  \begin{minipage}[b]{1\linewidth}
    \includegraphics[width=1\linewidth]{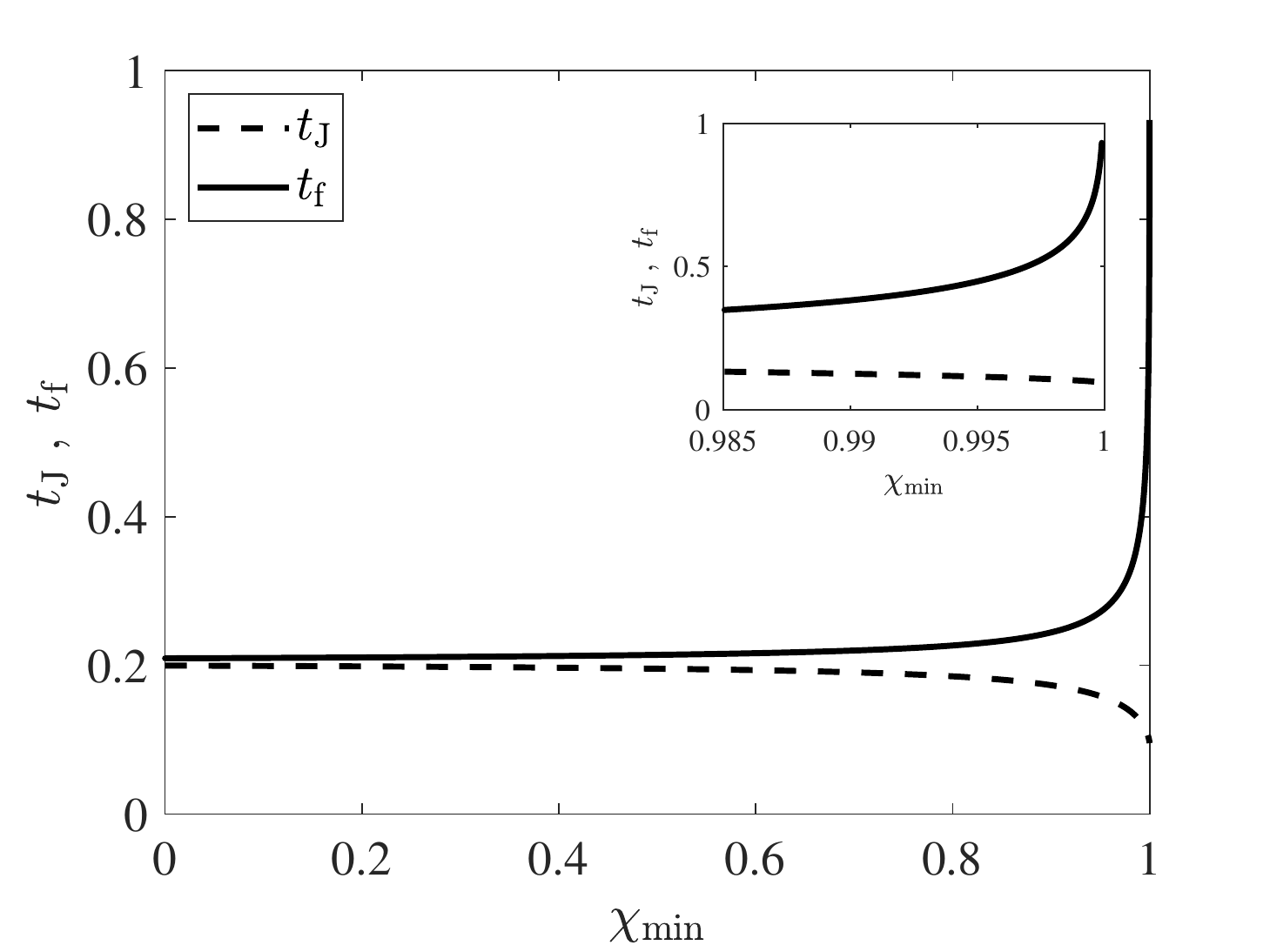} 
  \end{minipage}
  \hfill
  \begin{minipage}[b]{1\linewidth}
    \includegraphics[width=1\linewidth]{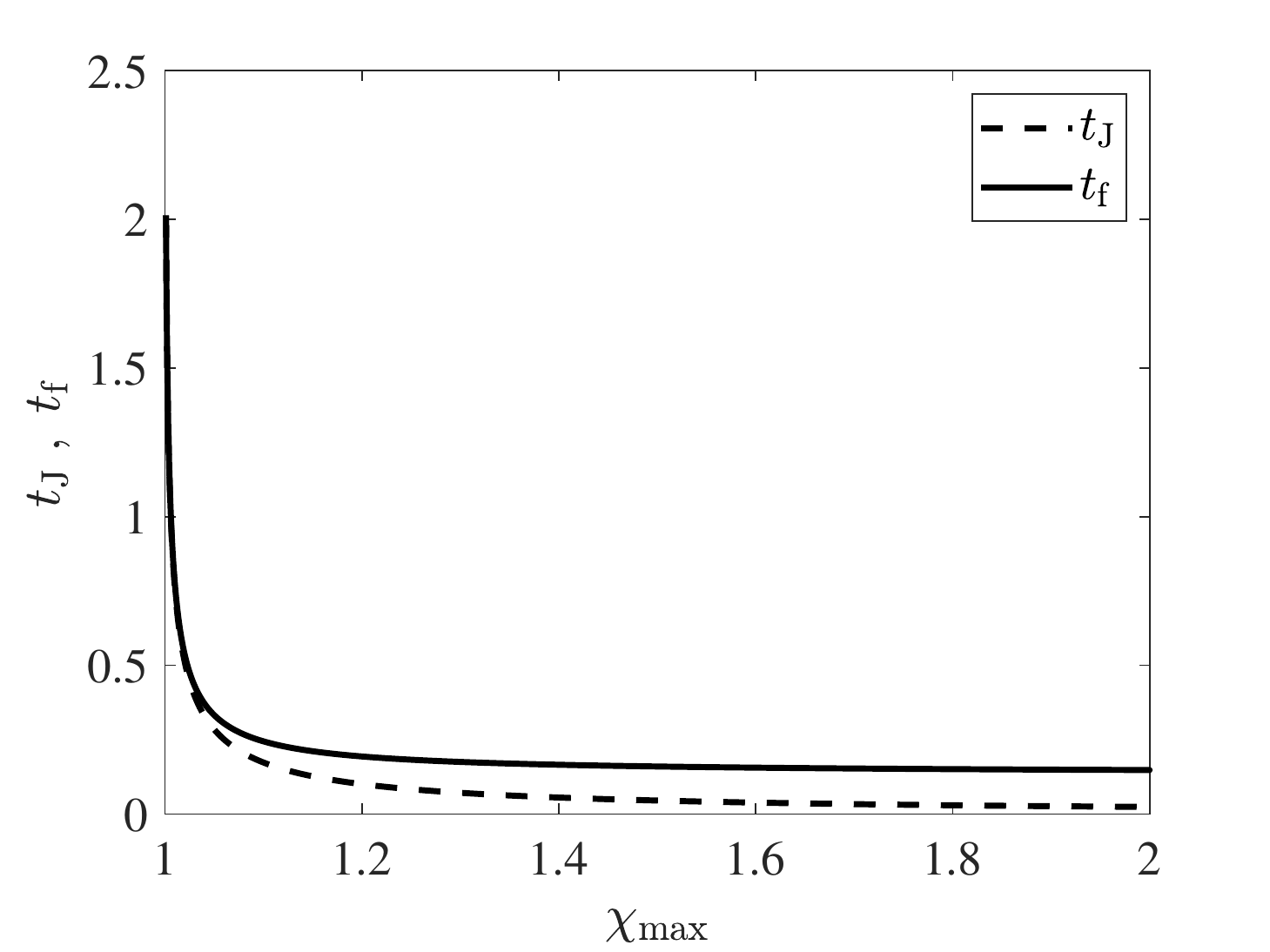} 
  \end{minipage} 
  \caption{Switching time $t_J$ and minimum connection time $t_{\fin}$ as functions of the thermostat limit values for the HC case. The initial temperature is now $T_{\ini}=0.99<1$. The reminder of the parameters are the same as in Fig.~\ref{fig:tj-tf-ch}. Again, the upper (lower) panel shows $t_J$ (dashed) and $t_{\fin}$ (solid) as functions of $\chi_{\min}$ ($\chi_{\max}$), for a fixed value of $\chi_{\max}=1.1$ ($\chi_{\min}=0.9$). The inset in the upper panel shows a zoom of the graph for $\chi_{\ini}\leq\chi_{\min}\leq 1$, $\chi_{\ini}=0.985$ for $T_{\ini}=1.01$, showing that there is no change of behaviour when $\chi_{\min}$ crosses $\chi_{\ini}$.}
	\label{fig:tj-tf-hc}
\end{figure}

Figure~\ref{fig:tj-tf-hc} illustrates the situation for the HC protocol ($T_{\ini}<1$ or $\delta T_{\ini}<0$). Note that the panels are basically the horizontal reflections of those in Fig.~\ref{fig:tj-tf-ch}, with the roles of $\chi_{\min}$ and $\chi_{\max}$ being exchanged. Therefore, the line of reasoning for physically understanding the observed behaviours is completely similar to the one in the previous section, and it will not be repeated here. We would only like to highlight the increase of the minimum connection time as the bounds become tighter, due to the shrinking of the set of admissible control functions, and its divergence for  $\chi_{\min}\to 1^-$ (fixed $\chi_{\max}$) and $\chi_{\max}\to 1^+$ (fixed $\chi_{\min}$), which marks the impossibility of reaching a target state with temperature $T_{\fin}=1$ lying outside the interval $[T_{\min},T_{\max}]$.


\section{Validity of the linear response approximation}
\label{sec:validity}

The results obtained and analysed in the previous sections are quite general. On the one hand, we have  derived expressions for the relevant physical quantities as functions of the bounds in the driving intensity $(\chi_{\min}, \chi_{\max})$ [or $(\delta\chi_{\min},\delta\chi_{\max})$]. On the other hand, the linear reponse approximation limits the results, since we have assumed that the system remains always close to the target NESS. Therefore, it is relevant to investigate the possible validity of our results beyond the strictly linear framework. 

In Ref.~\cite{prados_optimizing_2021}, it was  shown that 
the minimum connection time in the non-linear case---for a full-strength thermostat $0\leq\chi<\infty$, which we denote here by $t_{\fin}^{n\ell}$, is given by
\begin{equation}
	\label{eq:tf-teo-unbounded}
	    t_{\fin}^{n\ell} \sim   \left(\frac{2} {3B}\right)^{1/2} |\delta \chi_{\ini}|^{1/2}, \quad |\delta \chi_{\ini}|\ll 1.
\end{equation}
when the initial and final states are close---as expressed by the condition $|\delta \chi_{\ini}|\ll 1$~\footnote{It must be remarked that the non-dimensionalisation of time in Ref.~\cite{prados_optimizing_2021}, $t^*=\zeta_0 T_{\ini}^{1/2} t$, differs from ours in Eq.~\eqref{eq:scaled-vars} by a factor $\sqrt{T_{\ini}/T_{\fin}}$. This factor does not affect the lowest order asymptotic expression in Eq.~\eqref{eq:tf-teo-unbounded}, since the introduced corrections are higher-order.}. Actually, Eq.~\eqref{eq:tf-teo-unbounded} does not have to hold for the linear case developed in this paper, because we are considering that the driving intensity $\chi$ is restricted to a small interval, $\chi_{\min}=1+\delta\chi_{\min}\leq \chi=1+\delta\chi \leq \chi_{\max}=1+\delta\chi_{\max}$. (Recall that $\delta\chi_{\max}\geq 0$ whereas $\delta\chi_{\min}\leq 0$.)  Notwithstanding, we may progressively separate the bounds from unity and compare our linear response predictions with Eq.~\eqref{eq:tf-teo-unbounded}. More specifically, it is  interesting to take the limit $ \chi_{\min}\to 0$ and $\chi_{\max}\to\infty$ and analyse the possible convergence of our minimum connection time (for both the CH and HC cases) to the time given by Eq.~\eqref{eq:tf-teo-unbounded}.
\begin{figure}
	\centering
	\includegraphics[width = 1\linewidth]{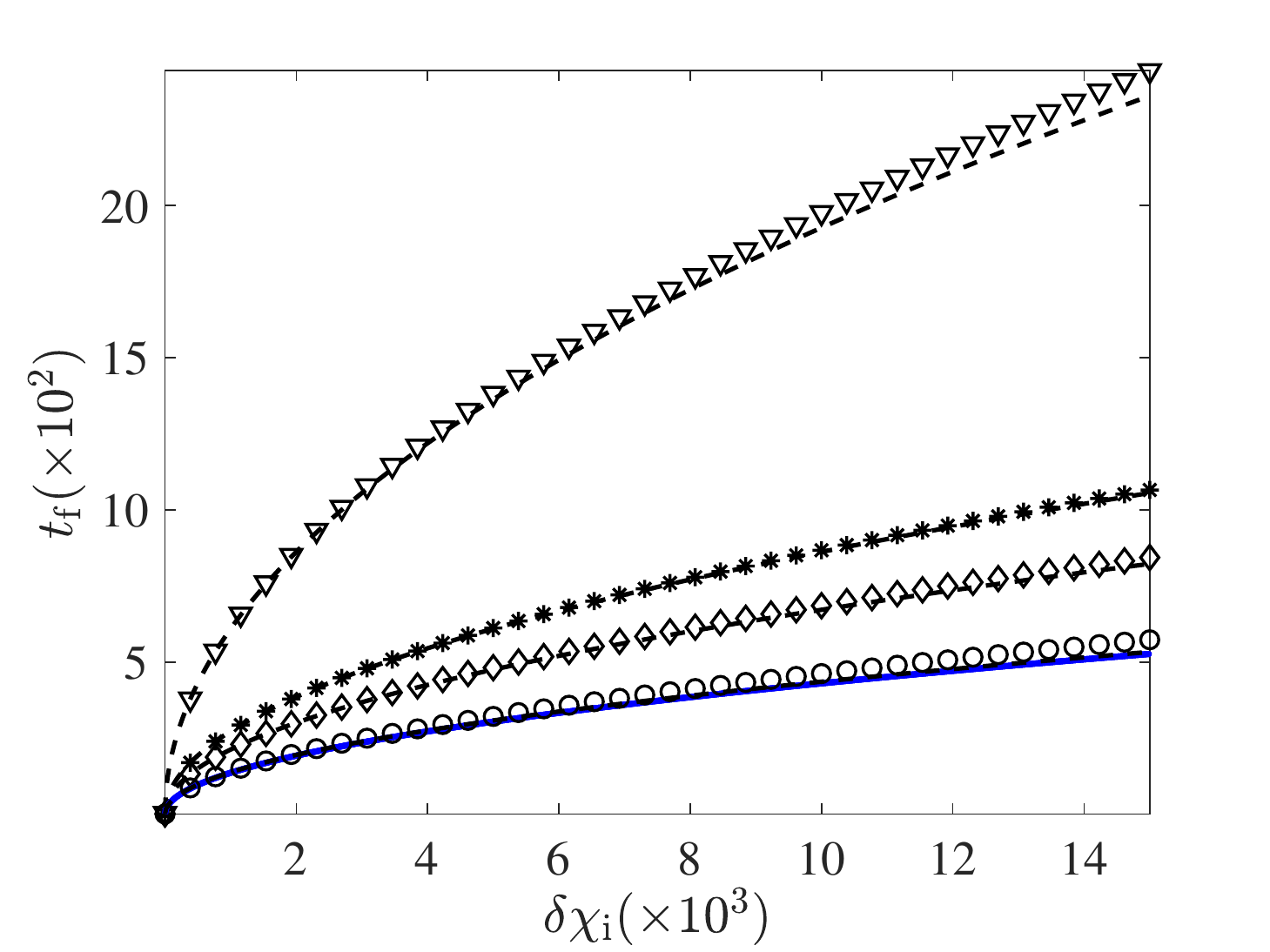}
	\caption{Minimum connection time $t_{\fin}$ versus the initial control $\delta \chi_{\ini}$ for the CH case.  Symbols represent the linear response prediction for $t_{\fin}$, as  given by Eq.~\eqref{1}, for different values of the bounds (from top to bottom: $\delta \chi_{\max}=0.1$ and $\delta \chi_{\min}=-0.1$ (triangles), $\delta \chi_{\max}=0.3$ and $\delta \chi_{\min}=-0.3$ (stars), $\delta \chi_{\max}=1$ and $\delta \chi_{\min}=-0.7$ (diamonds), $\delta \chi_{\max}=99$ and $\delta\chi_{\min}=-0.99$) (circles). Dashed lines correspond to Eq.~\eqref{eq:tf-short-approx} for each case, which shows the soundness of this approximate expression. The solid line corresponds to Eq.~\eqref{eq:tf-teo-unbounded}, which is basically superimposed with the dashed line for $\delta \chi_{\max}=99$ and $\delta \chi_{\min}=-0.99$. Other parameters are $\alpha=0.9$ and $d=2$.}
	\label{fig:comp_t_CH}
\end{figure}
\begin{figure}
	\centering
	\includegraphics[width = 1\linewidth]{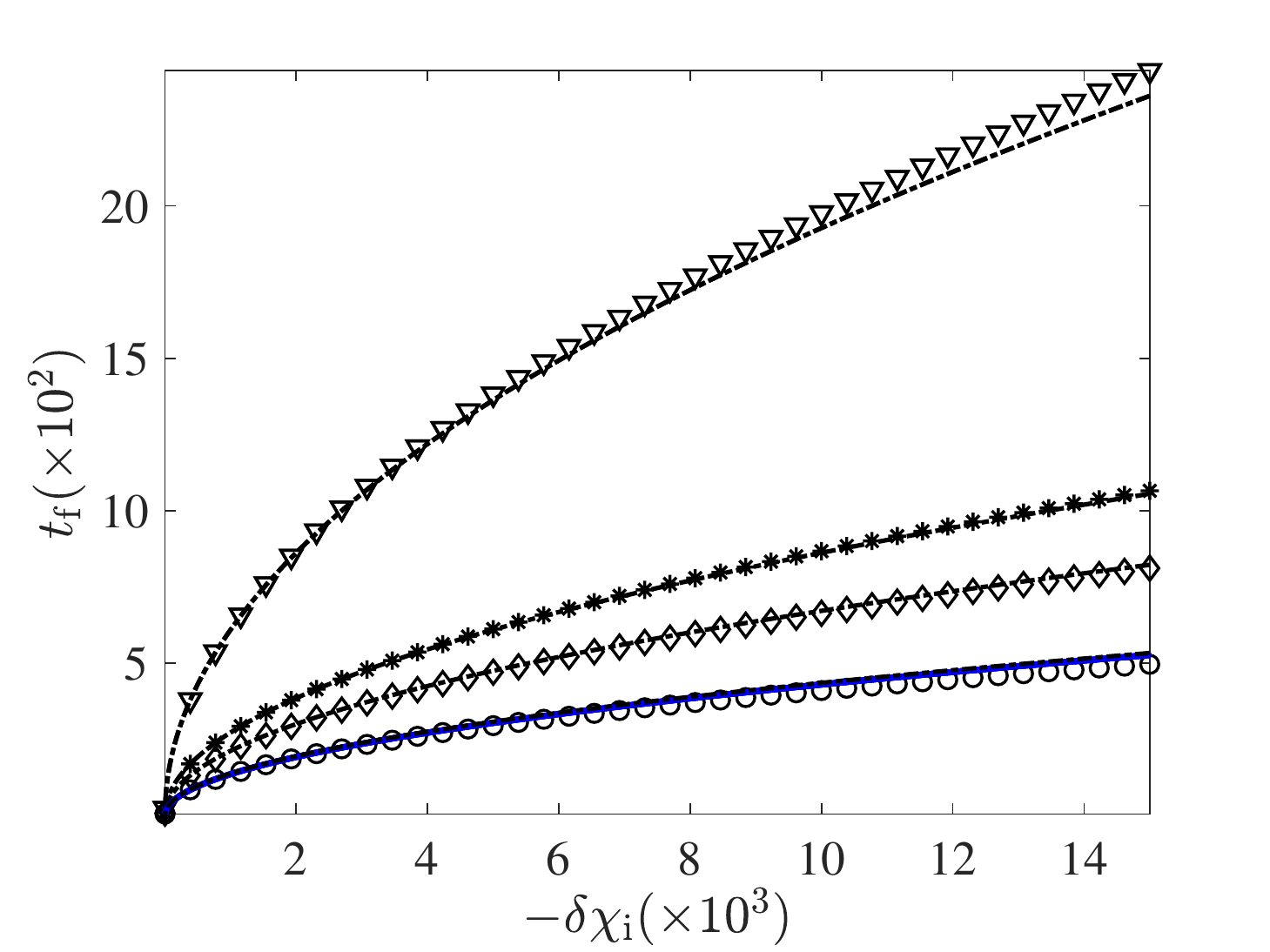}
	\caption{Minimum connection time $t_{\fin}$ versus the initial control $\delta \chi_{\ini}$ for the HC case. The line code is the same as in Fig.~\ref{fig:comp_t_CH}. Again, the solid line corresponding to Eq.~\eqref{eq:tf-teo-unbounded} is basically superimposed with the linear response prediction for the further from unity bounds. Once more, $\alpha=0.9$ and $d=2$.}
	\label{fig:comp_t_HC}
\end{figure}

In order to further explore this possible convergence, we have approximated $t_{\fin}$ to first order in $\delta \chi_{\ini}$. In Appendix~\ref{sec:tf-approximation}, it is shown that for short connecting times $t_{\fin}\ll 1$ one has
\begin{equation}
	\label{eq:tf-short-approx}
	    t_{\fin} \sim \left(\frac{2}{\lambda_1 \lambda_2}\right)^{1/2} \left(\frac{\delta\chi_{\tot}|\delta \chi_{\ini}|}{-\delta \chi_{\min}\delta \chi_{\max}}\right)^{1/2}, 
\end{equation}
where 
\begin{equation}
   \lambda_1\lambda_2=3\beta B+3(\beta-1)=3B\left(1+\frac{3}{16}a_2^{\st}\right)+\frac{9}{16}a_2^{\st}. 
\end{equation}
Note that Eq.~\eqref{eq:tf-short-approx} is valid to the lowest order in $t_{\fin}$---terms of the order of $t_{\fin}^2$ have been neglected---but no assumption has been made with regard to $\delta\chi_{\min}$ and $\delta\chi_{\max}$.

In order to make the comparison between the non-linear (with full-strength thermostat) and linear (with bounds in the driving) expressions above, we have represented the minimum connecting time for different values of the bounds $\chi_{\min}$ and $ \chi_{\max}$ in Figs.~\ref{fig:comp_t_CH} and \ref{fig:comp_t_HC}---for the CH and HC cases, respectively. Therein, we show the linear response expressions Eq.~\eqref{1} (CH case) and~\eqref{2} (HC case), together with the approximate linear expression \eqref{eq:tf-short-approx}, and the non-linear expression \eqref{eq:tf-teo-unbounded} for a full-strength thermostat. We observe how the times given by Eqs.~\eqref{1} and \eqref{2}, as well as their approximations \eqref{eq:tf-short-approx}, rapidly converge  to Eq.~\eqref{eq:tf-teo-unbounded} as the bounds separate from unity. This convergence is qualitatively similar in the CH and HC cases, there are no significant differences between them up to this point.

\begin{figure} 
    \includegraphics[width=1\linewidth]{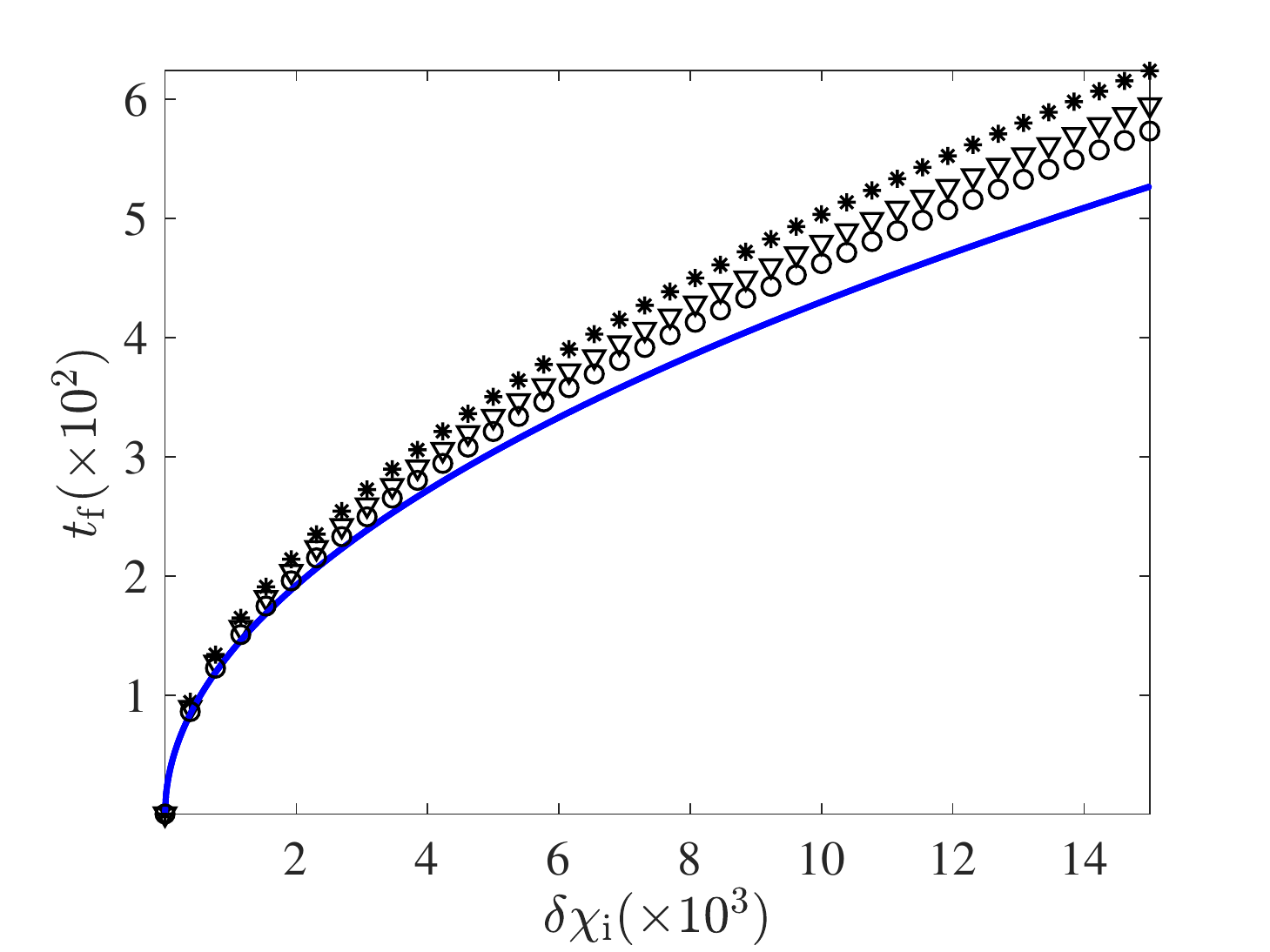} 
  	\caption{Convergence to the non-linear expression \eqref{eq:tf-teo-unbounded} as the bounds go to more extreme values, for the CH case. We plot the connection time $t_{\fin}$ versus the initial control $\delta \chi_{\ini}$ for $\alpha=0.9$ and $d=2$. Several sets of data are plotted: (i) the non-linear expression \eqref{eq:tf-teo-unbounded} (blue solid line), and (ii) the linear prediction, as given by  Eq.~\eqref{1}, for several values of the bounds, namely $\delta \chi_{\max}=9$ and $\delta  \chi_{\min}=-0.9$ (stars), $\delta \chi_{\max}=19$ and $\delta \chi_{\min}=-0.95$ (triangles), $\delta \chi_{\max}=99$ and $\delta \chi_{\min}=-0.99$ (circles). The time $t_{\fin}$ given by Eq.~\eqref{1} converges to that in Eq.~\eqref{eq:tf-teo-unbounded} ``from above''.
}
\label{fit-compar-CH}
  \end{figure}

Let us look at the convergence towards the non-linear expression \eqref{eq:tf-teo-unbounded} in more detail. First, we consider the CH case in Fig.~\ref{fit-compar-CH}, which can be seen as a zoom of Fig.~\ref{fig:comp_t_CH}---for values of the bounds such that the linear time is close to the non-linear one. It is clearly observed that, as the bounds of the driving separate from unity, the linear response prediction approaches the non-linear expression \eqref{eq:tf-teo-unbounded} ``from above'': the connection times of the linear theory are longer than those for the non-linear case. This is consistent, since Eq.~\eqref{eq:tf-teo-unbounded} was obtained for the largest possible set of control functions, i.e. $(\chi_{\min}=0,\chi_{\max}=\infty)$: the loosest the restrictions on the control functions are, the shortest the minimum connection time is.

\begin{figure} 
    \includegraphics[width=1\linewidth]{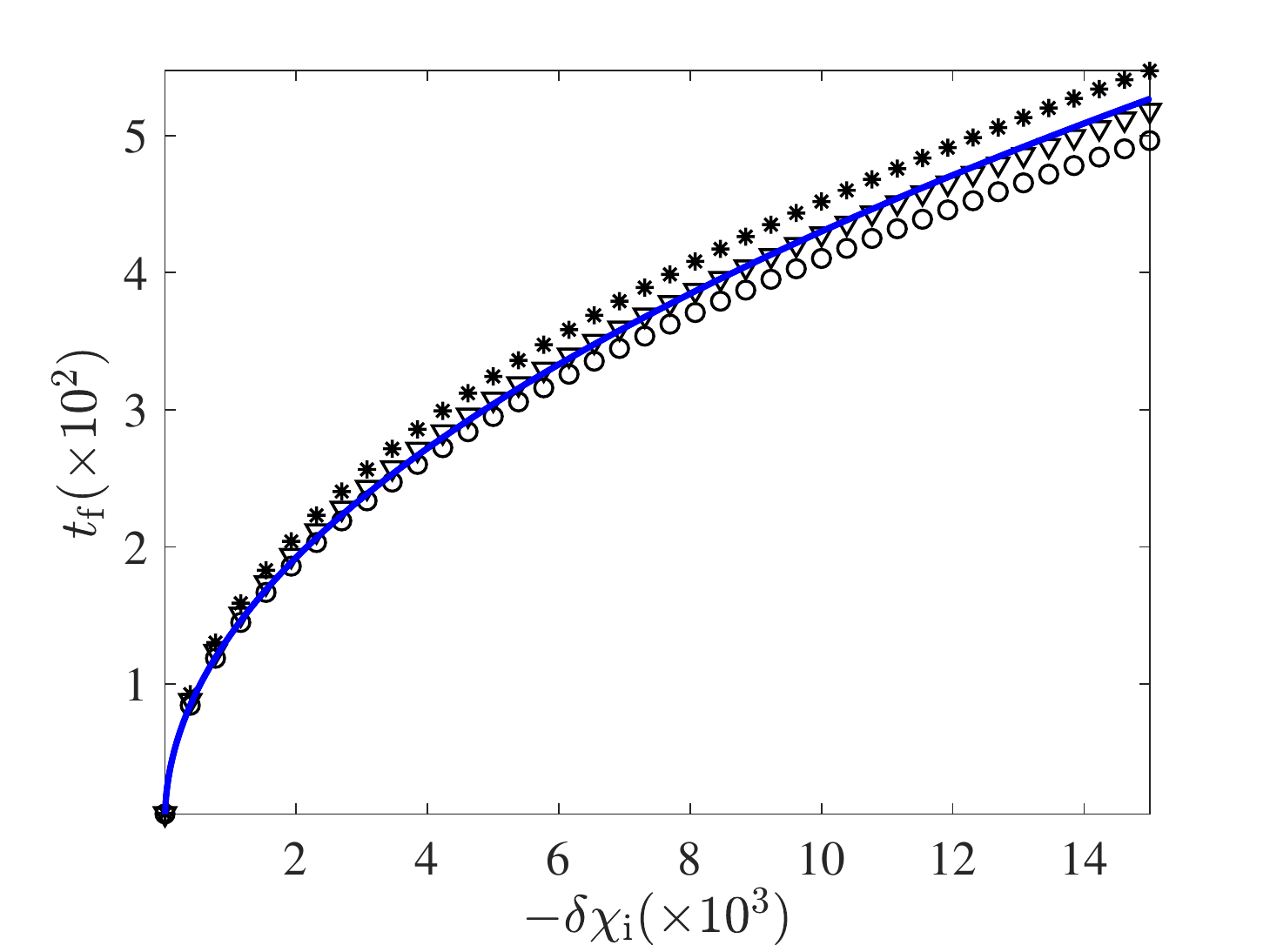} 
  	\caption{Convergence to the non-linear expression \eqref{eq:tf-teo-unbounded} as the bounds go to more extreme values, for the HC case. Symbols code of the data shown are the same as in Fig.~\ref{fit-compar-CH}. The breakage of the convergence ``from above'' to the non-linear result is clearly seen: for extreme enough values of the bounds, the linear time becomes smaller than the non-linear prediction for a full-strength thermostat.}
\label{fit-compar-HC}
\end{figure}

Now we have a closer look at the HC case in Fig.~\ref{fit-compar-HC}. As the bound in the controls move away from unity, the minimum connection time is also very close to the non-linear expression \eqref{eq:tf-teo-unbounded}. However, the convergence ``from above'' observed in the CH case is  broken. In fact,  the linear prediction is neatly below the non-linear one for the data corresponding to the most extreme values of the bounds. This marks a first physical limit for the range of controls that can be used in the linear approach: beyond the values $\chi_{\min}$ and $\chi_{\max}$ such that the linear prediction for the minimum connection time become smaller than that provided by the non-linear prediction \eqref{eq:tf-teo-unbounded}, the linear theory is clearly not valid. Recall that the latter was obtained for the full strength of the thermostat, $(\chi_{\min}=0,\chi_{\max}=\infty)$, so for a smaller set of controls the minimum connection time must be longer.
  
We have illustrated the breakage of the convergence ``from above'' in the HC protocol for the particular case $\alpha=0.9$ and $d=2$. This behaviour is robust: it occurs for all $\alpha$, and also for $d=3$. On the other hand, in the CH protocol, the inversion of the natural convergence ``from above'' never comes about. This asymmetry between the CH and HC protocols stems from the physical limit that $\chi_{\min}$ has: while $\chi_{\max}$ can be as large as desired, $\chi_{\min}$ must always be non-negative. This entails that, when applying a CH protocol, the granular temperature $T$ evolves between the values $0$ and $1$ for all times, which prevents the system from presenting important deviations from the  linear response behaviour. However, for the HC protocol, the temperature can reach arbitrarily large values under the action of a high enough driving $\chi_{\max}$, which makes the linear response approximation no longer valid. In fact, if we had studied the system from a purely mathematical point of view and removed the physical restriction $\chi_{\min}\geq 0$ (letting it vary between $-\infty$ and $+\infty$),  this asymmetry between the CH and HC cases would have disappeared.

It is interesting to remark that as $\alpha$ decreases (i.e. as the inelasticity increases) decreases, less extreme values of $\chi_{\max}$ and $ \chi_{\min}$ are needed to provoke the inversion. In other words, the linear approximation breaks down for less extreme bounds. For example, let us consider $\alpha=0.1$, which can be regarded as a high-inelasticity case---as opposed to the low-inelasticity case $\alpha=0.9$.  Fixing $\chi_{\min}=0$, the connection times of the linear approximation become shorter than those given by Eq.~\eqref{eq:tf-teo-unbounded} for $ \delta \chi_{\max} \geq 9.2 $ in the range of $\delta T_{\ini} \in [-0.01,0]$ (or, equivalently $\delta \chi_{\ini} \in[-0.015,0]$), smaller than the value $\delta \chi_{\max}\geq 9.8$   for $\alpha=0.9$. This trend with $\alpha$ of the bounds leading to the inversion of the convergence ``from above'' can be understood by recalling that $a_2^{\st}$ is a  decreasing function of $\alpha$. Consequently, the importance of the heating term in the evolution equation of the temperature~\eqref{dta}, $\chi\left(1+\frac{3}{16}a_2^{\st}\right)$, increases as $\alpha$ is lowered: a smaller value of $\chi$ is needed to get the same value of the heating term.



\section{Discussion}
\label{sec:discussion}

Our work improves the understanding of the optimal control of driven granular gases. The results obtained in this paper complement and enrich those obtained  in Ref.~\cite{prados_optimizing_2021} for a full-strength thermostat. The inclusion of bounds in the driving, $\chi_{\min}\leq\chi\leq\chi_{\max}$ raises non-trivial questions that have been answered by our study, like the range of initial and target temperatures that can be connected. Our investigation has been carried out in the linear response regime, i.e.  the initial and target states are close enough---and so are the bounds of the driving $\chi_{\min}$ and $\chi_{\max}$. This allows us to linearise the evolution equations around the final (target) NESS.

The linear response approximation leads to a set of evolution equations that are linear both in the control function and the dynamical variables---more precisely, in their deviations from their target values. Therefore we get a linear control problem that can be completely solved. A rigorous mathematical theorem ensures that the optimal control, minimising the connection time, is of bang-bang type with at most one switching: i.e., the optimal control comprises two time windows $[0,t_J)$ and $[t_J,t_{\fin})$, with the control being equal to one of its limiting values, either $\chi_{\min}$ or $\chi_{\max}$, in the first time window $[0,t_J)$ and changing to the other limiting value at the switching time $t_J$. Therefore, two types of bang-bang protocols arise, depending on the order of the bangs: $\chi_{\min}$ followed by $\chi_{\max}$, which we have termed CH, or $\chi_{\max}$ followed by $\chi_{\min}$, which we have termed HC. We have shown that the CH protocol is the optimal one when the initial temperature $T_{\ini}$ is larger than the final one $T_{\fin}$, whereas the HC protocol is the optimal one in the reverse situation, $T_{\ini}<T_{\fin}$.

We have investigated the behaviour of the connection time as a function of the bounds in the driving intensity. This study has allowed us to elucidate the range of initial and final temperatures that can be connected. We have shown that the final temperature has to lie between the temperatures $T_{\min}$ and $T_{\max}$, where $T_{\min}$ ($T_{\max}$) is the steady temperature corresponding to the constant driving $\chi_{\min}$ ($\chi_{\max}$). On the other hand, the initial temperature may lie outside the interval $[T_{\min},T_{\max}]$ and the connection is still possible: the minimum connection time is still finite when the upper bound $\chi_{\max}$ crosses the initial temperature $T_{\ini}>T_{\fin}$ (or the lower bound crosses the initial temperature $T_{\ini}<T_{\fin}$)~\footnote{It could be argued that, still, the most relevant physical situation corresponds to the case $T_{\ini}\in[T_{\min},T_{\max}]$, because one needs to prepare the system in the initial NESS.}.

Also, we have explored the limits of validity of the linear response approximation we have employed throughout. We have done this by loosening the restrictions on the bounds $\chi_{\min}$ and $\chi_{\max}$. Specifically, we have analysed the behaviour of our (linear response)  prediction for the minimum connection time, $t_{\fin}$, as $\chi_{\min}$ is decreased to very small values and $\chi_{\max}$ is increased to very large values. This behaviour has been compared to the minimum connecting time for the non-linear case $t_{\fin}^{n\ell}$, which was obtained  when the thermostat has its full strength, $\chi_{\min}=0$ and $\chi_{\max}=\infty$. Specifically, we have compared the linear time with the asymptotic expression for $t_{\fin}^{n\ell}$ for small temperature jumps---in which the corresponding connection times are also very small~\cite{prados_optimizing_2021}.  For the CH case, we have found that $t_{\fin}$ tends to $t_{\fin}^{n\ell}$ always ``from above'', $t_{\fin}>t_{\fin}^{n\ell}$. This is logical, since the largest set of controls---like that of the full-power thermostat---should lead to the shortest connection times. However, for the HC case, we have found that the tendency from above towards  $t_{\fin}$ is broken for large enough values of $\chi_{\max}$. This marks a limit of validity for the linear response approximation in this case. The asymmetry between the CH and HC protocols can be understood on a physical basis: in the latter case, heating precedes cooling and thus the temperature departs from the vicinity of $T_{\fin}$ for high enough $\chi_{\max}$---whereas in the former, cooling precedes heating and the system remains closer to the target state even when $\chi_{\max}$ becomes large.

In order to further look into the behaviour described in the previous paragraph, we have looked into the regime of short connecting times $t_{\fin}\ll 1$ within the linear response framework. Note that our linear response predictions for $t_{\fin}$, as given by Eqs.~\eqref{1} and \eqref{2},  contain all the powers of $\delta\chi_{\ini}/\delta{\chi}_{\tot}$, $\delta\chi_{\max}/\delta{\chi}_{\tot}$, $\delta\chi_{\min}/\delta{\chi}_{\tot}$. Linear response assumes that both $\delta \chi_{\ini}\ll 1$, $\delta{\chi}_{\max}\ll 1$, and $\delta\chi_{\min}\ll 1$, but the ratios between one another are in principle of the order of unity. It is only when $\delta\chi_{\ini}$ is much smaller than $\delta\chi_{\max}$ and $\delta\chi_{\min}$ that the connection time becomes small. In this regime, we have obtained a simple approximate expression for $t_{\fin}$ valid to the lowest order in $\delta\chi_{\ini}$, which also depends on the bounds in the driving. This approximate expression always gives connection times that are longer than that for the full-power thermostat, both for the CH and HC cases. This means that the inversion of the tendency ``from above'' towards $t_{\fin}^{n\ell}$ comes from higher-order terms in the ratios $\delta\chi_{\ini,\max,\min}/\delta\chi_{\tot}$. 

Our work also opens the door to finding new optimal controls for other non-equilibrium systems. For example, let us look at a colloidal particle moving in the vicinity of a minimum of the trapping potential---which can be thus considered to be harmonic. In that case, the temperature of the thermal bath in which the particle is immersed plays the role of the driving intensity. Interestingly, the temperature of the bath can be effectively increased by adding a random forcing that can be modelled as a Gaussian white noise~\cite{martinez_effective_2013,ciliberto_experiments_2017}. In this way, the effective temperature changes from $T_{\min}$ (room temperature) to $T_{\max}$ (thousands of kelvins). The similitude of the mathematical framework, linear evolution equations and bounded control, makes it appealing to analyse the optimal connection---also in the sense of minimising the connection time---in that case and compare the corresponding results with those derived here.

\begin{acknowledgments}
We acknowledge financial support from project PGC2018-093998-B-I00, funded by: FEDER/Ministerio de Ciencia e Innovación--Agencia Estatal de Investigación (Spain).
\end{acknowledgments}

\appendix


\section {Maximum principle for linear systems: verifying hypothesis}\label{sec:verifying-hypothesis}

Let us consider the linear, both in the variables $x$ and the controls $u$, control system 
\begin{equation}\label{eq:gen-lin-control}
\frac {dx} {dt} = A x + B u,
\end{equation}
in which $x: \mathbb{R} \to \mathbb{R}^n$, $u: \mathbb{R} \to U$, where the control set $U$ is a $m$-dimensional parallelepiped, and $A$ and $B$ are two matrices of suitable dimensions. Now we analyse the problem of bringing the system from $x_{\ini}$ to $x_{\fin}$ in the minimum possible time $t_{\fin}$, which is known as the time optimisation problem. The columns of the matrix $B$ are denoted by $b_j$, and we introduce the assumption that the set of vectors $\{b_j,Ab_j,A^2 b_j,... A^{n-1} b_j\}$ constitutes  a basis of $ \mathbb{R}^n$ for each $j=1,\ldots, m$. Under this \textit{controllability} hypothesis, we can formulate the following theorem:\\\\

\textbf{Theorem.} \textit{If all the eigenvalues of $A$ are real, then the optimal controls are bang-bang, i.e. they take the most extreme values of their definition domain and present, at most, $n-1$ switchings.}
\\

\begin{figure}
	\centering
	\includegraphics[width = 1\linewidth]{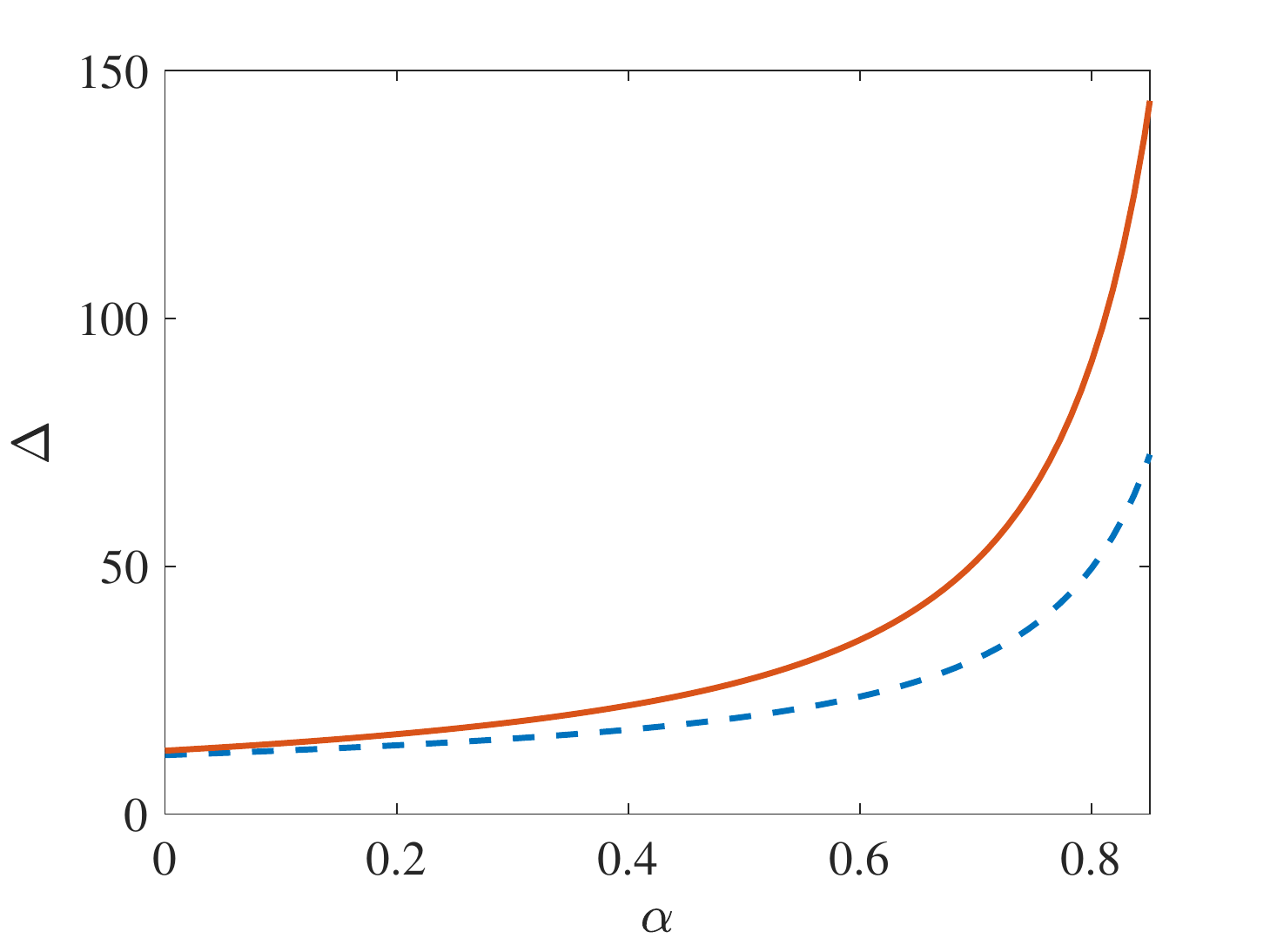}
	\caption{Discriminant $\Delta$ as a function of the restitution coefficient $\alpha$. Both the $d=2$ (solid line) and $d=3$ (dashed) cases are shown. The discriminant remains positive for all values of $ \alpha$, guaranteeing that the optimal control is of the bang-bang type with at most one switching.}
	\label{fig:discriminant}
\end{figure}

Our system \eqref{eq:sl} perfectly fits into the framework given by Eq.~\eqref{eq:gen-lin-control}, with the identifications
\begin{subequations}
\begin{align}
    x=&\begin{pmatrix}
    \delta T \\ \delta A_2
    \end{pmatrix}, \quad u=\delta\chi, \\
    A&=\begin{pmatrix}
-\frac{3}{2}  \beta & 1-\beta\\
3   & -2B
\end{pmatrix} 
, \quad B=\begin{pmatrix}
\beta \\ -2 \end{pmatrix}.
\end{align}
\end{subequations}
Therefore $n=2$ and $m=1$, with $U\equiv[\delta\chi_{\min},\delta\chi_{\max}]$. We know that the eigenvalues of the matrix $A$ are real, since they are given by $(-\lambda_1,-\lambda_2)$ in Eq.~\eqref{eq:eigenvalues}. Thus, the theorem above applies, and the controls are bang-bang with at most one switching if the vectors $\{b_1,A b_1\}$ form a basis of $\mathbb{R}^2$. The determinant of the matrix with columns $b_1$ and $Ab_1$ is $\Delta=4\left(\beta B-1\right)>0$, since  $\beta B>1$ for all $\alpha$. In fact, as seen in Fig.~\ref{fig:discriminant}, $\Delta$ increases with $\alpha$.

\section{Order for the bangs}
\label{sec:order-bangs}

In this Appendix, we prove that the optimal protocol for the case $T_{\ini}<T_{\fin}=1$ is of HC type. A completely analogous proof links the case $T_{\ini}>T_{\fin}=1$ to the CH protocol. We proceed by showing that one can only reach NESSs with $T_{\ini}<T_{\fin}=1$, i.e. with $\delta T_{\ini}<0$, making use of a HC protocol, it is impossible with a CH bang-bang.  

\begin{figure}
	\centering
	\includegraphics[width = 0.8\linewidth]{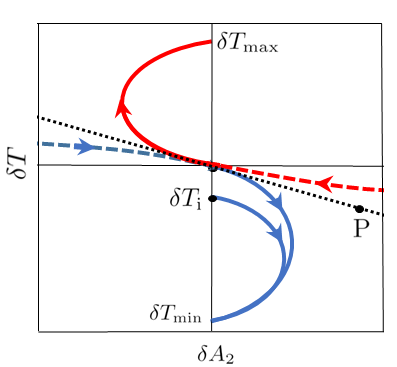}	\caption{Qualitative picture of the heating and cooling trajectories in the phase plane. Curves for $t>0$ are drawn with solid lines, curves for $t<0$ with dashed lines; heating ones ($\delta\chi_{\max})$ in red, cooling ones ($\delta\chi_{\min})$ in blue. The common tangent to the heating and cooling curves at the origin is represented by a black dotted line. The system starts cooling  from an initial state $\delta T_{\ini}<0$. Let us assume that the bang-bang protocol is of CH type. In the first step of the bang, the system follows the blue cooling curve: if it is allowed to relax during an infinite time, it reaches the NESS $(0,\delta T_{\min}=\frac{2}{3}\delta \chi_{min})$ over the vertical axis. The cooling must be interrupted at some time $t_J>0$, where the driving is switched to $\delta\chi_{\max}$: since the system must reach the target NESS at the origin, it needs to move over the branch of the heating curve corresponding to $t<0$.  However, this is impossible since this heating curve is always above the tangent line and does not intersect the cooling curve. Therefore, it is not feasible to drive the system to the origin using a CH protocol for $\delta T_{\ini}<0$.}
	\label{fig:order-bangs}
\end{figure}

The idea of the proof is based on rigorously establishing that the qualitative behaviour of the motion of the system in the phase plane $(\delta A_2,\delta T)$ is the one depicted in Fig.~\ref{fig:order-bangs}. We start by analysing the shape of the cooling curve which starts from a NESS $(0,\delta T_{\ini})$. For the cooling steps, $\delta T(t)$ and $\delta A_2(t)$ are given by their respective expressions in Eq.~\eqref{eq:cooling-CH}, with the substitution $t_J \to t$. Therefore, their time derivatives are
\begin{align}
      & \frac{d}{dt}\delta T=2\frac{ \delta\chi_{\ini}-\,\delta \chi_{\min}}{k}\left(\lambda_1 \bm{v}_1(1) e^{-\lambda_1 t}-\lambda_2  \bm{v}_2(1)e^{-\lambda_2 t}\right),\\& \frac{d}{dt}A_2=2\frac{ \delta\chi_{\ini}-\,\delta \chi_{\min}}{k}\left[\lambda_1  e^{-\lambda_1 t}-\lambda_2 e^{-\lambda_2 t}\right].
      \label{eq: derivatives}
\end{align}
Note that $dA_2/dt|_{t=0}=2(\delta\chi_{\ini}-\delta\chi_{\min})>0$. On the one hand, $\delta T(t)$ monotonically decreases from $\delta T_{\ini}=2\delta\chi_{\ini}/3$ for $t=0$ to $\delta T_{\min}=2\delta\chi_{\min}/3$ for $t\to\infty$,  because $d(\delta T)/dt$ does not vanish for $t>0$. In fact, the possible extremum of $\delta T$ occurs at a time $kt_1=\log {\frac {v_1(1) \lambda_1}{v_2(1) \lambda_2}}$, which either does not exist (for $\alpha>1/\sqrt{2}$) or is negative (for $\alpha<1/\sqrt{2}$). On the other hand, $d(\delta A_2)/dt$ vanishes at a time $t_0$ given by
\begin{equation}
    t_0=\frac{1}{k}\log {\frac {\lambda_1}{ \lambda_2}}>0,
\end{equation}
i.e. $\delta A_2$ increases from zero to positive values in the interval $[0,t_0)$, then reaches a maximum at $t=t_0$,  and decreases back to zero for $t_0<t$.
\begin{figure}
	\centering
	\includegraphics[width = 0.8\linewidth]{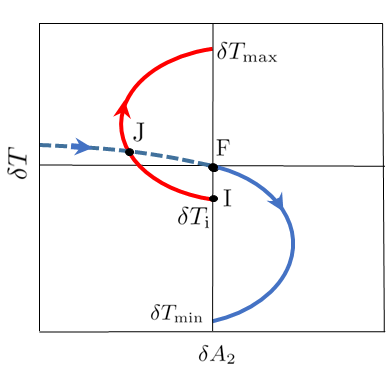}	\caption{Qualitative picture of the HC protocol in the phase plane. Curves for $t>0$ are drawn with solid lines, curves for $t<0$ with dashed lines; heating ones ($\delta\chi_{\max})$ in red, cooling ones ($\delta\chi_{\min})$ in blue. The system starts heating from an initial state $\delta T_{\ini}<0$, corresponding to point $I$. In the first step of the bang, the system follows the red heating curve: if allowed to relax during an infinite time, it reaches the NESS $(0,\delta T_{\max}=\frac{2}{3}\delta \chi_{\max})$ over the vertical axis. The heating is interrupted at the point $J$ over the cooling curve for $t<0$ (blue dashed line), where the driving is switched to $\delta\chi_{\min}$. The optimal connection thus comprises the arcs $IJ$ and $JF$.
	}
	\label{fig: HC_protocolo_fases}
\end{figure}

The above discussion entails that the motion of the point $(\delta A_2,\delta T)$ along the cooling curve in the phase plane follows indeed the shape depicted by the  blue solid line in Fig.~\ref{fig:order-bangs}. An analogous study shows that the shape of the heating trajectories that starts from a NESS, i.e. from the vertical axis $\delta A_2=0$,  must be like the red solid line in Fig.~\ref{fig:order-bangs}: $\delta T$ increases monotonically for $t>0$ and $\delta A_2$ stars decreasing, reaches a minimum, and afterwards increases back to zero.

Once we know the qualitative behaviour of the curves (heating and cooling) that start from an NESS, let us apply a cooling process to an initial state $(0,\delta T_{\ini}<0)$. We know that all of these cooling trajectories must be contained in the region between the vertical axis and the cooling curve starting from the origin, since phase plane trajectories for the same value of $\delta\chi$ cannot intersect. The only way for the system to reach the point $(0,0)$ with a CH protocol is that the cooling trajectory beginning at $(0,\delta T_{\ini}<0)$ intersect the heating trajectory ($\delta\chi_{\max}$) that crosses the origin $(0,0)$, i.e. to the branch of the red solid line corresponding to negative times (red dashed line). Our aim is to prove that this trajectory cannot enter the region described above, where all the cooling curves starting from $(0,\delta T_{\ini}<0)$  are confined. To prove this, note that the cooling and heating trajectories going through the origin have a common tangent, since
 \begin{align}
     \frac{\frac{d}{dt} \delta T}{\frac {d}{dt}\delta A_2}= \frac{-\beta \delta \chi}{2 \delta \chi}= \frac{-\beta}{2}<0
 \end{align}
is independent of $\delta \chi$ (black dotted line). We proceed to prove that the heating curve that goes through a point P over the tangent line is always above this tangent line and, therefore, cannot intersect any cooling curve. It is enough to show that the slope of the heating curve that goes through an arbitrary point over the tangent line is always larger (lower in absolute value) than the slope of the tangent. To do this, we take the point $P$ over the tangent
\begin{equation}
    \delta T_P=-\frac{\beta}{2} \delta A_{2P},
\end{equation}
and making use of Eq.~\eqref{eq:sl}, we have that
\begin{align}
    \frac{d}{dt} \delta T_P&=\beta \delta\chi_{\max} -\frac{3}{2} \beta \delta T_P+(1-\beta)\delta A_{2P}\nonumber \\&=\beta \delta\chi_{\max}+\left(\frac{3}{4}\beta^2+1-\beta\right)\delta A_{2P},
\end{align}
and 
\begin{align}
 \frac{d}{dt} \delta A_{2P}&=- 2\delta\chi_{\max} +3  \delta T_P-2B\delta A_{2P} \nonumber \\&=-2 \delta\chi_{\max}-\left(\frac{3}{2}\beta+2B\right)\delta A_{2P} .  
\end{align}
We have that $\frac{3}{4}\beta^2+1-\beta>0$,
\begin{align}
    &\frac{3}{4}\beta^2+1-\beta>0=\frac{3}{4} \left(1+\frac{3}{16}a_2^{\st}\right)^2-\frac{3}{16}a_2^{\st} \nonumber \\&=\frac{3}{4}\left(1+\frac{1}{8}a_2^{\st} \right)+\frac{27}{1024} (a_2^{\st})^2>0 \;\;\forall (\alpha,d),
\end{align}
and $\frac{3}{2}\beta+2B>0$. Therefore $\frac{d}{dt}\delta T_P>0$ and $\frac{d}{dt}\delta A_{2P}<0$  over the tangent line, and the slope of the heating curve on this point is negative,
\begin{align}
    m_P=\frac{\frac{d}{dt}\delta T_P}{\frac{d}{dt}A_{2P}}=\frac{\beta \delta\chi_{\max}+(\frac{3}{4}\beta^2+1-\beta)\delta A_{2P}}{-2 \delta\chi_{\max}-(\frac{3}{2}\beta+2B)\delta A_{2P}}<0.
\end{align}
Now we compare it with the slope of the tangent,
\begin{align}
    m_P+\frac{\beta}{2}&=\frac{\frac{3}{2}\left(\beta^2+1-\beta\right)\delta A_{2P}-\beta\left(\frac{3}{2}\beta+2B\right)\delta A_{2P}}{2\left(-2 \delta\chi_{\max}-(\frac{3}{2}\beta+2B)\delta A_{2P}\right)}\nonumber \\&=\frac{\left[4\beta B -3(1-\beta)\right]\delta A_{2P}}{2 \delta\chi_{\max}+(\frac{3}{2}\beta+2B)\delta A_{2P}} \geq 0
\end{align}
for $\delta A_{2P}\geq 0$, because $4\beta B -3(1-\beta)>0$. Then,
\begin{equation*}
    m_P>-\frac{\beta}{2}.
\end{equation*}
and the heating curve can not cross any cooling curve (since all of the are under the tangent line). We conclude that it not possible to drive the system from $(0,\delta T_{\ini}<0)$ to the origin with a CH protocol. 

On the other hand, the HC protocol starting from $(0,\delta T_{\ini}<0)$ and ending up at the origin is indeed feasible, as shown in Fig.~\ref{fig: HC_protocolo_fases}. The cooling curve that goes through the origin---specifically its dashed branch---divides the semi-plane $\delta A_2<0$ into two parts. The initial point $(0,\delta T_{\ini}<0)$ and the NESS for the heating part of the protocol, $(0,\delta T_{\max})$, lie at different sides thereof. As a consequence, the heating curve for the first bang---over which $\delta T$ monotonically increases and $\delta A_2$ has only one minimum---intersects at only one point the cooling curve for the second bang, giving rise to the optimal connection.


\section{Limit values for the intensity of the thermostat}
\label{sec:limit-values}

In this Appendix, we are interested in studying the range of values for the intensity of the thermostat that make it possible to connect the initial and target NESS (for both the CH and HC protocols). In particular, we would like to discern whether the connection is possible for any pair of values $(\chi_{\min},\chi_{\max})$ or there appears a region in  parameter space that make the connection impossible. We analyse the CH case (cooling, $T_{\ini}>1$) in detail, since the analysis of the HC case follows completely analogous lines.

Physically it seems clear that, in order to reach the final temperature $T_{\fin}=1$ from an initial temperature $T_{\ini}>1$,  it is imperative that the minimum intensity of the driving verifies $\chi_{\min}<1$---the minimum value of the thermostat intensity has to be smaller than that corresponding to the final temperature. But, what about $ \chi_{\max}$? Is the connection always possible as long as $\chi_{\max}>1$? Or, on the contrary, is there a lower bound that makes it impossible to connect the two NESS? The limit values $(\chi_{\min},\chi_{\max})$, beyond that the connection is no longer possible, are those that bring about a divergent minimum connection time $t_{\fin}$. 

In the following, we show how the line of reasoning above gives answers to the questions posed: the physical intuition on the limit value of $\chi_{\min}$ is correct, $t_{\fin}$ diverges in the limit as $\delta\chi_{\min}\to 0^-$, and the limit value of $\chi_{\max}$ is also unity, $t_{\fin}$ diverges in the limit as $\delta\chi_{\max}\to 0^+$. In order to prove these statements, we will follow the following procedure: to elucidate the behaviour with $\delta\chi_{\min}$ ($\delta\chi_{\max}$), we keep $\delta\chi_{\max}$ ($\delta\chi_{\min}$) fixed and progressively increase $\delta\chi_{\min}$ (decrease $\delta\chi_{\max}$) from negative (positive) values until the minimum connection time diverges.

From Eq.~\eqref{1}, which gives $t_{\fin}$ as a function of $\delta\chi_{\ini}$, $\delta\chi_{\min}$ and $\delta\chi_{\max}$ in the CH case, we can infer the values of $\delta\chi_{\min}$ that make $t_{\fin}$ diverge. This divergence only comes about when either the numerator of the logarithm  tends to $\infty$ or the denominator tends to $0$. We analyse both possibilities in the following. For the numerator to diverge, either $\delta\chi_{\min}=\delta\chi_{\ini}>0$ (recall that we are studying the CH protocol) or $t_J \to \infty$. In the latter case, $t_{\fin}$  always diverges~\footnote{Physically, it is evident that the connection time cannot be shorter than the switching time, $t_{\fin}\geq t_J$. Mathematically,  the divergence of the numerator always wins because $\lambda_1 > \lambda_2$ and $t_{\fin} \sim t_J$; the time spent in the second part of the bang-bang becomes negligible as compared with $t_J$.}. The switching time $t_J$ is determined by Eq.~\eqref{1.2}, which tells us that when $t_J \to \infty$ 
\begin{equation}
    \delta\chi_{\tot}\to \delta\chi_{\max}.
    \label{eq: limite_tj}
\end{equation}
Therefore, $t_J \to \infty$ when $\delta\chi_{\min} \to 0^-$, in agreement with the physical intuition described above. This makes it unnecessary to study the other possibility of divergence of $t_{\fin}$, $\delta\chi_{\min} \to\delta\chi_{\ini}>0$. Let us explore the second possibility, i.e the vanishing of the denominator of the logarithm in Eq.~\eqref{1}, which occurs when 
\begin{equation}\label{eq:denom-log-zero}
    t_J\to \frac {1} {\lambda_2} \ln \left(\frac{\delta\chi_{\ini}-\delta\chi_{\min}}{\delta\chi_{\tot}} \right).
\end{equation}
Note that the numerator of the logarithm is positive for this value of $t_J$, because $\lambda_1>\lambda_2$. When Eq.~\eqref{eq:denom-log-zero} holds, the left hand side (lhs) of Eq.~\eqref{1.2} vanishes. As a consequence, it is $\delta\chi_{\max}\to 0^+$, since the factor accompanying it on the rhs of Eq.~\eqref{eq:denom-log-zero} is basically the numerator of the logarithm in Eq.~\eqref{1}. Therefore, $t_{\fin} \to \infty$ when $\delta\chi_{\max}\to 0^+$.

Wrapping things up, our analysis above implies that it is always possible to connect two non-equilibrium steady states as long as $\delta \chi_{\min} <0$ and $\delta \chi_{\max} > 0$, that is, the lower (upper) bound of the thermostat intensity, $ \chi_{\min}$ ($ \chi_{\max}$) is below (above) the one corresponding to the final state---i.e. unity, with our choice of variables. Therefore, there are no additional regions in parameter space that do not allow for connecting the two NESS. As already said above, the HC case (heating, $T_{\ini}<1$) is  treated in a completely analogous way, with the same conclusion: $\chi_{\min}<1$ and $ \chi_{\max}>1$, with the roles of these limitations exchanged with respect to the case CH.


\section{Bang-bang for the non-linear case}
\label{sec:bang-bang-non-linear}

In the linear case, Eq.~\eqref{eq:sl}, the number of switchings of the bang-bang control is given by the theorem in Appendix~\ref{sec:verifying-hypothesis}. In the first Sonine approximation employed to describe the granular gas, we have two variables and thus only one switching. In this way, either the CH protocol or the HC protocol is that minimising the connection time between the initial and final NESS.

In the non-linear case, Eq.~\eqref{eq:evol-eqs-non-linear}, the optimal connection is also of bang-bang type~\cite{prados_optimizing_2021}. The evolution equations---despite being non-linear in the temperature---are linear in the intensity of the driving, and Pontryagin's maximum principle~\cite{pontryagin_mathematical_1987,liberzon_calculus_2012} ensures that the optimal control minimising the connection time is bang-bang. However, the number of switchings from one extreme value of $\chi$ to the other is not known. The simplest two-step bang-bang protocols were investigated in Ref.~\cite{prados_optimizing_2021}, but it was not proved that the two-step bang-bangs led to the minimum time. Here we present such a proof.

\begin{figure}
	\centering
	\includegraphics[width = 1\linewidth]{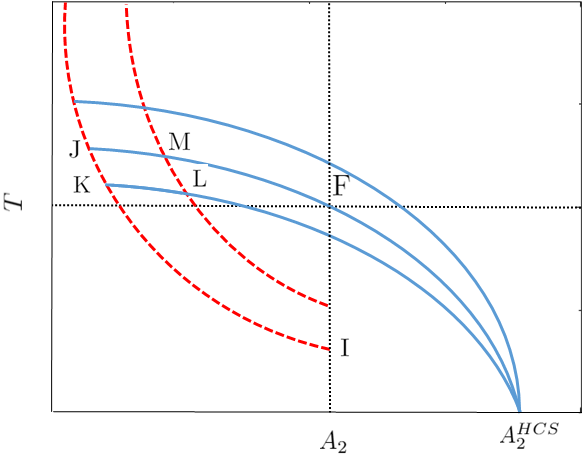}
	\caption{Comparison of the two- and four-step bang-bang processes for $T_{\ini}<1$. The two-step bang-bang connects  $I$ and $F$ with the arcs $IJ$ and $JF$, whereas the four-step one  comprises the arcs  $IK$, $KL$, $LM$ and $MF$. Dotted lines represent the axis. The latter gets $I$ and $F$ connected, but it takes it longer to complete it---for all the possible points $T_K<T_J$. Any four-step protocol whose first  cooling arc ends at a point with $T_K>T_J$ cannot drive the system to the target NESS $F=(1,1)$.}
	\label{fig: four-protocol-hc}
\end{figure}

In this Appendix, we consider more complex bang-bang processes, with more than two steps. We show that the connection time for this more complex protocols is always longer than that for the two-step bang-bangs. For the non-linear case, we look into the movement of the system in the phase plane $(A_2,T)$ [instead of $(\delta A_2,\delta T)]$. Let us first focus on the case $T_{\ini}<1$, illustrated by Fig.~\ref{fig: four-protocol-hc}. The system starts from a point $I=(1,T_{\ini})$ and ends up at the target point $F=(1,T_{\fin}=1)$. The heating curve---with $\chi=\chi_{\max}=\infty$---passing through $I$ (red solid line) and the cooling curve---with $\chi_{\min}=0$---passing through $F$ (blue solid line) intersect at the point $J$. The two-step bang-bang is formed by joining the portion of the heating curve joining $I$ and $J$ and the portion of the cooling curve joining $J$ and $F$, i.e. the arcs $IJ$ and $JF$.

A four-step bang-bang is shaped as follows. Let us consider a point $K$ belonging to the portion of the heating curve from $I$ to $F$, interrupt the heating at this point and switch the driving to $\chi_{\min}=0$. Then the system starts to sweep the cooling curve passing through $K$ (blue dashed line). At the point $L$, the cooling is interrupted and the driving switched to $\chi_{\max}$. Then the system starts to sweep the heating curve passing through $L$ (red dashed curve). The heating is interrupted when the latter heating curve reaches the point $M$, which belongs to the cooling curve passing through $F$. The arcs $IK$, $KL$, $LM$, and $MF$ build a four-step bang-bang. The point $K$ must verify $T_K<T_J$, otherwise it is easy to show that the four-step bang-bang cannot reach the target NESS.

The above picture entails that, in order to show the optimality of the two-step bang-bang, we have to establish that the time needed for going from $K$ to $L$, $t_{KL}$, is longer than the time needed for going from $J$ to $M$, $t_{JM}$. Making use of Eq.~\eqref{eq:evol-eqs-non-linear}, the equation of motion in the phase plane with $\chi_{\min}=0$ is given by 
\begin{align}\label{eq:evol-eqs-nl-chizero}
    &\dot{T}=-T^{3/2} (1+ \frac {3} {16} a_2^{\st} A_2), & \dot{A_2}= 2T^{1/2}\frac{A_2^{HCS}-A_2}{A_2^{HCS}-1},
\end{align}
where $A_2^{\HCS}=a_2^{\HCS}/a_2^{\st}>1$. Thus, in the phase plane one has
\begin{align}\label{eq:phase-plane-eqs-nl-chizero}
    \frac{dT}{dA_2}=-\frac{(1+ \frac{3}{16}a_2^{\st} A_2)T}{2 \frac{A_2^{\HCS}-A_2}{A_2^{\HCS}-1}},
\end{align}
which can be integrated to obtain an explicit expression for $T(A_2)$ over the cooling curve~\cite{prados_optimizing_2021}.
From Eq.~\eqref{eq:evol-eqs-nl-chizero}, we have
\begin{align}
    dt=\frac{A_2^{\HCS}-1}{2} \frac{dA_2}{(T(A_2))^{1/2}(A_2^{\HCS}-A_2)},
\end{align}
and integrating it we obtain the expressions for $t_{JM}$ and $t_{KL}$
\begin{subequations}
\begin{align}
   & t_{JM}=\frac{A_2^{\HCS}-1}{2} \int_{A_{2J}}^{A_{2M}} \frac {dA_2}{[T(A_2)]^{1/2}(A_2^{\HCS}-A_2)}\\ 
   & t_{KL}=\frac{A_2^{\HCS}-1}{2} \int_{A_{2K}}^{A_{2L}} \frac {dA'_2}{[T(A'_2)]^{1/2}(A_2^{\HCS}-A'_2)}
\end{align}
\end{subequations}
In order to compare $t_{JM}$ and $t_{KL}$, it is convenient to rewrite them in terms of a common variable. This variable can be naturally defined by employing the bijection that the heating curves establish between the points belonging to the $JM$ and $LK$ arcs. In the limit as $\chi\to\infty$, Eq.~\eqref{eq:evol-eqs-non-linear} leads to
\begin{equation}
    \frac{dT}{dA_2}=-\frac{\beta T}{2A_2} \implies T^2 A_2^\beta=\text{const.}
\end{equation}
Therefore, we define a variable $\xi$ in the following way,
\begin{align}
    \xi=T^2(A_2) A_2^{\beta}, \quad \xi_{\ini}\equiv T_{\ini}^2 \leq \xi &\leq \xi_{L}\equiv T^2_L A_{2L}^{\beta} \nonumber \\
    &=\xi_{M}\equiv T^2_M A_{2M}^{\beta},
    \label{eq: common-variable}
\end{align}
and we have
\begin{subequations}
\begin{align}
   &t_{JM}=\frac{A_2^{\HCS}-1}{2} \int_{\xi_{\ini}}^{\xi_{L}} \frac{d\xi}{\xi^{1/4}} \frac {dA_2}{d \xi} \frac{A_2^{\beta/4}}{A_2^{\HCS}-A_2(\xi)} \\ &t_{KL}=\frac{A_2^{\HCS}-1}{2} \int_{\xi_{\ini}}^{\xi_{L}} \frac{d\xi}{\xi^{1/4}} \frac {dA_2'}{d \xi} \frac{{A_2'}^{\beta/4}}{A_2^{\HCS}-A_2'(\xi)} 
\end{align}
\end{subequations}
In view of Fig.~\ref{fig: four-protocol-hc} it is straightforward to see that 
\begin{align}\label{eq:inequal-ch-a2s}
    \frac{A_2^{\beta/4}}{A_2^{\HCS}-A_2(\xi)}<\frac{{A_2'}^{\beta/4}}{A_2^{\HCS}-A_2'(\xi)},
\end{align}
since $A_2(\xi)<A'_2(\xi)$, $\forall\xi$. Now we show that the next inequality also holds,
\begin{equation}
    \frac{d A_2(\xi)}{d \xi}< \frac{d A_2'(\xi)}{d \xi}.
\end{equation}
Taking logarithms in Eq.~\eqref{eq: common-variable}, one obtains
\begin{align}
    & \left(2 \frac{d\ln T(A_2)}{dA_2}+ \frac{\beta}{A_2}\right) dA_2= \frac{d \xi}{\xi}.
\end{align}
Plugging Eq.~\eqref{eq:phase-plane-eqs-nl-chizero} into the last equation, we get
\begin{align}
    \Xi(A_2) dA_2=\frac{d \xi}{\xi},
\end{align}
where
\begin{align}
    \Xi(A_2) \equiv -(A_2^{\HCS}-1)\frac{1+ \frac{3}{16}a_2^{\st} A_2}{A_2^{\HCS}-A_2} + \frac{\beta}{A_2}
\end{align}
is a decreasing function of $A_2$ (all of the terms in the sum of its derivative are negative), positive for $A_2<1$ and negative for $A_2>1$. The resulting integrals can be written as 
\begin{subequations}\label{eq:time-integral}
\begin{align}
   &t_{JM}=\frac{A_2^{\HCS}-1}{2} \int_{\xi_{\ini}}^{\xi_{L}} \frac{d\xi}{\xi^{5/4}} \frac{1}{\Xi(A_2(\xi))} \frac{A_2^{\beta/4}}{A_2^{\HCS}-A_2(\xi)},
   \\ &t_{KL}=\frac{A_2^{\HCS}-1}{2} \int_{\xi_{\ini}}^{\xi_{L}} \frac{d\xi}{\xi^{5/4}} \frac{1}{\Xi(A'_2(\xi))} \frac{{A_2'}^{\beta/4}}{A_2^{\HCS}-A_2'(\xi)}.
\end{align}
\end{subequations}
These expressions let us conclude that $t_{JM}<t_{KL}$: $A_2(\xi)<A'_2(\xi)$ and thus the integrand in $t_{JM}$ is smaller than the integrand in $t_{KL}$. Therefore, a two-step protocol is better than a four-step one. This line of reasoning can be easily extended two six-step, eight-step, etc. protocols,  showing each of them to be worst than the previous one. Protocols with an odd number of switches can be described as a limiting case of protocols with an even number of switching: for example, the three-step protocol can be seen as a four-step protocol for which $M$ coincides with $F$.

Let us inspect the case $T_{\ini}>1$. Again, we start by considering a two-step protocol, comprising the arcs $IJ$ and $JF$. Also, we build a  four-step protocol, comprising the  arcs $IK$, $KL$, $LM$ and $MF$. The expression for the times are formally equal to those for the case $T_{\ini}>1$, given by Eq.~\eqref{eq:time-integral}, 
\begin{align}
   &t_{KJ}=\frac{A_2^{\HCS}-1}{2} \int_{1}^{\xi_{K}} \frac{d \xi}{\xi^{5/4}}\frac{A_2(\xi)^{\beta/4}}{|\Xi(A_2(\xi))|(A_2^{\HCS}-A_2(\xi))},\nonumber\\ &t_{LM}=\frac{A_2^{\HCS}-1}{2} \int_{1}^{\xi_{K}} \frac{d \xi}{\xi^{5/4}}\frac{A_2'(\xi)^{\beta/4}}{|\Xi(A_2'(\xi))|(A_2^{\HCS}-A_2'(\xi))},\nonumber
\end{align}
where $1=\xi_M=\xi_J$ and $\xi_L=\xi_K$. However, the inequality in Eq.~\eqref{eq:inequal-ch-a2s} is reversed, so it is not direct that $t_{KJ}<t_{LM}$. Still, it is possible to inspect the behaviour of $F(A_2)$, defined as
\begin{align}
    F(A_2)=\frac{A_2^{\beta/4}}{|\Xi(A_2)|(A_2^{\HCS}-A_2)}
\end{align}
in the interval $1<A_2<A_2^{\HCS}$: therein, $F(A_2)$ monotonically decreases with $A_2$, and thus $t_{KJ}<t_{LM}$. As a consequence, the four-step bang-bang lasts longer than the two-step bang-bang. Six-step, eight-step, etc. protocols are even worse, so the two-step bang-bang is the optimal one also for $T_{\ini}>1$.
\begin{figure}
	\centering
	\includegraphics[width = 1\linewidth]{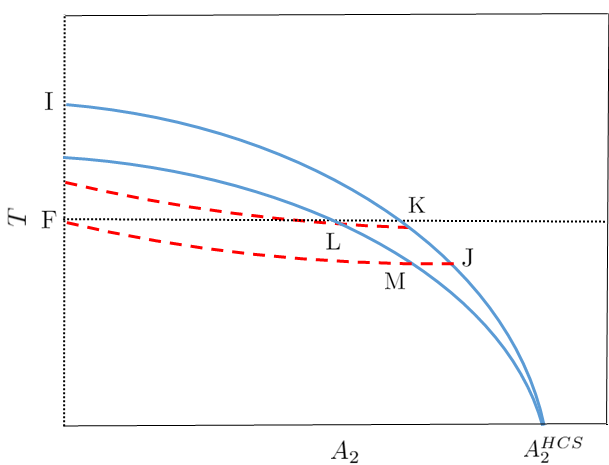}
	\caption{Comparison of the two- and four-step bang-bang processes for $T_{\ini}>1$. Analogously to the case $T_{\ini}<1$, he two-step bang-bang connects  $I$ and $F$ with the arcs $IJ$ and $JF$, whereas the four-step one  comprises the arcs  $IK$, $KL$, $LM$ and $MF$. Dotted lines represent the axis. It takes it longer to complete the latter---but now $T_K>T_J$. In this case, if $T_K<T_J$ the system cannot be driven to the target NESS.}
	\label{fig:four-protocol-ch}
\end{figure}


\section{Approximate expressions for short connection time}
\label{sec:tf-approximation}

Let us inspect the CH case, since the analysis of the HC case follows along similar lines---as usual. Note that, despite the closeness of the initial and final states, both  the switching time $t_J$ and the connecting time $t_{\fin}$ are in principle of the order of unity. This may be surprising at first sight, but we have to take into account that the intensity of the driving is also bounded in a small  interval, $\delta\chi\in[\delta \chi_{\min},\delta \chi_{\max}]$ $\forall t\geq 0$. In fact, $t_J$ and $t_{\fin}$ depend on the ratios $\delta\chi_{\ini}/\delta\chi_{\tot}$, $\delta\chi_{\min}/\delta\chi_{\tot}$ and $\delta\chi_{\max}/\delta\chi_{\tot}$---all of them order of unity quantities, in principle.

Our purpose in this Appendix is to find an approximate expression for the minimum connecting time when it is very short, i.e. when $t_{\fin}\ll 1$---and so is $t_J$, because $t_J\leq t_{\fin}\ll 1$. Note that $t_J$ and $t_{\fin}$ vanish simultaneously when $\delta\chi_{\ini}=0$. First, we can make use of Eq.~\eqref{1} to find a relation between $t_{\fin}$ and $t_J$ in this regime, 
\begin{align}
 t_{\fin}= &\frac{1}{k} \ln\left[1+ \frac  {\frac {\delta\chi_{\tot}}{\delta\chi_{\ini}-\delta\chi_{\min}}\left(e^{\lambda_1 t_{J}}-e^{\lambda_2 t_J}\right)}{\frac {\delta\chi_{\tot}}{\delta\chi_{\ini}-\delta\chi_{\min}}e^{\lambda_2 t_{J}}-1} \right]\nonumber \\
 = &\frac{1}{k}\frac{\frac {\delta\chi_{\tot}}{\delta\chi_{\ini}-\delta\chi_{\min}}(\lambda_1-\lambda_2)t_J}{\frac {\delta\chi_{\tot}}{\delta\chi_{\ini}-\delta\chi_{\min}}-1}+\mathcal{O}(t_J^2),
\end{align}
i.e.
\begin{equation}\label{eq:tf-tJ-small}
    t_{\fin}\sim \frac{\delta\chi_{\tot}}{\delta\chi_{\max}-\delta\chi_{\ini}}t_J\sim \frac{\delta\chi_{\tot}}{\delta\chi_{\max}}t_J.
\end{equation}
We have neglected $\delta\chi_{\ini}$ in the denominator, because it gives higher-order corrections that have been already neglected. 

In the same vein, we can expand Eq.~\eqref{1.2} in powers of $t_J$. First, we rewrite Eq.~\eqref{1.2} in the equivalent form
\begin{align}
    \left(\delta\chi_{\max}-\delta\chi_{\ini}\right) &\left[1+ a \left(e^{\lambda_2 t_J}-1\right)\right]^{\frac{\lambda_1}{k}} \nonumber \\
    &=\delta\chi_{\max}\left[1+ a \left(e^{\lambda_1 t_J}-1\right)\right]^{\frac{\lambda_2}{k}},
\label{eq:tJ-CH-v2}
\end{align}
where we have introduced a parameter $a$ defined by
\begin{equation}
    a=\frac{\delta\chi_{\tot}}{\delta\chi_{\max}-\delta\chi_{\ini}}.
\end{equation}
Equation~\eqref{eq:tJ-CH-v2} makes clear that $t_J\to 0$ when $\delta\chi_{\ini}\to 0$. Now, we expand it in powers of $t_J$ using
\begin{align}
    \left[1+ a \left(e^{\lambda_2 t_J}-1\right)\right]&^{\frac{\lambda_1}{k}}=
    1+a\frac{\lambda_1\lambda_2}{k}t_J \nonumber \\ & +\frac{a\lambda_2^2\lambda_1}{2k}
    \left(1-a+a\frac{\lambda_1}{k}\right)t_J^2+\mathcal{O}(t_J^3)
\end{align}
and an analogous expression---exchanging $\lambda_1$ and $\lambda_2$---for the term in brackets on the rhs.  In this way, we get to the lowest order 
\begin{equation}
    \delta\chi_{\ini}=\delta\chi_{\max}\frac{1}{2}a(a-1)\lambda_1\lambda_2 t_J^2+\mathcal{O}(t_J^3).
\end{equation}
Terms of the order of $t_J^3$ also include the contributions proportional to $\delta\chi_{\ini} t_J$ in the expansion---note that $\delta\chi_{\ini}=\mathcal{O}(t_J^2)$. Bringing to bear the definition of $a$, to the lowest order we have to substitute $\delta\chi_{\ini}$ with zero, i.e. $a\sim \delta\chi_{\tot}/\delta\chi_{\max}$ and
\begin{equation}
    \frac{1}{2}\lambda_1\lambda_2 t_J^2 \sim \frac{\delta\chi_{\ini}}{\delta\chi_{\tot}}\frac{\delta\chi_{\max}}{-\delta\chi_{\min}}.
\end{equation}
For the connection time, we thus get
\begin{equation}
    \frac{1}{2}\lambda_1\lambda_2 t_{\fin}^2 \sim \frac{\delta\chi_{\ini}}{\delta\chi_{\max}}\frac{\delta\chi_{\tot}}{-\delta\chi_{\min}},
\end{equation}
which is equivalent to Eq.~\eqref{eq:tf-short-approx} in the main text.


\bibliography{Mi-biblioteca-06-oct-2021}

\begin{thebibliography}{60}%
\makeatletter
\providecommand \@ifxundefined [1]{%
 \@ifx{#1\undefined}
}%
\providecommand \@ifnum [1]{%
 \ifnum #1\expandafter \@firstoftwo
 \else \expandafter \@secondoftwo
 \fi
}%
\providecommand \@ifx [1]{%
 \ifx #1\expandafter \@firstoftwo
 \else \expandafter \@secondoftwo
 \fi
}%
\providecommand \natexlab [1]{#1}%
\providecommand \enquote  [1]{``#1''}%
\providecommand \bibnamefont  [1]{#1}%
\providecommand \bibfnamefont [1]{#1}%
\providecommand \citenamefont [1]{#1}%
\providecommand \href@noop [0]{\@secondoftwo}%
\providecommand \href [0]{\begingroup \@sanitize@url \@href}%
\providecommand \@href[1]{\@@startlink{#1}\@@href}%
\providecommand \@@href[1]{\endgroup#1\@@endlink}%
\providecommand \@sanitize@url [0]{\catcode `\\12\catcode `\$12\catcode
  `\&12\catcode `\#12\catcode `\^12\catcode `\_12\catcode `\%12\relax}%
\providecommand \@@startlink[1]{}%
\providecommand \@@endlink[0]{}%
\providecommand \url  [0]{\begingroup\@sanitize@url \@url }%
\providecommand \@url [1]{\endgroup\@href {#1}{\urlprefix }}%
\providecommand \urlprefix  [0]{URL }%
\providecommand \Eprint [0]{\href }%
\providecommand \doibase [0]{https://doi.org/}%
\providecommand \selectlanguage [0]{\@gobble}%
\providecommand \bibinfo  [0]{\@secondoftwo}%
\providecommand \bibfield  [0]{\@secondoftwo}%
\providecommand \translation [1]{[#1]}%
\providecommand \BibitemOpen [0]{}%
\providecommand \bibitemStop [0]{}%
\providecommand \bibitemNoStop [0]{.\EOS\space}%
\providecommand \EOS [0]{\spacefactor3000\relax}%
\providecommand \BibitemShut  [1]{\csname bibitem#1\endcsname}%
\let\auto@bib@innerbib\@empty
\bibitem [{\citenamefont {Jaeger}\ \emph {et~al.}(1996)\citenamefont {Jaeger},
  \citenamefont {Nagel},\ and\ \citenamefont
  {Behringer}}]{jaeger_granular_1996}%
  \BibitemOpen
  \bibfield  {author} {\bibinfo {author} {\bibfnamefont {H.~M.}\ \bibnamefont
  {Jaeger}}, \bibinfo {author} {\bibfnamefont {S.~R.}\ \bibnamefont {Nagel}},\
  and\ \bibinfo {author} {\bibfnamefont {R.~P.}\ \bibnamefont {Behringer}},\
  }\bibfield  {title} {\bibinfo {title} {Granular solids, liquids, and gases},\
  }\href@noop {} {\bibfield  {journal} {\bibinfo  {journal} {Reviews of Modern
  Physics}\ }\textbf {\bibinfo {volume} {68}},\ \bibinfo {pages} {1259}
  (\bibinfo {year} {1996})}\BibitemShut {NoStop}%
\bibitem [{\citenamefont {Haff}(1983)}]{haff_grain_1983}%
  \BibitemOpen
  \bibfield  {author} {\bibinfo {author} {\bibfnamefont {P.~K.}\ \bibnamefont
  {Haff}},\ }\bibfield  {title} {\bibinfo {title} {Grain flow as a
  fluid-mechanical phenomenon},\ }\href@noop {} {\bibfield  {journal} {\bibinfo
   {journal} {Journal of Fluid Mechanics}\ }\textbf {\bibinfo {volume} {134}},\
  \bibinfo {pages} {401} (\bibinfo {year} {1983})}\BibitemShut {NoStop}%
\bibitem [{\citenamefont {Goldshtein}\ and\ \citenamefont
  {Shapiro}(1995)}]{goldshtein_mechanics_1995}%
  \BibitemOpen
  \bibfield  {author} {\bibinfo {author} {\bibfnamefont {A.}~\bibnamefont
  {Goldshtein}}\ and\ \bibinfo {author} {\bibfnamefont {M.}~\bibnamefont
  {Shapiro}},\ }\bibfield  {title} {\bibinfo {title} {Mechanics of collisional
  motion of granular materials. {Part} 1. {General} hydrodynamic equations},\
  }\href@noop {} {\bibfield  {journal} {\bibinfo  {journal} {Journal of Fluid
  Mechanics}\ }\textbf {\bibinfo {volume} {282}},\ \bibinfo {pages} {75}
  (\bibinfo {year} {1995})}\BibitemShut {NoStop}%
\bibitem [{\citenamefont {Brey}\ \emph {et~al.}(1996)\citenamefont {Brey},
  \citenamefont {Ruiz-Montero},\ and\ \citenamefont
  {Cubero}}]{brey_homogeneous_1996}%
  \BibitemOpen
  \bibfield  {author} {\bibinfo {author} {\bibfnamefont {J.~J.}\ \bibnamefont
  {Brey}}, \bibinfo {author} {\bibfnamefont {M.~J.}\ \bibnamefont
  {Ruiz-Montero}},\ and\ \bibinfo {author} {\bibfnamefont {D.}~\bibnamefont
  {Cubero}},\ }\bibfield  {title} {\bibinfo {title} {Homogeneous cooling state
  of a low-density granular flow},\ }\href@noop {} {\bibfield  {journal}
  {\bibinfo  {journal} {Physical Review E}\ }\textbf {\bibinfo {volume} {54}},\
  \bibinfo {pages} {3664} (\bibinfo {year} {1996})}\BibitemShut {NoStop}%
\bibitem [{\citenamefont {Huthmann}\ \emph {et~al.}(2000)\citenamefont
  {Huthmann}, \citenamefont {Orza},\ and\ \citenamefont
  {Brito}}]{huthmann_dynamics_2000}%
  \BibitemOpen
  \bibfield  {author} {\bibinfo {author} {\bibfnamefont {M.}~\bibnamefont
  {Huthmann}}, \bibinfo {author} {\bibfnamefont {J.~A.~G.}\ \bibnamefont
  {Orza}},\ and\ \bibinfo {author} {\bibfnamefont {R.}~\bibnamefont {Brito}},\
  }\bibfield  {title} {\bibinfo {title} {Dynamics of deviations from the
  {Gaussian} state in a freely cooling homogeneous system of smooth inelastic
  particles},\ }\href {https://doi.org/10.1007/s100350000047} {\bibfield
  {journal} {\bibinfo  {journal} {Granular Matter}\ }\textbf {\bibinfo {volume}
  {2}},\ \bibinfo {pages} {189} (\bibinfo {year} {2000})}\BibitemShut {NoStop}%
\bibitem [{\citenamefont {Brey}\ \emph {et~al.}(2004)\citenamefont {Brey},
  \citenamefont {Ruiz-Montero},\ and\ \citenamefont
  {Moreno}}]{brey_steady-state_2004}%
  \BibitemOpen
  \bibfield  {author} {\bibinfo {author} {\bibfnamefont {J.~J.}\ \bibnamefont
  {Brey}}, \bibinfo {author} {\bibfnamefont {M.~J.}\ \bibnamefont
  {Ruiz-Montero}},\ and\ \bibinfo {author} {\bibfnamefont {F.}~\bibnamefont
  {Moreno}},\ }\bibfield  {title} {\bibinfo {title} {Steady-state
  representation of the homogeneous cooling state of a granular gas},\
  }\href@noop {} {\bibfield  {journal} {\bibinfo  {journal} {Physical Review
  E}\ }\textbf {\bibinfo {volume} {69}},\ \bibinfo {pages} {051303} (\bibinfo
  {year} {2004})}\BibitemShut {NoStop}%
\bibitem [{\citenamefont {Brey}\ \emph {et~al.}(2007)\citenamefont {Brey},
  \citenamefont {Prados}, \citenamefont {García~de Soria},\ and\ \citenamefont
  {Maynar}}]{brey_scaling_2007}%
  \BibitemOpen
  \bibfield  {author} {\bibinfo {author} {\bibfnamefont {J.~J.}\ \bibnamefont
  {Brey}}, \bibinfo {author} {\bibfnamefont {A.}~\bibnamefont {Prados}},
  \bibinfo {author} {\bibfnamefont {M.~I.}\ \bibnamefont {García~de Soria}},\
  and\ \bibinfo {author} {\bibfnamefont {P.}~\bibnamefont {Maynar}},\
  }\bibfield  {title} {\bibinfo {title} {Scaling and aging in the homogeneous
  cooling state of a granular fluid of hard particles},\ }\href@noop {}
  {\bibfield  {journal} {\bibinfo  {journal} {Journal of Physics A:
  Mathematical and Theoretical}\ }\textbf {\bibinfo {volume} {40}},\ \bibinfo
  {pages} {14331} (\bibinfo {year} {2007})}\BibitemShut {NoStop}%
\bibitem [{\citenamefont {Van~Noije}\ and\ \citenamefont
  {Ernst}(1998)}]{van_noije_velocity_1998}%
  \BibitemOpen
  \bibfield  {author} {\bibinfo {author} {\bibfnamefont {T.~P.~C.}\
  \bibnamefont {Van~Noije}}\ and\ \bibinfo {author} {\bibfnamefont {M.~H.}\
  \bibnamefont {Ernst}},\ }\bibfield  {title} {\bibinfo {title} {Velocity
  distributions in homogeneous granular fluids: the free and the heated case},\
  }\href@noop {} {\bibfield  {journal} {\bibinfo  {journal} {Granul. Matter}\
  }\textbf {\bibinfo {volume} {1}},\ \bibinfo {pages} {57} (\bibinfo {year}
  {1998})}\BibitemShut {NoStop}%
\bibitem [{\citenamefont {Montanero}\ and\ \citenamefont
  {Santos}(2000)}]{montanero_computer_2000}%
  \BibitemOpen
  \bibfield  {author} {\bibinfo {author} {\bibfnamefont {J.~M.}\ \bibnamefont
  {Montanero}}\ and\ \bibinfo {author} {\bibfnamefont {A.}~\bibnamefont
  {Santos}},\ }\bibfield  {title} {\bibinfo {title} {Computer simulation of
  uniformly heated granular fluids},\ }\href
  {https://doi.org/10.1007/s100350050035} {\bibfield  {journal} {\bibinfo
  {journal} {Granular Matter}\ }\textbf {\bibinfo {volume} {2}},\ \bibinfo
  {pages} {53} (\bibinfo {year} {2000})}\BibitemShut {NoStop}%
\bibitem [{\citenamefont {van Noije}\ \emph {et~al.}(1999)\citenamefont {van
  Noije}, \citenamefont {Ernst}, \citenamefont {Trizac},\ and\ \citenamefont
  {Pagonabarraga}}]{van_noije_randomly_1999}%
  \BibitemOpen
  \bibfield  {author} {\bibinfo {author} {\bibfnamefont {T.~P.~C.}\
  \bibnamefont {van Noije}}, \bibinfo {author} {\bibfnamefont {M.~H.}\
  \bibnamefont {Ernst}}, \bibinfo {author} {\bibfnamefont {E.}~\bibnamefont
  {Trizac}},\ and\ \bibinfo {author} {\bibfnamefont {I.}~\bibnamefont
  {Pagonabarraga}},\ }\bibfield  {title} {\bibinfo {title} {Randomly driven
  granular fluids: {Large}-scale structure},\ }\href
  {https://doi.org/10.1103/PhysRevE.59.4326} {\bibfield  {journal} {\bibinfo
  {journal} {Physical Review E}\ }\textbf {\bibinfo {volume} {59}},\ \bibinfo
  {pages} {4326} (\bibinfo {year} {1999})}\BibitemShut {NoStop}%
\bibitem [{\citenamefont {García~de Soria}\ \emph {et~al.}(2009)\citenamefont
  {García~de Soria}, \citenamefont {Maynar},\ and\ \citenamefont
  {Trizac}}]{garcia_de_soria_energy_2009}%
  \BibitemOpen
  \bibfield  {author} {\bibinfo {author} {\bibfnamefont {M.~I.}\ \bibnamefont
  {García~de Soria}}, \bibinfo {author} {\bibfnamefont {P.}~\bibnamefont
  {Maynar}},\ and\ \bibinfo {author} {\bibfnamefont {E.}~\bibnamefont
  {Trizac}},\ }\bibfield  {title} {\bibinfo {title} {Energy fluctuations in a
  randomly driven granular fluid},\ }\href
  {https://doi.org/10.1080/00268970902794842} {\bibfield  {journal} {\bibinfo
  {journal} {Molecular Physics}\ }\textbf {\bibinfo {volume} {107}},\ \bibinfo
  {pages} {383} (\bibinfo {year} {2009})}\BibitemShut {NoStop}%
\bibitem [{\citenamefont {Maynar}\ \emph {et~al.}(2009)\citenamefont {Maynar},
  \citenamefont {García~de Soria},\ and\ \citenamefont
  {Trizac}}]{maynar_fluctuating_2009}%
  \BibitemOpen
  \bibfield  {author} {\bibinfo {author} {\bibfnamefont {P.}~\bibnamefont
  {Maynar}}, \bibinfo {author} {\bibfnamefont {M.~I.}\ \bibnamefont {García~de
  Soria}},\ and\ \bibinfo {author} {\bibfnamefont {E.}~\bibnamefont {Trizac}},\
  }\bibfield  {title} {\bibinfo {title} {Fluctuating hydrodynamics for driven
  granular gases},\ }\href@noop {} {\bibfield  {journal} {\bibinfo  {journal}
  {Eur. Phys. J. Spec. Top.}\ }\textbf {\bibinfo {volume} {179}},\ \bibinfo
  {pages} {123} (\bibinfo {year} {2009})}\BibitemShut {NoStop}%
\bibitem [{\citenamefont {García~de Soria}\ \emph {et~al.}(2012)\citenamefont
  {García~de Soria}, \citenamefont {Maynar},\ and\ \citenamefont
  {Trizac}}]{garcia_de_soria_universal_2012}%
  \BibitemOpen
  \bibfield  {author} {\bibinfo {author} {\bibfnamefont {M.~I.}\ \bibnamefont
  {García~de Soria}}, \bibinfo {author} {\bibfnamefont {P.}~\bibnamefont
  {Maynar}},\ and\ \bibinfo {author} {\bibfnamefont {E.}~\bibnamefont
  {Trizac}},\ }\bibfield  {title} {\bibinfo {title} {Universal reference state
  in a driven homogeneous granular gas},\ }\href
  {https://doi.org/10.1103/PhysRevE.85.051301} {\bibfield  {journal} {\bibinfo
  {journal} {Physical Review E}\ }\textbf {\bibinfo {volume} {85}},\ \bibinfo
  {pages} {051301} (\bibinfo {year} {2012})}\BibitemShut {NoStop}%
\bibitem [{\citenamefont {S{\'a}nchez-Rey}\ and\ \citenamefont
  {Prados}(2021)}]{sanchez-rey_linear_2021}%
  \BibitemOpen
  \bibfield  {author} {\bibinfo {author} {\bibfnamefont {B.}~\bibnamefont
  {S{\'a}nchez-Rey}}\ and\ \bibinfo {author} {\bibfnamefont {A.}~\bibnamefont
  {Prados}},\ }\bibfield  {title} {\bibinfo {title} {Linear response in the
  uniformly heated granular gas},\ }\href
  {https://doi.org/10.1103/PhysRevE.104.024903} {\bibfield  {journal} {\bibinfo
   {journal} {Physical Review E}\ }\textbf {\bibinfo {volume} {104}},\ \bibinfo
  {pages} {024903} (\bibinfo {year} {2021})}\BibitemShut {NoStop}%
\bibitem [{\citenamefont {Prados}\ and\ \citenamefont
  {Trizac}(2014)}]{prados_kovacs-like_2014}%
  \BibitemOpen
  \bibfield  {author} {\bibinfo {author} {\bibfnamefont {A.}~\bibnamefont
  {Prados}}\ and\ \bibinfo {author} {\bibfnamefont {E.}~\bibnamefont
  {Trizac}},\ }\bibfield  {title} {\bibinfo {title} {Kovacs-{Like} {Memory}
  {Effect} in {Driven} {Granular} {Gases}},\ }\href
  {https://doi.org/10.1103/PhysRevLett.112.198001} {\bibfield  {journal}
  {\bibinfo  {journal} {Physical Review Letters}\ }\textbf {\bibinfo {volume}
  {112}},\ \bibinfo {pages} {198001} (\bibinfo {year} {2014})}\BibitemShut
  {NoStop}%
\bibitem [{\citenamefont {Trizac}\ and\ \citenamefont
  {Prados}(2014)}]{trizac_memory_2014}%
  \BibitemOpen
  \bibfield  {author} {\bibinfo {author} {\bibfnamefont {E.}~\bibnamefont
  {Trizac}}\ and\ \bibinfo {author} {\bibfnamefont {A.}~\bibnamefont
  {Prados}},\ }\bibfield  {title} {\bibinfo {title} {Memory effect in uniformly
  heated granular gases},\ }\href {https://doi.org/10.1103/PhysRevE.90.012204}
  {\bibfield  {journal} {\bibinfo  {journal} {Physical Review E}\ }\textbf
  {\bibinfo {volume} {90}},\ \bibinfo {pages} {012204} (\bibinfo {year}
  {2014})}\BibitemShut {NoStop}%
\bibitem [{\citenamefont {Lasanta}\ \emph {et~al.}(2017)\citenamefont
  {Lasanta}, \citenamefont {Vega~Reyes}, \citenamefont {Prados},\ and\
  \citenamefont {Santos}}]{lasanta_when_2017}%
  \BibitemOpen
  \bibfield  {author} {\bibinfo {author} {\bibfnamefont {A.}~\bibnamefont
  {Lasanta}}, \bibinfo {author} {\bibfnamefont {F.}~\bibnamefont {Vega~Reyes}},
  \bibinfo {author} {\bibfnamefont {A.}~\bibnamefont {Prados}},\ and\ \bibinfo
  {author} {\bibfnamefont {A.}~\bibnamefont {Santos}},\ }\bibfield  {title}
  {\bibinfo {title} {When the {Hotter} {Cools} {More} {Quickly}: {Mpemba}
  {Effect} in {Granular} {Fluids}},\ }\href
  {https://doi.org/10.1103/PhysRevLett.119.148001} {\bibfield  {journal}
  {\bibinfo  {journal} {Physical Review Letters}\ }\textbf {\bibinfo {volume}
  {119}},\ \bibinfo {pages} {148001} (\bibinfo {year} {2017})}\BibitemShut
  {NoStop}%
\bibitem [{\citenamefont {Prados}(2021)}]{prados_optimizing_2021}%
  \BibitemOpen
  \bibfield  {author} {\bibinfo {author} {\bibfnamefont {A.}~\bibnamefont
  {Prados}},\ }\bibfield  {title} {\bibinfo {title} {Optimizing the relaxation
  route with optimal control},\ }\href
  {https://doi.org/10.1103/PhysRevResearch.3.023128} {\bibfield  {journal}
  {\bibinfo  {journal} {Physical Review Research}\ }\textbf {\bibinfo {volume}
  {3}},\ \bibinfo {pages} {023128} (\bibinfo {year} {2021})}\BibitemShut
  {NoStop}%
\bibitem [{\citenamefont {Chen}\ \emph
  {et~al.}(2010{\natexlab{a}})\citenamefont {Chen}, \citenamefont {Ruschhaupt},
  \citenamefont {Schmidt}, \citenamefont {del Campo}, \citenamefont
  {Guéry-Odelin},\ and\ \citenamefont {Muga}}]{chen_fast_2010}%
  \BibitemOpen
  \bibfield  {author} {\bibinfo {author} {\bibfnamefont {X.}~\bibnamefont
  {Chen}}, \bibinfo {author} {\bibfnamefont {A.}~\bibnamefont {Ruschhaupt}},
  \bibinfo {author} {\bibfnamefont {S.}~\bibnamefont {Schmidt}}, \bibinfo
  {author} {\bibfnamefont {A.}~\bibnamefont {del Campo}}, \bibinfo {author}
  {\bibfnamefont {D.}~\bibnamefont {Guéry-Odelin}},\ and\ \bibinfo {author}
  {\bibfnamefont {J.~G.}\ \bibnamefont {Muga}},\ }\bibfield  {title} {\bibinfo
  {title} {Fast {Optimal} {Frictionless} {Atom} {Cooling} in {Harmonic}
  {Traps}: {Shortcut} to {Adiabaticity}},\ }\href
  {https://doi.org/10.1103/PhysRevLett.104.063002} {\bibfield  {journal}
  {\bibinfo  {journal} {Physical Review Letters}\ }\textbf {\bibinfo {volume}
  {104}},\ \bibinfo {pages} {063002} (\bibinfo {year}
  {2010}{\natexlab{a}})}\BibitemShut {NoStop}%
\bibitem [{\citenamefont {Chen}\ \emph
  {et~al.}(2010{\natexlab{b}})\citenamefont {Chen}, \citenamefont {Lizuain},
  \citenamefont {Ruschhaupt}, \citenamefont {Guéry-Odelin},\ and\
  \citenamefont {Muga}}]{chen_shortcut_2010}%
  \BibitemOpen
  \bibfield  {author} {\bibinfo {author} {\bibfnamefont {X.}~\bibnamefont
  {Chen}}, \bibinfo {author} {\bibfnamefont {I.}~\bibnamefont {Lizuain}},
  \bibinfo {author} {\bibfnamefont {A.}~\bibnamefont {Ruschhaupt}}, \bibinfo
  {author} {\bibfnamefont {D.}~\bibnamefont {Guéry-Odelin}},\ and\ \bibinfo
  {author} {\bibfnamefont {J.~G.}\ \bibnamefont {Muga}},\ }\bibfield  {title}
  {\bibinfo {title} {Shortcut to {Adiabatic} {Passage} in {Two}- and
  {Three}-{Level} {Atoms}},\ }\href
  {https://doi.org/10.1103/PhysRevLett.105.123003} {\bibfield  {journal}
  {\bibinfo  {journal} {Physical Review Letters}\ }\textbf {\bibinfo {volume}
  {105}},\ \bibinfo {pages} {123003} (\bibinfo {year}
  {2010}{\natexlab{b}})}\BibitemShut {NoStop}%
\bibitem [{\citenamefont {Deffner}\ and\ \citenamefont
  {Campbell}(2017)}]{deffner_quantum_2017}%
  \BibitemOpen
  \bibfield  {author} {\bibinfo {author} {\bibfnamefont {S.}~\bibnamefont
  {Deffner}}\ and\ \bibinfo {author} {\bibfnamefont {S.}~\bibnamefont
  {Campbell}},\ }\bibfield  {title} {\bibinfo {title} {Quantum speed limits:
  from {Heisenberg}’s uncertainty principle to optimal quantum control},\
  }\href {https://doi.org/10.1088/1751-8121/aa86c6} {\bibfield  {journal}
  {\bibinfo  {journal} {Journal of Physics A: Mathematical and Theoretical}\
  }\textbf {\bibinfo {volume} {50}},\ \bibinfo {pages} {453001} (\bibinfo
  {year} {2017})}\BibitemShut {NoStop}%
\bibitem [{\citenamefont {Guéry-Odelin}\ \emph {et~al.}(2019)\citenamefont
  {Guéry-Odelin}, \citenamefont {Ruschhaupt}, \citenamefont {Kiely},
  \citenamefont {Torrontegui}, \citenamefont {Martínez-Garaot},\ and\
  \citenamefont {Muga}}]{guery-odelin_shortcuts_2019}%
  \BibitemOpen
  \bibfield  {author} {\bibinfo {author} {\bibfnamefont {D.}~\bibnamefont
  {Guéry-Odelin}}, \bibinfo {author} {\bibfnamefont {A.}~\bibnamefont
  {Ruschhaupt}}, \bibinfo {author} {\bibfnamefont {A.}~\bibnamefont {Kiely}},
  \bibinfo {author} {\bibfnamefont {E.}~\bibnamefont {Torrontegui}}, \bibinfo
  {author} {\bibfnamefont {S.}~\bibnamefont {Martínez-Garaot}},\ and\ \bibinfo
  {author} {\bibfnamefont {J.}~\bibnamefont {Muga}},\ }\bibfield  {title}
  {\bibinfo {title} {Shortcuts to adiabaticity: {Concepts}, methods, and
  applications},\ }\href {https://doi.org/10.1103/RevModPhys.91.045001}
  {\bibfield  {journal} {\bibinfo  {journal} {Reviews of Modern Physics}\
  }\textbf {\bibinfo {volume} {91}},\ \bibinfo {pages} {045001} (\bibinfo
  {year} {2019})}\BibitemShut {NoStop}%
\bibitem [{\citenamefont {Schmiedl}\ and\ \citenamefont
  {Seifert}(2007)}]{schmiedl_optimal_2007}%
  \BibitemOpen
  \bibfield  {author} {\bibinfo {author} {\bibfnamefont {T.}~\bibnamefont
  {Schmiedl}}\ and\ \bibinfo {author} {\bibfnamefont {U.}~\bibnamefont
  {Seifert}},\ }\bibfield  {title} {\bibinfo {title} {Optimal {Finite}-{Time}
  {Processes} {In} {Stochastic} {Thermodynamics}},\ }\href
  {https://doi.org/10.1103/PhysRevLett.98.108301} {\bibfield  {journal}
  {\bibinfo  {journal} {Physical Review Letters}\ }\textbf {\bibinfo {volume}
  {98}},\ \bibinfo {pages} {108301} (\bibinfo {year} {2007})}\BibitemShut
  {NoStop}%
\bibitem [{\citenamefont {Aurell}\ \emph {et~al.}(2011)\citenamefont {Aurell},
  \citenamefont {Mejía-Monasterio},\ and\ \citenamefont
  {Muratore-Ginanneschi}}]{aurell_optimal_2011}%
  \BibitemOpen
  \bibfield  {author} {\bibinfo {author} {\bibfnamefont {E.}~\bibnamefont
  {Aurell}}, \bibinfo {author} {\bibfnamefont {C.}~\bibnamefont
  {Mejía-Monasterio}},\ and\ \bibinfo {author} {\bibfnamefont
  {P.}~\bibnamefont {Muratore-Ginanneschi}},\ }\bibfield  {title} {\bibinfo
  {title} {Optimal {Protocols} and {Optimal} {Transport} in {Stochastic}
  {Thermodynamics}},\ }\href {https://doi.org/10.1103/PhysRevLett.106.250601}
  {\bibfield  {journal} {\bibinfo  {journal} {Physical Review Letters}\
  }\textbf {\bibinfo {volume} {106}},\ \bibinfo {pages} {250601} (\bibinfo
  {year} {2011})}\BibitemShut {NoStop}%
\bibitem [{\citenamefont {Machta}(2015)}]{machta_dissipation_2015}%
  \BibitemOpen
  \bibfield  {author} {\bibinfo {author} {\bibfnamefont {B.~B.}\ \bibnamefont
  {Machta}},\ }\bibfield  {title} {\bibinfo {title} {A dissipation bound for
  thermodynamic control},\ }\href@noop {} {\bibfield  {journal} {\bibinfo
  {journal} {arXiv preprint arXiv:1508.04150}\ } (\bibinfo {year}
  {2015})}\BibitemShut {NoStop}%
\bibitem [{\citenamefont {Martínez}\ \emph
  {et~al.}(2016{\natexlab{a}})\citenamefont {Martínez}, \citenamefont
  {Roldán}, \citenamefont {Dinis}, \citenamefont {Petrov}, \citenamefont
  {Parrondo},\ and\ \citenamefont {Rica}}]{martinez_brownian_2016}%
  \BibitemOpen
  \bibfield  {author} {\bibinfo {author} {\bibfnamefont {I.~A.}\ \bibnamefont
  {Martínez}}, \bibinfo {author} {\bibfnamefont {E.}~\bibnamefont {Roldán}},
  \bibinfo {author} {\bibfnamefont {L.}~\bibnamefont {Dinis}}, \bibinfo
  {author} {\bibfnamefont {D.}~\bibnamefont {Petrov}}, \bibinfo {author}
  {\bibfnamefont {J.~M.~R.}\ \bibnamefont {Parrondo}},\ and\ \bibinfo {author}
  {\bibfnamefont {R.~A.}\ \bibnamefont {Rica}},\ }\bibfield  {title} {\bibinfo
  {title} {Brownian {Carnot} engine},\ }\href
  {https://doi.org/10.1038/nphys3518} {\bibfield  {journal} {\bibinfo
  {journal} {Nature Physics}\ }\textbf {\bibinfo {volume} {12}},\ \bibinfo
  {pages} {67} (\bibinfo {year} {2016}{\natexlab{a}})}\BibitemShut {NoStop}%
\bibitem [{\citenamefont {Muratore-Ginanneschi}\ and\ \citenamefont
  {Schwieger}(2017)}]{muratore-ginanneschi_application_2017}%
  \BibitemOpen
  \bibfield  {author} {\bibinfo {author} {\bibfnamefont {P.}~\bibnamefont
  {Muratore-Ginanneschi}}\ and\ \bibinfo {author} {\bibfnamefont
  {K.}~\bibnamefont {Schwieger}},\ }\bibfield  {title} {\bibinfo {title} {An
  {Application} of {Pontryagin}’s {Principle} to {Brownian} {Particle}
  {Engineered} {Equilibration}},\ }\href {https://doi.org/10.3390/e19070379}
  {\bibfield  {journal} {\bibinfo  {journal} {Entropy}\ }\textbf {\bibinfo
  {volume} {19}},\ \bibinfo {pages} {379} (\bibinfo {year} {2017})}\BibitemShut
  {NoStop}%
\bibitem [{\citenamefont {Van~Vu}\ and\ \citenamefont
  {Hasegawa}(2020)}]{van_vu_thermodynamic_2020}%
  \BibitemOpen
  \bibfield  {author} {\bibinfo {author} {\bibfnamefont {T.}~\bibnamefont
  {Van~Vu}}\ and\ \bibinfo {author} {\bibfnamefont {Y.}~\bibnamefont
  {Hasegawa}},\ }\bibfield  {title} {\bibinfo {title} {Thermodynamic
  uncertainty relations under arbitrary control protocols},\ }\href
  {https://doi.org/10.1103/PhysRevResearch.2.013060} {\bibfield  {journal}
  {\bibinfo  {journal} {Physical Review Research}\ }\textbf {\bibinfo {volume}
  {2}},\ \bibinfo {pages} {013060} (\bibinfo {year} {2020})}\BibitemShut
  {NoStop}%
\bibitem [{\citenamefont {Martínez}\ \emph
  {et~al.}(2016{\natexlab{b}})\citenamefont {Martínez}, \citenamefont
  {Petrosyan}, \citenamefont {Guéry-Odelin}, \citenamefont {Trizac},\ and\
  \citenamefont {Ciliberto}}]{martinez_engineered_2016}%
  \BibitemOpen
  \bibfield  {author} {\bibinfo {author} {\bibfnamefont {I.~A.}\ \bibnamefont
  {Martínez}}, \bibinfo {author} {\bibfnamefont {A.}~\bibnamefont
  {Petrosyan}}, \bibinfo {author} {\bibfnamefont {D.}~\bibnamefont
  {Guéry-Odelin}}, \bibinfo {author} {\bibfnamefont {E.}~\bibnamefont
  {Trizac}},\ and\ \bibinfo {author} {\bibfnamefont {S.}~\bibnamefont
  {Ciliberto}},\ }\bibfield  {title} {\bibinfo {title} {Engineered swift
  equilibration of a {Brownian} particle},\ }\href
  {https://doi.org/10.1038/nphys3758} {\bibfield  {journal} {\bibinfo
  {journal} {Nature Physics}\ }\textbf {\bibinfo {volume} {12}},\ \bibinfo
  {pages} {843} (\bibinfo {year} {2016}{\natexlab{b}})}\BibitemShut {NoStop}%
\bibitem [{\citenamefont {Plata}\ \emph {et~al.}(2019)\citenamefont {Plata},
  \citenamefont {Guéry-Odelin}, \citenamefont {Trizac},\ and\ \citenamefont
  {Prados}}]{plata_optimal_2019}%
  \BibitemOpen
  \bibfield  {author} {\bibinfo {author} {\bibfnamefont {C.~A.}\ \bibnamefont
  {Plata}}, \bibinfo {author} {\bibfnamefont {D.}~\bibnamefont
  {Guéry-Odelin}}, \bibinfo {author} {\bibfnamefont {E.}~\bibnamefont
  {Trizac}},\ and\ \bibinfo {author} {\bibfnamefont {A.}~\bibnamefont
  {Prados}},\ }\bibfield  {title} {\bibinfo {title} {Optimal work in a harmonic
  trap with bounded stiffness},\ }\href
  {https://doi.org/10.1103/PhysRevE.99.012140} {\bibfield  {journal} {\bibinfo
  {journal} {Physical Review E}\ }\textbf {\bibinfo {volume} {99}},\ \bibinfo
  {pages} {012140} (\bibinfo {year} {2019})}\BibitemShut {NoStop}%
\bibitem [{\citenamefont {Li}\ \emph {et~al.}(2017)\citenamefont {Li},
  \citenamefont {Quan},\ and\ \citenamefont {Tu}}]{li_shortcuts_2017}%
  \BibitemOpen
  \bibfield  {author} {\bibinfo {author} {\bibfnamefont {G.}~\bibnamefont
  {Li}}, \bibinfo {author} {\bibfnamefont {H.~T.}\ \bibnamefont {Quan}},\ and\
  \bibinfo {author} {\bibfnamefont {Z.~C.}\ \bibnamefont {Tu}},\ }\bibfield
  {title} {\bibinfo {title} {Shortcuts to isothermality and nonequilibrium work
  relations},\ }\href {https://doi.org/10.1103/PhysRevE.96.012144} {\bibfield
  {journal} {\bibinfo  {journal} {Physical Review E}\ }\textbf {\bibinfo
  {volume} {96}},\ \bibinfo {pages} {012144} (\bibinfo {year}
  {2017})}\BibitemShut {NoStop}%
\bibitem [{\citenamefont {Chupeau}\ \emph
  {et~al.}(2018{\natexlab{a}})\citenamefont {Chupeau}, \citenamefont
  {Ciliberto}, \citenamefont {Guéry-Odelin},\ and\ \citenamefont
  {Trizac}}]{chupeau_engineered_2018}%
  \BibitemOpen
  \bibfield  {author} {\bibinfo {author} {\bibfnamefont {M.}~\bibnamefont
  {Chupeau}}, \bibinfo {author} {\bibfnamefont {S.}~\bibnamefont {Ciliberto}},
  \bibinfo {author} {\bibfnamefont {D.}~\bibnamefont {Guéry-Odelin}},\ and\
  \bibinfo {author} {\bibfnamefont {E.}~\bibnamefont {Trizac}},\ }\bibfield
  {title} {\bibinfo {title} {Engineered swift equilibration for {Brownian}
  objects: from underdamped to overdamped dynamics},\ }\href
  {https://doi.org/10.1088/1367-2630/aac875} {\bibfield  {journal} {\bibinfo
  {journal} {New Journal of Physics}\ }\textbf {\bibinfo {volume} {20}},\
  \bibinfo {pages} {075003} (\bibinfo {year} {2018}{\natexlab{a}})}\BibitemShut
  {NoStop}%
\bibitem [{\citenamefont {Albay}\ \emph {et~al.}(2019)\citenamefont {Albay},
  \citenamefont {Wulaningrum}, \citenamefont {Kwon}, \citenamefont {Lai},\ and\
  \citenamefont {Jun}}]{albay_thermodynamic_2019}%
  \BibitemOpen
  \bibfield  {author} {\bibinfo {author} {\bibfnamefont {J.~A.~C.}\
  \bibnamefont {Albay}}, \bibinfo {author} {\bibfnamefont {S.~R.}\ \bibnamefont
  {Wulaningrum}}, \bibinfo {author} {\bibfnamefont {C.}~\bibnamefont {Kwon}},
  \bibinfo {author} {\bibfnamefont {P.-Y.}\ \bibnamefont {Lai}},\ and\ \bibinfo
  {author} {\bibfnamefont {Y.}~\bibnamefont {Jun}},\ }\bibfield  {title}
  {\bibinfo {title} {Thermodynamic cost of a shortcuts-to-isothermal transport
  of a {Brownian} particle},\ }\href
  {https://doi.org/10.1103/PhysRevResearch.1.033122} {\bibfield  {journal}
  {\bibinfo  {journal} {Physical Review Research}\ }\textbf {\bibinfo {volume}
  {1}},\ \bibinfo {pages} {033122} (\bibinfo {year} {2019})}\BibitemShut
  {NoStop}%
\bibitem [{\citenamefont {Albay}\ \emph {et~al.}(2020)\citenamefont {Albay},
  \citenamefont {Lai},\ and\ \citenamefont {Jun}}]{albay_realization_2020}%
  \BibitemOpen
  \bibfield  {author} {\bibinfo {author} {\bibfnamefont {J.~A.~C.}\
  \bibnamefont {Albay}}, \bibinfo {author} {\bibfnamefont {P.-Y.}\ \bibnamefont
  {Lai}},\ and\ \bibinfo {author} {\bibfnamefont {Y.}~\bibnamefont {Jun}},\
  }\bibfield  {title} {\bibinfo {title} {Realization of finite-rate isothermal
  compression and expansion using optical feedback trap},\ }\href
  {https://doi.org/10.1063/1.5143602} {\bibfield  {journal} {\bibinfo
  {journal} {Applied Physics Letters}\ }\textbf {\bibinfo {volume} {116}},\
  \bibinfo {pages} {103706} (\bibinfo {year} {2020})}\BibitemShut {NoStop}%
\bibitem [{\citenamefont {Plata}\ \emph
  {et~al.}(2020{\natexlab{a}})\citenamefont {Plata}, \citenamefont
  {Guéry-Odelin}, \citenamefont {Trizac},\ and\ \citenamefont
  {Prados}}]{plata_finite-time_2020}%
  \BibitemOpen
  \bibfield  {author} {\bibinfo {author} {\bibfnamefont {C.~A.}\ \bibnamefont
  {Plata}}, \bibinfo {author} {\bibfnamefont {D.}~\bibnamefont
  {Guéry-Odelin}}, \bibinfo {author} {\bibfnamefont {E.}~\bibnamefont
  {Trizac}},\ and\ \bibinfo {author} {\bibfnamefont {A.}~\bibnamefont
  {Prados}},\ }\bibfield  {title} {\bibinfo {title} {Finite-time adiabatic
  processes: {Derivation} and speed limit},\ }\href
  {https://doi.org/10.1103/PhysRevE.101.032129} {\bibfield  {journal} {\bibinfo
   {journal} {Physical Review E}\ }\textbf {\bibinfo {volume} {101}},\ \bibinfo
  {pages} {032129} (\bibinfo {year} {2020}{\natexlab{a}})}\BibitemShut
  {NoStop}%
\bibitem [{\citenamefont {Mart\'{\i}nez}\ \emph {et~al.}(2013)\citenamefont
  {Mart\'{\i}nez}, \citenamefont {Rold\'an}, \citenamefont {Parrondo},\ and\
  \citenamefont {Petrov}}]{martinez_effective_2013}%
  \BibitemOpen
  \bibfield  {author} {\bibinfo {author} {\bibfnamefont {I.~A.}\ \bibnamefont
  {Mart\'{\i}nez}}, \bibinfo {author} {\bibfnamefont {E.}~\bibnamefont
  {Rold\'an}}, \bibinfo {author} {\bibfnamefont {J.~M.~R.}\ \bibnamefont
  {Parrondo}},\ and\ \bibinfo {author} {\bibfnamefont {D.}~\bibnamefont
  {Petrov}},\ }\bibfield  {title} {\bibinfo {title} {Effective heating to
  several thousand kelvins of an optically trapped sphere in a liquid},\
  }\href@noop {} {\bibfield  {journal} {\bibinfo  {journal} {Physical Review
  E}\ }\textbf {\bibinfo {volume} {87}},\ \bibinfo {pages} {032159} (\bibinfo
  {year} {2013})}\BibitemShut {NoStop}%
\bibitem [{\citenamefont {Ciliberto}(2017)}]{ciliberto_experiments_2017}%
  \BibitemOpen
  \bibfield  {author} {\bibinfo {author} {\bibfnamefont {S.}~\bibnamefont
  {Ciliberto}},\ }\bibfield  {title} {\bibinfo {title} {Experiments in
  {Stochastic} {Thermodynamics}: {Short} {History} and {Perspectives}},\ }\href
  {https://doi.org/10.1103/PhysRevX.7.021051} {\bibfield  {journal} {\bibinfo
  {journal} {Physical Review X}\ }\textbf {\bibinfo {volume} {7}},\ \bibinfo
  {pages} {021051} (\bibinfo {year} {2017})}\BibitemShut {NoStop}%
\bibitem [{\citenamefont {Filliger}\ and\ \citenamefont
  {Reimann}(2007)}]{filliger_brownian_2007}%
  \BibitemOpen
  \bibfield  {author} {\bibinfo {author} {\bibfnamefont {R.}~\bibnamefont
  {Filliger}}\ and\ \bibinfo {author} {\bibfnamefont {P.}~\bibnamefont
  {Reimann}},\ }\bibfield  {title} {\bibinfo {title} {Brownian {Gyrator}: {A}
  {Minimal} {Heat} {Engine} on the {Nanoscale}},\ }\href
  {https://doi.org/10.1103/PhysRevLett.99.230602} {\bibfield  {journal}
  {\bibinfo  {journal} {Physical Review Letters}\ }\textbf {\bibinfo {volume}
  {99}},\ \bibinfo {pages} {230602} (\bibinfo {year} {2007})}\BibitemShut
  {NoStop}%
\bibitem [{\citenamefont {Argun}\ \emph {et~al.}(2017)\citenamefont {Argun},
  \citenamefont {Soni}, \citenamefont {Dabelow}, \citenamefont {Bo},
  \citenamefont {Pesce}, \citenamefont {Eichhorn},\ and\ \citenamefont
  {Volpe}}]{argun_experimental_2017}%
  \BibitemOpen
  \bibfield  {author} {\bibinfo {author} {\bibfnamefont {A.}~\bibnamefont
  {Argun}}, \bibinfo {author} {\bibfnamefont {J.}~\bibnamefont {Soni}},
  \bibinfo {author} {\bibfnamefont {L.}~\bibnamefont {Dabelow}}, \bibinfo
  {author} {\bibfnamefont {S.}~\bibnamefont {Bo}}, \bibinfo {author}
  {\bibfnamefont {G.}~\bibnamefont {Pesce}}, \bibinfo {author} {\bibfnamefont
  {R.}~\bibnamefont {Eichhorn}},\ and\ \bibinfo {author} {\bibfnamefont
  {G.}~\bibnamefont {Volpe}},\ }\bibfield  {title} {\bibinfo {title}
  {Experimental realization of a minimal microscopic heat engine},\ }\href
  {https://doi.org/10.1103/PhysRevE.96.052106} {\bibfield  {journal} {\bibinfo
  {journal} {Physical Review E}\ }\textbf {\bibinfo {volume} {96}},\ \bibinfo
  {pages} {052106} (\bibinfo {year} {2017})}\BibitemShut {NoStop}%
\bibitem [{\citenamefont {Chiang}\ \emph {et~al.}(2017)\citenamefont {Chiang},
  \citenamefont {Lee}, \citenamefont {Lai},\ and\ \citenamefont
  {Chen}}]{chiang_electrical_2017}%
  \BibitemOpen
  \bibfield  {author} {\bibinfo {author} {\bibfnamefont {K.-H.}\ \bibnamefont
  {Chiang}}, \bibinfo {author} {\bibfnamefont {C.-L.}\ \bibnamefont {Lee}},
  \bibinfo {author} {\bibfnamefont {P.-Y.}\ \bibnamefont {Lai}},\ and\ \bibinfo
  {author} {\bibfnamefont {Y.-F.}\ \bibnamefont {Chen}},\ }\bibfield  {title}
  {\bibinfo {title} {Electrical autonomous {Brownian} gyrator},\ }\href
  {https://doi.org/10.1103/PhysRevE.96.032123} {\bibfield  {journal} {\bibinfo
  {journal} {Physical Review E}\ }\textbf {\bibinfo {volume} {96}},\ \bibinfo
  {pages} {032123} (\bibinfo {year} {2017})}\BibitemShut {NoStop}%
\bibitem [{\citenamefont {Baldassarri}\ \emph {et~al.}(2020)\citenamefont
  {Baldassarri}, \citenamefont {Puglisi},\ and\ \citenamefont
  {Sesta}}]{baldassarri_engineered_2020}%
  \BibitemOpen
  \bibfield  {author} {\bibinfo {author} {\bibfnamefont {A.}~\bibnamefont
  {Baldassarri}}, \bibinfo {author} {\bibfnamefont {A.}~\bibnamefont
  {Puglisi}},\ and\ \bibinfo {author} {\bibfnamefont {L.}~\bibnamefont
  {Sesta}},\ }\bibfield  {title} {\bibinfo {title} {Engineered swift
  equilibration of a {Brownian} gyrator},\ }\href
  {https://doi.org/10.1103/PhysRevE.102.030105} {\bibfield  {journal} {\bibinfo
   {journal} {Physical Review E}\ }\textbf {\bibinfo {volume} {102}},\ \bibinfo
  {pages} {030105} (\bibinfo {year} {2020})}\BibitemShut {NoStop}%
\bibitem [{\citenamefont {Plata}\ \emph
  {et~al.}(2020{\natexlab{b}})\citenamefont {Plata}, \citenamefont
  {Guéry-Odelin}, \citenamefont {Trizac},\ and\ \citenamefont
  {Prados}}]{plata_building_2020}%
  \BibitemOpen
  \bibfield  {author} {\bibinfo {author} {\bibfnamefont {C.~A.}\ \bibnamefont
  {Plata}}, \bibinfo {author} {\bibfnamefont {D.}~\bibnamefont
  {Guéry-Odelin}}, \bibinfo {author} {\bibfnamefont {E.}~\bibnamefont
  {Trizac}},\ and\ \bibinfo {author} {\bibfnamefont {A.}~\bibnamefont
  {Prados}},\ }\bibfield  {title} {\bibinfo {title} {Building an irreversible
  {Carnot}-like heat engine with an overdamped harmonic oscillator},\ }\href
  {https://doi.org/10.1088/1742-5468/abb0e1} {\bibfield  {journal} {\bibinfo
  {journal} {Journal of Statistical Mechanics: Theory and Experiment}\ }\textbf
  {\bibinfo {volume} {2020}},\ \bibinfo {pages} {093207} (\bibinfo {year}
  {2020}{\natexlab{b}})}\BibitemShut {NoStop}%
\bibitem [{\citenamefont {Lu}\ and\ \citenamefont
  {Raz}(2017)}]{lu_nonequilibrium_2017}%
  \BibitemOpen
  \bibfield  {author} {\bibinfo {author} {\bibfnamefont {Z.}~\bibnamefont
  {Lu}}\ and\ \bibinfo {author} {\bibfnamefont {O.}~\bibnamefont {Raz}},\
  }\bibfield  {title} {\bibinfo {title} {Nonequilibrium thermodynamics of the
  {Markovian} {Mpemba} effect and its inverse},\ }\href
  {https://doi.org/10.1073/pnas.1701264114} {\bibfield  {journal} {\bibinfo
  {journal} {Proceedings of the National Academy of Sciences}\ }\textbf
  {\bibinfo {volume} {114}},\ \bibinfo {pages} {5083} (\bibinfo {year}
  {2017})}\BibitemShut {NoStop}%
\bibitem [{\citenamefont {Baity-Jesi}\ \emph {et~al.}(2019)\citenamefont
  {Baity-Jesi}, \citenamefont {Calore}, \citenamefont {Cruz}, \citenamefont
  {Fernandez}, \citenamefont {Gil-Narvión}, \citenamefont {Gordillo-Guerrero},
  \citenamefont {Iñiguez}, \citenamefont {Lasanta}, \citenamefont {Maiorano},
  \citenamefont {Marinari}, \citenamefont {Martin-Mayor}, \citenamefont
  {Moreno-Gordo}, \citenamefont {Muñoz~Sudupe}, \citenamefont {Navarro},
  \citenamefont {Parisi}, \citenamefont {Perez-Gaviro}, \citenamefont
  {Ricci-Tersenghi}, \citenamefont {Ruiz-Lorenzo}, \citenamefont {Schifano},
  \citenamefont {Seoane}, \citenamefont {Tarancón}, \citenamefont
  {Tripiccione},\ and\ \citenamefont {Yllanes}}]{baity-jesi_mpemba_2019}%
  \BibitemOpen
  \bibfield  {author} {\bibinfo {author} {\bibfnamefont {M.}~\bibnamefont
  {Baity-Jesi}}, \bibinfo {author} {\bibfnamefont {E.}~\bibnamefont {Calore}},
  \bibinfo {author} {\bibfnamefont {A.}~\bibnamefont {Cruz}}, \bibinfo {author}
  {\bibfnamefont {L.~A.}\ \bibnamefont {Fernandez}}, \bibinfo {author}
  {\bibfnamefont {J.~M.}\ \bibnamefont {Gil-Narvión}}, \bibinfo {author}
  {\bibfnamefont {A.}~\bibnamefont {Gordillo-Guerrero}}, \bibinfo {author}
  {\bibfnamefont {D.}~\bibnamefont {Iñiguez}}, \bibinfo {author}
  {\bibfnamefont {A.}~\bibnamefont {Lasanta}}, \bibinfo {author} {\bibfnamefont
  {A.}~\bibnamefont {Maiorano}}, \bibinfo {author} {\bibfnamefont
  {E.}~\bibnamefont {Marinari}}, \bibinfo {author} {\bibfnamefont
  {V.}~\bibnamefont {Martin-Mayor}}, \bibinfo {author} {\bibfnamefont
  {J.}~\bibnamefont {Moreno-Gordo}}, \bibinfo {author} {\bibfnamefont
  {A.}~\bibnamefont {Muñoz~Sudupe}}, \bibinfo {author} {\bibfnamefont
  {D.}~\bibnamefont {Navarro}}, \bibinfo {author} {\bibfnamefont
  {G.}~\bibnamefont {Parisi}}, \bibinfo {author} {\bibfnamefont
  {S.}~\bibnamefont {Perez-Gaviro}}, \bibinfo {author} {\bibfnamefont
  {F.}~\bibnamefont {Ricci-Tersenghi}}, \bibinfo {author} {\bibfnamefont
  {J.~J.}\ \bibnamefont {Ruiz-Lorenzo}}, \bibinfo {author} {\bibfnamefont
  {S.~F.}\ \bibnamefont {Schifano}}, \bibinfo {author} {\bibfnamefont
  {B.}~\bibnamefont {Seoane}}, \bibinfo {author} {\bibfnamefont
  {A.}~\bibnamefont {Tarancón}}, \bibinfo {author} {\bibfnamefont
  {R.}~\bibnamefont {Tripiccione}},\ and\ \bibinfo {author} {\bibfnamefont
  {D.}~\bibnamefont {Yllanes}},\ }\bibfield  {title} {\bibinfo {title} {The
  {Mpemba} effect in spin glasses is a persistent memory effect},\ }\href
  {https://doi.org/10.1073/pnas.1819803116} {\bibfield  {journal} {\bibinfo
  {journal} {Proceedings of the National Academy of Sciences}\ }\textbf
  {\bibinfo {volume} {116}},\ \bibinfo {pages} {15350} (\bibinfo {year}
  {2019})}\BibitemShut {NoStop}%
\bibitem [{\citenamefont {Santos}\ and\ \citenamefont
  {Prados}(2020)}]{santos_mpemba_2020}%
  \BibitemOpen
  \bibfield  {author} {\bibinfo {author} {\bibfnamefont {A.}~\bibnamefont
  {Santos}}\ and\ \bibinfo {author} {\bibfnamefont {A.}~\bibnamefont
  {Prados}},\ }\bibfield  {title} {\bibinfo {title} {Mpemba effect in molecular
  gases under nonlinear drag},\ }\href {https://doi.org/10.1063/5.0016243}
  {\bibfield  {journal} {\bibinfo  {journal} {Physics of Fluids}\ }\textbf
  {\bibinfo {volume} {32}},\ \bibinfo {pages} {072010} (\bibinfo {year}
  {2020})}\BibitemShut {NoStop}%
\bibitem [{\citenamefont {Gal}\ and\ \citenamefont
  {Raz}(2020)}]{gal_precooling_2020}%
  \BibitemOpen
  \bibfield  {author} {\bibinfo {author} {\bibfnamefont {A.}~\bibnamefont
  {Gal}}\ and\ \bibinfo {author} {\bibfnamefont {O.}~\bibnamefont {Raz}},\
  }\bibfield  {title} {\bibinfo {title} {Precooling {Strategy} {Allows}
  {Exponentially} {Faster} {Heating}},\ }\href
  {https://doi.org/10.1103/PhysRevLett.124.060602} {\bibfield  {journal}
  {\bibinfo  {journal} {Physical Review Letters}\ }\textbf {\bibinfo {volume}
  {124}},\ \bibinfo {pages} {060602} (\bibinfo {year} {2020})}\BibitemShut
  {NoStop}%
\bibitem [{\citenamefont {Kumar}\ and\ \citenamefont
  {Bechhoefer}(2020)}]{kumar_exponentially_2020}%
  \BibitemOpen
  \bibfield  {author} {\bibinfo {author} {\bibfnamefont {A.}~\bibnamefont
  {Kumar}}\ and\ \bibinfo {author} {\bibfnamefont {J.}~\bibnamefont
  {Bechhoefer}},\ }\bibfield  {title} {\bibinfo {title} {Exponentially faster
  cooling in a colloidal system},\ }\href
  {http://www.nature.com/articles/s41586-020-2560-x} {\bibfield  {journal}
  {\bibinfo  {journal} {Nature}\ }\textbf {\bibinfo {volume} {584}},\ \bibinfo
  {pages} {64} (\bibinfo {year} {2020})}\BibitemShut {NoStop}%
\bibitem [{\citenamefont {Lapolla}\ and\ \citenamefont
  {Godec}(2020)}]{lapolla_faster_2020}%
  \BibitemOpen
  \bibfield  {author} {\bibinfo {author} {\bibfnamefont {A.}~\bibnamefont
  {Lapolla}}\ and\ \bibinfo {author} {\bibfnamefont {A.}~\bibnamefont
  {Godec}},\ }\bibfield  {title} {\bibinfo {title} {Faster {Uphill}
  {Relaxation} in {Thermodynamically} {Equidistant} {Temperature} {Quenches}},\
  }\href {https://link.aps.org/doi/10.1103/PhysRevLett.125.110602} {\bibfield
  {journal} {\bibinfo  {journal} {Physical Review Letters}\ }\textbf {\bibinfo
  {volume} {125}},\ \bibinfo {pages} {110602} (\bibinfo {year}
  {2020})}\BibitemShut {NoStop}%
\bibitem [{Note1()}]{Note1}%
  \BibitemOpen
  \bibinfo {note} {Both $a_2^{\protect \text {HCS}}$ and $a_2^{\protect \text
  {s}}$ change sign for $ \alpha =1/ \protect \sqrt {2}$, so that $a_2$
  typically changes sign with the inelasticity. On the other hand, the scaled
  variable $A_2$ always remains positive.}\BibitemShut {Stop}%
\bibitem [{\citenamefont {Pontryagin}(1987)}]{pontryagin_mathematical_1987}%
  \BibitemOpen
  \bibfield  {author} {\bibinfo {author} {\bibfnamefont {L.~S.}\ \bibnamefont
  {Pontryagin}},\ }\href@noop {} {\emph {\bibinfo {title} {Mathematical
  {Theory} of {Optimal} {Processes}}}}\ (\bibinfo  {publisher} {CRC Press},\
  \bibinfo {year} {1987})\BibitemShut {NoStop}%
\bibitem [{Note2()}]{Note2}%
  \BibitemOpen
  \bibinfo {note} {More specifically, this result stems from theorem 10 in
  section 17 of Ref.~\cite {pontryagin_mathematical_1987}, We check that the
  hypotheses of this theorem are fulfilled in Appendix~\ref
  {sec:verifying-hypothesis}.}\BibitemShut {Stop}%
\bibitem [{\citenamefont {Liberzon}(2012)}]{liberzon_calculus_2012}%
  \BibitemOpen
  \bibfield  {author} {\bibinfo {author} {\bibfnamefont {D.}~\bibnamefont
  {Liberzon}},\ }\href@noop {} {\emph {\bibinfo {title} {Calculus of
  {Variations} and {Optimal} {Control} {Theory}: {A} {Concise}
  {Introduction}}}}\ (\bibinfo  {publisher} {Princeton University Press},\
  \bibinfo {year} {2012})\BibitemShut {NoStop}%
\bibitem [{\citenamefont {Ding}\ \emph {et~al.}(2020)\citenamefont {Ding},
  \citenamefont {Huang}, \citenamefont {Paul}, \citenamefont {Hao},\ and\
  \citenamefont {Chen}}]{ding_smooth_2020}%
  \BibitemOpen
  \bibfield  {author} {\bibinfo {author} {\bibfnamefont {Y.}~\bibnamefont
  {Ding}}, \bibinfo {author} {\bibfnamefont {T.-Y.}\ \bibnamefont {Huang}},
  \bibinfo {author} {\bibfnamefont {K.}~\bibnamefont {Paul}}, \bibinfo {author}
  {\bibfnamefont {M.}~\bibnamefont {Hao}},\ and\ \bibinfo {author}
  {\bibfnamefont {X.}~\bibnamefont {Chen}},\ }\bibfield  {title} {\bibinfo
  {title} {Smooth bang-bang shortcuts to adiabaticity for atomic transport in a
  moving harmonic trap},\ }\href {https://doi.org/10.1103/PhysRevA.101.063410}
  {\bibfield  {journal} {\bibinfo  {journal} {Physical Review A}\ }\textbf
  {\bibinfo {volume} {101}},\ \bibinfo {pages} {063410} (\bibinfo {year}
  {2020})}\BibitemShut {NoStop}%
\bibitem [{\citenamefont {Martikyan}\ \emph {et~al.}(2020)\citenamefont
  {Martikyan}, \citenamefont {Guéry-Odelin},\ and\ \citenamefont
  {Sugny}}]{martikyan_comparison_2020}%
  \BibitemOpen
  \bibfield  {author} {\bibinfo {author} {\bibfnamefont {V.}~\bibnamefont
  {Martikyan}}, \bibinfo {author} {\bibfnamefont {D.}~\bibnamefont
  {Guéry-Odelin}},\ and\ \bibinfo {author} {\bibfnamefont {D.}~\bibnamefont
  {Sugny}},\ }\bibfield  {title} {\bibinfo {title} {Comparison between optimal
  control and shortcut to adiabaticity protocols in a linear control system},\
  }\href {https://doi.org/10.1103/PhysRevA.101.013423} {\bibfield  {journal}
  {\bibinfo  {journal} {Physical Review A}\ }\textbf {\bibinfo {volume}
  {101}},\ \bibinfo {pages} {013423} (\bibinfo {year} {2020})}\BibitemShut
  {NoStop}%
\bibitem [{\citenamefont {Chupeau}\ \emph
  {et~al.}(2018{\natexlab{b}})\citenamefont {Chupeau}, \citenamefont {Besga},
  \citenamefont {Guéry-Odelin}, \citenamefont {Trizac}, \citenamefont
  {Petrosyan},\ and\ \citenamefont {Ciliberto}}]{chupeau_thermal_2018}%
  \BibitemOpen
  \bibfield  {author} {\bibinfo {author} {\bibfnamefont {M.}~\bibnamefont
  {Chupeau}}, \bibinfo {author} {\bibfnamefont {B.}~\bibnamefont {Besga}},
  \bibinfo {author} {\bibfnamefont {D.}~\bibnamefont {Guéry-Odelin}}, \bibinfo
  {author} {\bibfnamefont {E.}~\bibnamefont {Trizac}}, \bibinfo {author}
  {\bibfnamefont {A.}~\bibnamefont {Petrosyan}},\ and\ \bibinfo {author}
  {\bibfnamefont {S.}~\bibnamefont {Ciliberto}},\ }\bibfield  {title} {\bibinfo
  {title} {Thermal bath engineering for swift equilibration},\ }\href
  {https://doi.org/10.1103/PhysRevE.98.010104} {\bibfield  {journal} {\bibinfo
  {journal} {Physical Review E}\ }\textbf {\bibinfo {volume} {98}},\ \bibinfo
  {pages} {010104} (\bibinfo {year} {2018}{\natexlab{b}})}\BibitemShut
  {NoStop}%
\bibitem [{\citenamefont {Kourbane-Houssene}\ \emph {et~al.}(2018)\citenamefont
  {Kourbane-Houssene}, \citenamefont {Erignoux}, \citenamefont {Bodineau},\
  and\ \citenamefont {Tailleur}}]{kourbane-houssene_exact_2018}%
  \BibitemOpen
  \bibfield  {author} {\bibinfo {author} {\bibfnamefont {M.}~\bibnamefont
  {Kourbane-Houssene}}, \bibinfo {author} {\bibfnamefont {C.}~\bibnamefont
  {Erignoux}}, \bibinfo {author} {\bibfnamefont {T.}~\bibnamefont {Bodineau}},\
  and\ \bibinfo {author} {\bibfnamefont {J.}~\bibnamefont {Tailleur}},\
  }\bibfield  {title} {\bibinfo {title} {Exact {Hydrodynamic} {Description} of
  {Active} {Lattice} {Gases}},\ }\href
  {https://doi.org/10.1103/PhysRevLett.120.268003} {\bibfield  {journal}
  {\bibinfo  {journal} {Physical Review Letters}\ }\textbf {\bibinfo {volume}
  {120}},\ \bibinfo {pages} {268003} (\bibinfo {year} {2018})}\BibitemShut
  {NoStop}%
\bibitem [{\citenamefont {Manacorda}\ and\ \citenamefont
  {Puglisi}(2017)}]{manacorda_lattice_2017}%
  \BibitemOpen
  \bibfield  {author} {\bibinfo {author} {\bibfnamefont {A.}~\bibnamefont
  {Manacorda}}\ and\ \bibinfo {author} {\bibfnamefont {A.}~\bibnamefont
  {Puglisi}},\ }\bibfield  {title} {\bibinfo {title} {Lattice {Model} to
  {Derive} the {Fluctuating} {Hydrodynamics} of {Active} {Particles} with
  {Inertia}},\ }\href {https://doi.org/10.1103/PhysRevLett.119.208003}
  {\bibfield  {journal} {\bibinfo  {journal} {Physical Review Letters}\
  }\textbf {\bibinfo {volume} {119}},\ \bibinfo {pages} {208003} (\bibinfo
  {year} {2017})}\BibitemShut {NoStop}%
\bibitem [{Note3()}]{Note3}%
  \BibitemOpen
  \bibinfo {note} {It must be remarked that the non-dimensionalisation of time
  in Ref.~\cite {prados_optimizing_2021}, $t^*=\zeta _0 T_{\protect \text
  {i}}^{1/2} t$, differs from ours in Eq.~\protect \textup {\hbox
  {\mathsurround \z@ \protect \normalfont (\ignorespaces \ref
  {eq:scaled-vars}\unskip \@@italiccorr )}} by a factor $\protect \sqrt
  {T_{\protect \text {i}}/T_{\protect \text {f}}}$. This factor does not affect
  the lowest order asymptotic expression in Eq.~\protect \textup {\hbox
  {\mathsurround \z@ \protect \normalfont (\ignorespaces \ref
  {eq:tf-teo-unbounded}\unskip \@@italiccorr )}}, since the introduced
  corrections are higher-order.}\BibitemShut {Stop}%
\bibitem [{Note4()}]{Note4}%
  \BibitemOpen
  \bibinfo {note} {It could be argued that, still, the most relevant physical
  situation corresponds to the case $T_{\protect \text {i}}\in [T_{\protect
  \qopname \relax m{min}},T_{\protect \qopname \relax m{max}}]$, because one
  needs to prepare the system in the initial NESS.}\BibitemShut {Stop}%
\bibitem [{Note5()}]{Note5}%
  \BibitemOpen
  \bibinfo {note} {Physically, it is evident that the connection time cannot be
  shorter than the switching time, $t_{\protect \text {f}}\geq t_J$.
  Mathematically, the divergence of the numerator always wins because $\lambda
  _1 > \lambda _2$ and $t_{\protect \text {f}} \sim t_J$; the time spent in the
  second part of the bang-bang becomes negligible as compared with
  $t_J$.}\BibitemShut {Stop}%
\end{thebibliography}%

\end{document}